\documentclass[sn-mathphys,Numbered]{sn-jnl}
\usepackage{appendix}
\usepackage{import}
\usepackage{float}
\usepackage{amsmath,amsfonts,amssymb,amsbsy,amsthm}
\usepackage{mathtools}
\usepackage{hyperref}
\usepackage{subfig}
\usepackage{multirow}
\usepackage{multicol}
\usepackage{etoolbox}
\usepackage{algorithm}
\usepackage[noend]{algpseudocode}
\usepackage[dvipsnames]{xcolor}
\usepackage{paralist}
\allowdisplaybreaks
\usepackage{booktabs}
\usepackage[group-separator={,}]{siunitx}
\usepackage{textcomp}
\usepackage{mathrsfs}
\usepackage{ifpdf}
\usepackage{lmodern}
\usepackage[LY1]{fontenc}

\usepackage{xcolor}%
\usepackage{manyfoot}%
\usepackage{booktabs}%
\usepackage{algorithmicx}%
\usepackage{algpseudocode}%
\usepackage{listings}%

\usepackage{makecell}

\lstdefinestyle{customcpp}{
  backgroundcolor=\color{gray!10},  
  basicstyle=\scriptsize\ttfamily,              
  breaklines=true,                  
  captionpos=b,                     
  commentstyle=\color{green!80!black},       
  keywordstyle=\color{blue},        
  stringstyle=\color{red},          
  language=C++,                     
}

\lstdefinestyle{custompython}{
  backgroundcolor=\color{gray!10},  
  basicstyle=\scriptsize\ttfamily,              
  breaklines=true,                  
  captionpos=b,                     
  commentstyle=\color{green!80!black},       
  keywordstyle=\color{blue},        
  stringstyle=\color{red},          
  language=Python,
}

\usepackage{graphicx}

\usepackage{caption}

\usepackage[english]{babel} 
\addto\extrasenglish{

}

\newcommand{\SGU}[1]{{{[SU]}}}

\hypersetup{hidelinks=true}

\renewcommand{\arraystretch}{3}

\definecolor{arsenic}{rgb}{0.23, 0.27, 0.29}
\definecolor{charcoal}{rgb}{0.21, 0.27, 0.31}
\definecolor{hanblue}{rgb}{0.27, 0.42, 0.81}
\definecolor{blue-ncs}{rgb}{0.0, 0.53, 0.74}
\definecolor{awesome}{rgb}{1.0, 0.13,0.32}
\definecolor{darkgreen}{rgb}{0, .4,0}

\newcommand{\vect}[1]{{\boldsymbol{\mathbf{#1}}}}

\newcommand{\sbelCode}[1]{\textsf{#1}}

\theoremstyle{thmstyleone}%
\theoremstyle{thmstyletwo}%

\theoremstyle{thmstylethree}%
\usepackage{soul}
\usepackage{xcolor}

\newcommand\Cpp{C\texttt{++}}
\newcommand\DEME{DEM-Engine}
\newcommand\Starccm{STAR-CCM\texttt{+}}
\DeclareMathOperator{\sign}{sign}

\begin{document}

\title[Chrono DEM-Engine: A Discrete Element Method dual-GPU simulator with customizable contact forces and element shape]{Chrono DEM-Engine: A Discrete Element Method dual-GPU simulator with customizable contact forces and element shape}

\author[1]{\fnm{Ruochun} \sur{Zhang}}\email{rzhang294@wisc.edu}

\author[2]{\fnm{Bonaventura} \sur{Tagliafierro}}\email{bonaventura.tagliafierro@upc.edu}

\author[1]{\fnm{Colin} \sur{Vanden Heuvel}}\email{colin.vandenheuvel@wisc.edu}

\author[1]{\fnm{Shlok} \sur{Sabarwal}}\email{ssabarwal@wisc.edu}

\author[1]{\fnm{Luning} \sur{Bakke}}\email{lfang9@wisc.edu}

\author[1]{\fnm{Yulong} \sur{Yue}}\email{yyue32@wisc.edu}

\author[1]{\fnm{Xin} \sur{Wei}}\email{xwei84@wisc.edu}

\author[1]{\fnm{Radu} \sur{Serban}}\email{serban@wisc.edu}

\author*[1]{\fnm{Dan} \sur{Negrut}}\email{negrut@wisc.edu}

\affil*[1]{\orgdiv{Department of Mechanical Engineering}, \orgname{University of Wisconsin-Madison}, \orgaddress{\street{1513 Engineering Dr}, \city{Madison}, \postcode{53706}, \state{WI}, \country{USA}}}

\affil[2]{\orgdiv{Maritime Engineering Laboratory}, \orgname{Universitat Politècnica de Catalunya - Barcelona Tech}, \orgaddress{\street{C. Jordi/Girona 1-3}, \city{Barcelona}, \postcode{08034}, \country{Spain}}}


\abstract{This paper introduces DEM-Engine, a new submodule of Project Chrono, that is designed to carry out Discrete Element Method (DEM) simulations. Based on spherical primitive shapes, DEM-Engine can simulate polydisperse granular materials and handle complex shapes generated as assemblies of primitives, referred to as \textit{clumps}. %
DEM-Engine has a multi-tier parallelized structure that is optimized to operate simultaneously on two GPUs. The code uses custom-defined data types to reduce memory footprint and increase bandwidth. A novel ``delayed contact detection'' algorithm allows the decoupling of the contact detection and force computation, thus splitting the workload into two asynchronous GPU streams. DEM-Engine uses just-in-time compilation to support user-defined contact force models. This paper discusses its \Cpp\ and Python interfaces and presents a variety of numerical tests, in which impact forces, complex-shaped particle flows, and a custom force model are validated considering well-known benchmark cases. Additionally, the full potential of the simulator is demonstrated for the investigation of extraterrestrial rover mobility on granular terrain. The chosen case study demonstrates that large-scale co-simulations (comprising 11 million elements) spanning 15 seconds, in conjunction with an external multi-body dynamics system, can be efficiently executed within a day. Lastly, a performance test suggests that \DEME\ displays linear scaling up to 150 million elements on two NVIDIA A100 GPUs.}


\maketitle

\newpage
\tableofcontents
\newpage

\section{Introduction}

The Discrete Element Method (DEM) is a numerical technique for predicting the mechanical behavior of granular materials~\cite{cundall1979discrete}. In DEM, the motion of each individual particle is monitored, and interactions between particles are modeled in a fully detailed manner. Over time, DEM has evolved and is now a popular method for examining the dynamics of extensive granular systems~\cite{poschelDEM-textbook2005}, ranging from mixing~\cite{lemieux2008large}, particulate flows~\cite{apostolou2008discrete}, geomechanics events \cite{tang2009tsaoling, salciarini2010discrete,OSullivan2011}, to astrophysical scenarios~\cite{sanchez2011simulating}. Applications of DEM include modeling soil dynamics~\cite{frederikOleDEM2022}, tire-soil interactions \cite{antonioVehicleTireGranMatSim2017}, and rover movement on extraterrestrial surfaces \cite{iagnema2015}. 

Two main challenges make DEM simulations computationally expensive. Firstly, the small and often stiff elements necessitate the time integrator to adopt very small time steps, e.g., \num{E-6}--\num{E-5} seconds, to ensure numerical stability. Secondly, the collision detection stage of the simulation is computationally demanding. To enhance computational speed, DEM has been accelerated using parallel computing with OpenMP~\cite{openMP} as seen in~\cite{amritkar2014efficient,knuth12}; MPI standard~\cite{mpi-3.0} for distributed memory clusters~\cite{yan2018comprehensive}; and combined MPI--OpenMP parallelism~\cite{checkaraou2018hybrid,liggghts2013,lammpsWebSite,starccm,raduNicDanGVSETS2018}. The Graphics Processing Unit (GPU) offers another avenue for parallel computations and has been incorporated into DEM, as in \cite{xu2011quasi, govender-BlazeDEMGPU2016,ganDEM-GPU2016,he-powderGPU2018,conlainBillion2019}. Regardless of the computational platform, reported DEM studies typically involve between $10^3$ and $10^5$ elements~\cite{iwashitaRolling1998,direnzo2004525,dacruz2005,bazant2006,Emden2008,wasfy2014coupled,lommen2014,Utili2015,Potticary2015,michael2015,Ciantia2016,zheng2017,parteli2016,kivugo2017,calvetti2016,he-powderGPU2018}, which is considerably smaller than real-world scenarios. For instance, a cubic meter of sand contains around two billion particles~\cite{japanDEMlarge2018}. 

LAMMPS (Large-scale Atomic/Molecular Massively Parallel Simulator)~\cite{lammpsWebSite} is a widely used open-source software package for molecular dynamics simulations and DEM simulations. LAMMPS is written in \Cpp\ and is designed to run efficiently on parallel computing architectures using both MPI and OpenMP, making it suitable for simulating large-scale systems.
LAMMPS provides a variety of built-in potentials for modeling interatomic and intermolecular interactions, as well as the ability to define custom potentials. LAMMPS also supports a range of boundary conditions, including periodic, reflecting, and fixed boundaries. The default time stepper in LAMMPS is the Verlet algorithm, which is a symplectic second-order method. LAMMPS supports a range of contact models, including Hertz--Mindlin, linear-spring, cohesive and inter-particle bond models. LIGGGHTS (LAMMPS Improved for General Granular and Granular Heat Transfer Simulation) is a DEM package that is based on the LAMMPS code.  Like LAMMPS, LIGGGHTS is optimized for parallel computing and leverages combined MPI--OpenMP parallelism. While LAMMPS is more versatile, e.g., \cite{LAMMPSDNA2018,dias2021molecular,MCDEM2022}, LIGGGHTS focuses specifically on granular material simulations, offering features and capabilities tailored to that end, such as neighbor lists and domain decomposition. These added utilities come into play in granular flows, heat transfer in granular materials, and other DEM-specific concerns.


\Starccm~\cite{starccm} is a commercial Computational Fluid Dynamics (CFD) software package that includes a DEM solver for simulating the behavior of granular materials. The software also supports a range of contact models.
One of the strengths of \Starccm\ is its ability to couple DEM simulations with fluid flow simulations, allowing for the simulation of complex multiphase flows. The coupling between the DEM and fluid flow simulations is typically achieved through a two-way coupling algorithm that exchanges information between the two simulations at each time step. The software also includes models for turbulence, heat transfer, and chemically reactive flows, and incorporates design exploration and optimization tools, allowing engineers to not just simulate a given design, but also explore a variety of design possibilities.

A DEM case study anchored by {\Starccm} is summarized in~\cite{en14133797}. The study aimed to investigate the sand-retention mechanisms that occur at the opening of sand filters under various conditions, such as particle shape, size, and concentration. A coupled CFD--DEM model was used to predict the retention mechanisms under steady flow conditions of the well-bore, where CFD was used to model the fluid flow, and DEM was used to model the particle flow. The coarse grid unresolved and the smoothed unresolved (refined grid unresolved) coupling approaches implemented in \Starccm\ were used to transfer data between the fluid and solid phases and calculate the forces.  Verification of the CFD--DEM model was then conducted by mesh sensitivity analysis. The growing trend in CFD--DEM coupling research underscores the community's heightened interest in integrating multi-physics into DEM simulations, likely driven by the rapid advancements in computational power.

Compared to the LAMMPS and \Starccm, Chrono::GPU~\cite{chronoGranular2021}, an open-source DEM simulator developed originally as the granular dynamics support for Chrono~\cite{chronoOverview2016}, takes a different path in that it emphasizes efficiency. To maximize performance, Chrono::GPU operates on GPUs and exclusively supports monodisperse spherical DEM elements. Additionally, a custom data type scheme is used to reduce its memory footprint. A recent independent study~\cite{dem-pbrIdaho2023} reveals that Chrono::GPU, while running on an RTX 2060 Mobile NVIDIA GPU card of a laptop, delivers performance that is two orders of magnitude faster than other well-regarded DEM packages operating on clusters with hundreds of CPU cores.

YADE (Yet Another Dynamic Engine) is an open-source DEM simulator for granular materials, powders, and other particulate systems. Written in \Cpp\ and Python, it is known for its ability to handle complex geometries and its Python scripting-imparted extensibility. One research study that used YADE for DEM simulations is~\cite{HAUSTEIN2017118}, in which the authors were interested in the deformation of the particles under stress. Therein, the particles are modeled as a collection of smaller particles connected by springs. The authors made additional developments to the DEM model, so the volume of the element overlapping area is uniformly redistributed over the particle, the radius of each contact partner is increased, and in the end, the volume and mass are kept constant. Large deformations and complex element geometries are used in this study. Another recent study that used YADE for DEM simulations is reported in~\cite{snowYadeFoam}. Therein, the authors simulated the process of icing using an Euler--Lagrangian approach. YADE was used to calculate the motion of snow crystals, while the open-source CFD package OpenFOAM was used in conjunction with YADE to simulate flow hydrodynamics. 

To circumvent extensive computation times, DEM packages often resort to simplistic element geometries to simplify collision detection. Predominantly, spheres of uniform size are chosen, significantly streamlining collision detection~\cite{Ericson05}. Yet, certain applications require more intricate geometries, necessitating the usage of nonspherical elements to ensure accurate system dynamics \cite{favierClumpsSpheres1999,particleShapeCleary2011,particleShapeKiangi2013,unionOf2Spheres2013,dem-cfdMonashAustralia2016,andradeDEM2018,marteauExperimental2021}. 
From the aforementioned packages, YADE can use superquadric shapes to represent particles. Superquadrics are a family of shapes that include ellipsoids, boxes, and more. They can represent a range of shapes with varying degrees of roundness or sharpness. YADE also supports polyhedral-shaped particles. Another approach YADE employs is the use of the ``multi-sphere method''~\cite{dem-cfdMonashAustralia2016}, meaning grouping simpler particles (like spheres) together to form more complex shapes. Likewise, LAMMPS supports this multi-sphere method, too. LAMMPS also supports ellipsoidal and spherical particles. \Starccm, being an established commercial DEM solution, offers a library of predefined shapes (spheres, cylinders, tetrahedra, etc.), while retaining a general-purpose custom shape support using triangulated surfaces. These custom shapes are treated as rigid bodies within the DEM framework.
When these methods to address nonspherical elements are employed, the number of elements in simulations tends to reduce significantly in order to manage the amount of time required to complete a simulation.

Recognizing the characteristics, strengths, and limitations of the existing DEM solvers, the solution presented here, Chrono DEM-Engine~\cite{RuochunDEMERepo}, aims to strike a balance: (i) it accommodates a large number of discrete elements (into tens of millions); (ii) it employs a composition of multiple spheres to represent nontrivial geometries; (iii) it integrates a rapid collision detection method as per~\cite{hammadTobyDan2012} and a novel asynchronous threads management algorithm to ensure a numerical performance ahead of state of the art; (iv) its API design leaves enough room and flexibility for easy integration in co-simulations (explained in Sec.~\ref{sec:workflow}~and~\ref{sec:rover}), and gives users the freedom to define explicitly the physics they wish to simulate using a custom force model script (explained in Sec.~\ref{sec:custom_force}).  In this contribution, our emphasis is to thoroughly document the numerical features of this package, and provide guidelines for the user to easily pick up this package and then fully take advantage of its potential.

The structure of this paper is laid out as follows. The literature survey, presented in this section, identifies a prevailing need within the DEM community for an adaptable, efficient solver capable of managing large-scale simulations. Section~\ref{sec:features} explores the distinct numerical capabilities of Chrono DEM-Engine and illustrates how it addresses this identified need. Section~\ref{sec:workflow} offers a breakdown of a sample simulation script, equipping the user with a foundational understanding of the package's operation. Section~\ref{sec:force_model} unravels the implementation of the default Hertz--Mindlin model and provides guidance on incorporating custom models. Section~\ref{sec:mixer} demonstrates the solver's efficiency, spotlighting a large-scale simulation involving up to 150 million sphere primitives. Section~\ref{sec:num_exp} underscores the validation endeavors, presenting a suite of DEM simulations that emphasize the impact and capabilities of varying element shape representations and force models. Section~\ref{sec:conclusion} reiterates the essence of the paper, accentuating the proposed future developments with language models.

\section{Implementation features}\label{sec:features}

Chrono DEM-Engine is open-source, can run on commodity hardware and it does so fast and at scale. It allows large-scale DEM simulations to be efficiently executed on desktops equipped with one or two graphic cards.  Its open-source nature and ability to embed user-defined contact models meet requirements often found in exploratory projects. This section introduces the simulator's key features.


\subsection{Multi-GPU solution and delayed active-contact set update}

In DEM, the contact detection process is needed to identify the contact pairs in the simulation system before the force calculation step takes place. The contact detection and force calculation are typically done consecutively in each time step.
DEM-Engine embraces a different strategy, which uses two distinct and parallel computational threads to update the active contacts set (done by the ``kinematics thread''), and the integration of the equations of motion (done by the ``dynamics thread''), respectively. The dynamics thread processes each contact in the Active-Contact Set (ACS) \textit{at each time step} to reassess the contact penetration $\delta_n$ and the ancillary information. The dynamics thread receives an ACS update when the kinematics thread finishes producing it, or if so desired, it can wait for the ACS update when the dynamics thread advances the system state too far ahead of the time stamp of the last ACS update from the kinematics thread. Through this collaboration pattern, the two threads work concurrently and the cost of contact detection is nearly ``hidden in the shadow'' of computation done by the dynamics thread, which continuously advances the state of the system. 
To avoid missing mutual contacts that might crop up between the moments the ACS is updated, we artificially enlarge all contact geometries in the DEM system to preemptively detect potential contact pairs that might emerge in the near future. Note that this is done only to include additional potential contacts in the ACS, and does not affect the shape of the elements that participate in the simulation.

It is worth noting that by adding this artificial margin to all contact geometries, the kinematics thread reports false positives, i.e., a contact between two elements might be in the ACS, yet the two elements are not in contact.
This fact will be identified by the dynamics thread when carrying out the force calculation.
The thickness of this added margin is determined by the simulation entities' velocity (which is bounded and known by the solver), the time step size, which is typically small, and $n_{\text{max}}$, the maximum number of time steps the dynamics thread is allowed to advance without receiving an ACS update from the kinematics thread. It usually assumes values of the order of tens of microns for millimeter-sized granular material. This is small compared to typical DEM element sizes. Overall, the overhead caused by the false-positive contacts does not offset the benefit of deferring the ACS update.

\begin{figure}[htp!]
	\centering
	\captionsetup{justification=centering}
	\includegraphics[width=.9\linewidth]{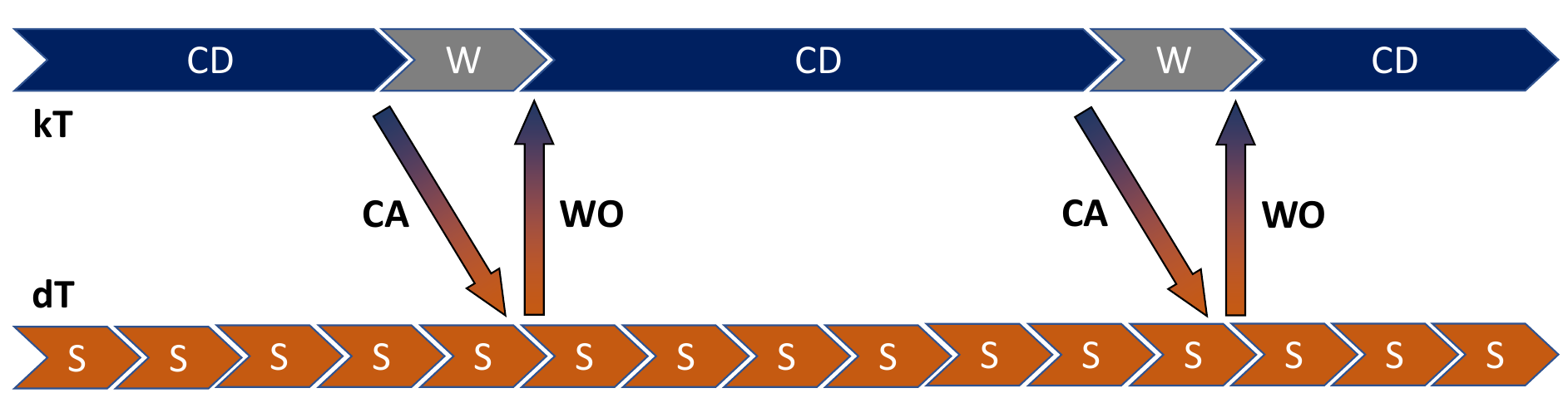}
	\caption{Ideal collaboration pattern, where the dynamics thread advances the physics continuously while the kinematics thread occasionally waits for updated state information to commence an ACS update.} \label{fig:collab_ideal}
\end{figure}

The synchronization pattern between the kinematics and dynamics threads is illustrated in Fig.~\ref{fig:collab_ideal}. There, ``\textbf{S}'' represents a time step that the dynamics thread executes, where the contact forces are calculated (see Sec.~\ref{sec:hertz_model}), and the system state is advanced in time. A contact detection step that the kinematics thread executes is marked with ``\textbf{CD}''. Periodically, the kinematics thread finishes a contact detection step and sends the signal to the dynamics thread, allowing the dynamics thread to receive the contact array, ``\textbf{CA}'', from the kinematics thread. Then the dynamics thread will send a work order ``\textbf{WO}'' with the current simulation system state, for the kinematics thread to pick up and continue the next contact detection step. Before the next ``\textbf{CA}'' update is received, the dynamics thread will use this ``\textbf{CA}'' to execute the time steps.

\begin{figure}[htp!]
	\centering
	\captionsetup{justification=centering}
	\includegraphics[width=.9\linewidth]{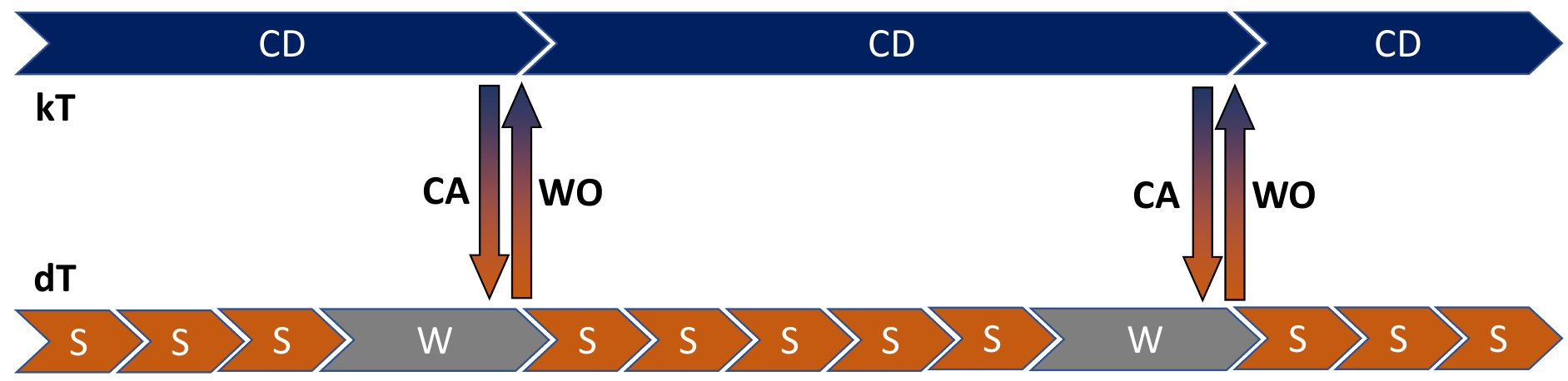}
	\caption{Non-optimal collaboration pattern, where the dynamics thread waits for the kinematics thread occasionally to generate the ACS. \DEME\ will automatically avoid this scenario.} \label{fig:collab_unideal}
\end{figure}

Because the dynamics thread only receives updates from and sends work orders to the kinematics thread after a time step is finished, the kinematics thread could stay idle between ACS update jobs. 
This is marked with ``\textbf{W}'' in Fig.~\ref{fig:collab_ideal}. Having the kinematics thread wait occasionally is considered an ideal collaboration pattern since in this case, the dynamics thread runs continuously, therefore the system marches in time uninterruptedly. A less-than-ideal collaboration pattern is illustrated in Fig.~\ref{fig:collab_unideal}. There, the dynamics thread occasionally waits for updates from the kinematics thread, reducing the overall efficiency of the solver. This happens when the dynamics thread advances the simulation beyond $n_{\text{max}}$ time steps without receiving an update from the kinematics thread, and is therefore forced to idle.
One could avoid this scenario by increasing $n_{\text{max}}$. However, as discussed before this would consequently increase the thickness of the artificial margin added to contact geometries, leading to more undesirable false-positive contacts.
Hence, $n_{\text{max}}$ should be kept at the smallest value that does not cause the dynamics thread to wait. 
DEM-Engine will automatically use this principle and the execution timing history to adapt $n_{\text{max}}$ to an appropriate value, and moderate itself so that the collaboration pattern stays as depicted in Fig.~\ref{fig:collab_ideal}. 

\begin{figure}[htp!]
	\centering
	\captionsetup{justification=centering}
	\includegraphics[width=.9\linewidth]{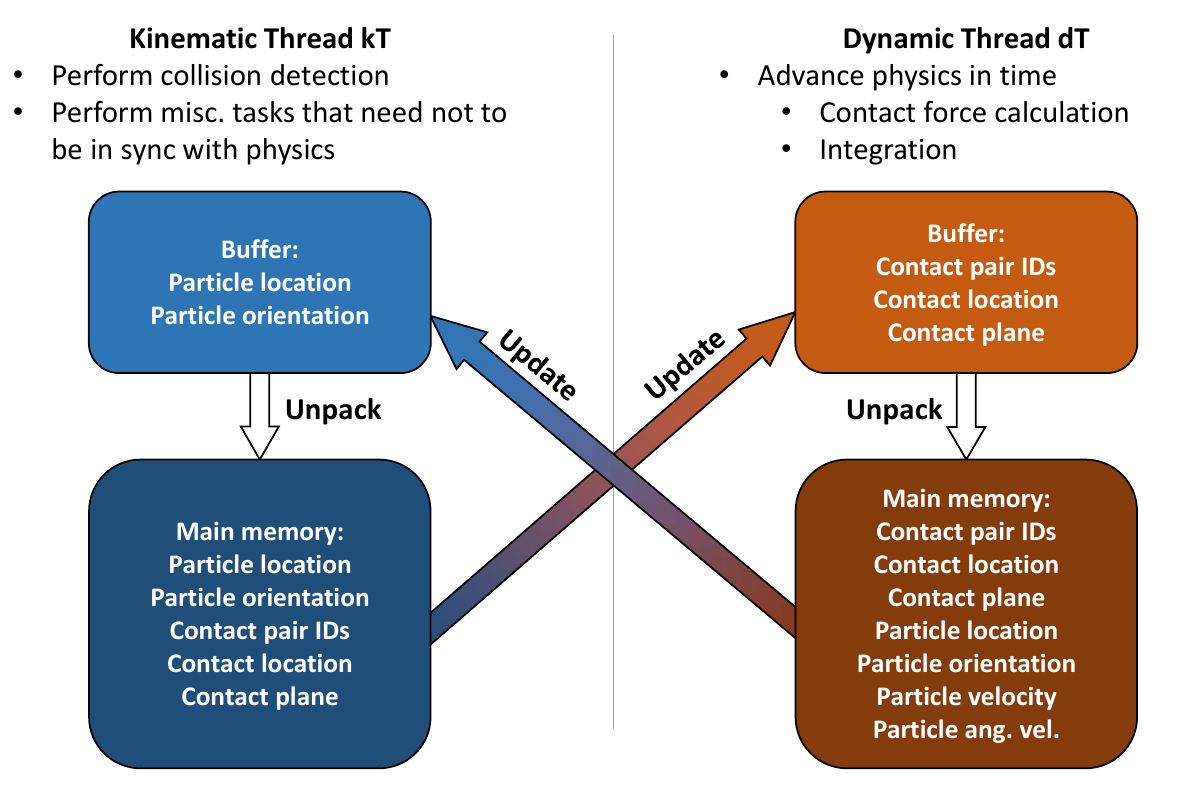}
	\caption{The collaboration pattern of the kinematic and dynamics thread. They each runs on a dedicated GPU.} \label{fig:kTdT}
\end{figure}

At the implementation level, DEM-Engine is currently optimized for using two GPUs, as each of the two host CPU threads is mapped to a GPU device respectively.
The kinematic and dynamics thread collaboration pattern is summarized in Fig.~\ref{fig:kTdT}.
After being produced by the kinematics thread, the contact information is transferred to a buffer memory. Then the dynamics thread will be notified and copy the contact information to its working memory. 
The dynamics thread carries out a similar routine when updating the kinematics thread with new element positions.
Neither of them directly modifies the working memory of the other to avoid race conditions. 
Although logically there are two buffer memory pools and each thread owns one, physically, they are both allocated on the GPU mapped to the dynamics thread. This allows the dynamics thread to spend minimum time on copying from its buffer, speeding up the computation.

\subsection{Just-in-time CUDA kernel compilation} \label{sec:jit}
The CUDA kernels in DEM-Engine are compiled when the simulation starts being executed, rather than being statically compiled. This is done by leveraging the CUDA runtime compilation tool Jitify~\cite{NVRTC}. Several benefits come with this design choice. 

With Jitify, the solver can detect the capabilities of the GPU on which it is running and generate code specifically tailored for that device. For instance, if a program is designed to be used across a variety of architectures, just-in-time compilation ensures the utilization of the optimal instruction set for each device, ensuring the generated CUDA code is optimized for an end user's specific hardware and requirements. At the same time, since the compilation occurs at runtime, the code is not bound to a specific version of the CUDA toolkit. This characteristic can make applications more resilient against changes or updates in the CUDA environment.

It should be mentioned that just-in-time compilation introduces an overhead. The first time a kernel is run, there is a delay due to its compilation. However, assuming the DEM simulations with DEM-Engine are generally large and invoke a time span of typically hours, this cost is negligible.

\subsubsection{Custom force model}\label{sec:custom_force}

Since Jitify allows for dynamic code generation, we use it for implementing custom force models. The intricate and evolving nature of DEM simulations often requires a higher degree of adaptability to cater to the multifaceted modeling needs of its users, namely the expanding list of approaches in contact and cohesion force modeling~\cite{BERRY2023118209,pr11010005}. Rather than constraining the user to a predefined set of force models, DEM-Engine allows, if so desired, for the force models to be supplied via a user-supplied \Cpp\ script, greatly increasing the solver's applicability. A walk-through of a model implementation can be found in Sec.~\ref{sec:custom_model}. 

\subsubsection{Family tag} \label{sec:family}

Jitify also allows for a low-cost implementation of prescribed motion. This is done through the family tag utility. Every simulation entity can be assigned an integer family tag between 0 and 255 (this is implemented through a \sbelCode{uint8\_t}; though rarely needed, it can be changed to a different data type such as \sbelCode{uint16\_t} to expand the range), then the solver can be notified to apply prescribed motions to this family tag. This prescription information is just-in-time compiled as a part of the integration CUDA kernel, thus no branching overhead is introduced.
The sample script in Sec.~\ref{sec:workflow} showcases this functionality with the usage of the \sbelCode{SetFamilyPrescribedAngVel} method. On the other hand, if the use case calls for more fine-grain motion control, such as when the velocity of a simulation object is determined by some external process, then the ``motion injection'' approach detailed in Sec.~\ref{sec:tracker} should be used.

As a side note, the family tags can also be used to mask contacts. The user is allowed to specify whether the solver should detect and resolve contacts between simulation entities in certain families. This is a utility used throughout the demos provided along with this package at~\cite{RuochunDEMERepo}.

\subsection{Custom and mixed data type} \label{sec:data_type}

In high-performance computing, memory footprint and bandwidth play a crucial role in determining a code's performance. As the complexities of the simulations grow, it becomes evident that relying solely on standard data types--such as {\texttt{double}}--might inadvertently lead to sub-optimal memory usage and consequently, potential performance bottlenecks. For instance, a deformation of a DEM body is of the order of \num{E-9} to \num{E-5} m. Why would one use a budget of 64 bits, which is provisioned for the {\texttt{double}} type to capture an extremely broad range of numbers, to represent a very narrow range of the positive real axis that hosts an element's \num{E-9} to \num{E-5} deformation? This would be a waste of bits, which leads to less accuracy and/or lower bandwidth.
Given the hierarchical memory architecture in CUDA, from global to shared memory, the significance of ensuring that the memory bandwidth is utilized effectively and that latency is minimized becomes even more critical.

To this end, DEM-Engine introduces the utilization of custom and mixed data types. Unlike stock data types that come with a predefined bit budget, e.g., 64 bits for {\texttt{double}}, custom data types offer finer control over memory use. For instance, the spatial coordinate in DEM-Engine is represented using integers rather than floating-point numbers. The entire simulation domain is decomposed into cubes with a known edge length, which is adapted based on the domain size. Each of these cubes is termed a ``voxel'' and is assigned an index represented by a \sbelCode{uint64\_t} data type. Additionally, to specify the location of a body within a voxel, three \sbelCode{uint16\_t}s are employed, each dividing the voxel uniformly into \(2^{16}\) parts in its respective direction.

\begin{figure}[htp!]
	\centering
	\captionsetup{justification=centering}
	\includegraphics[width=.95\linewidth]{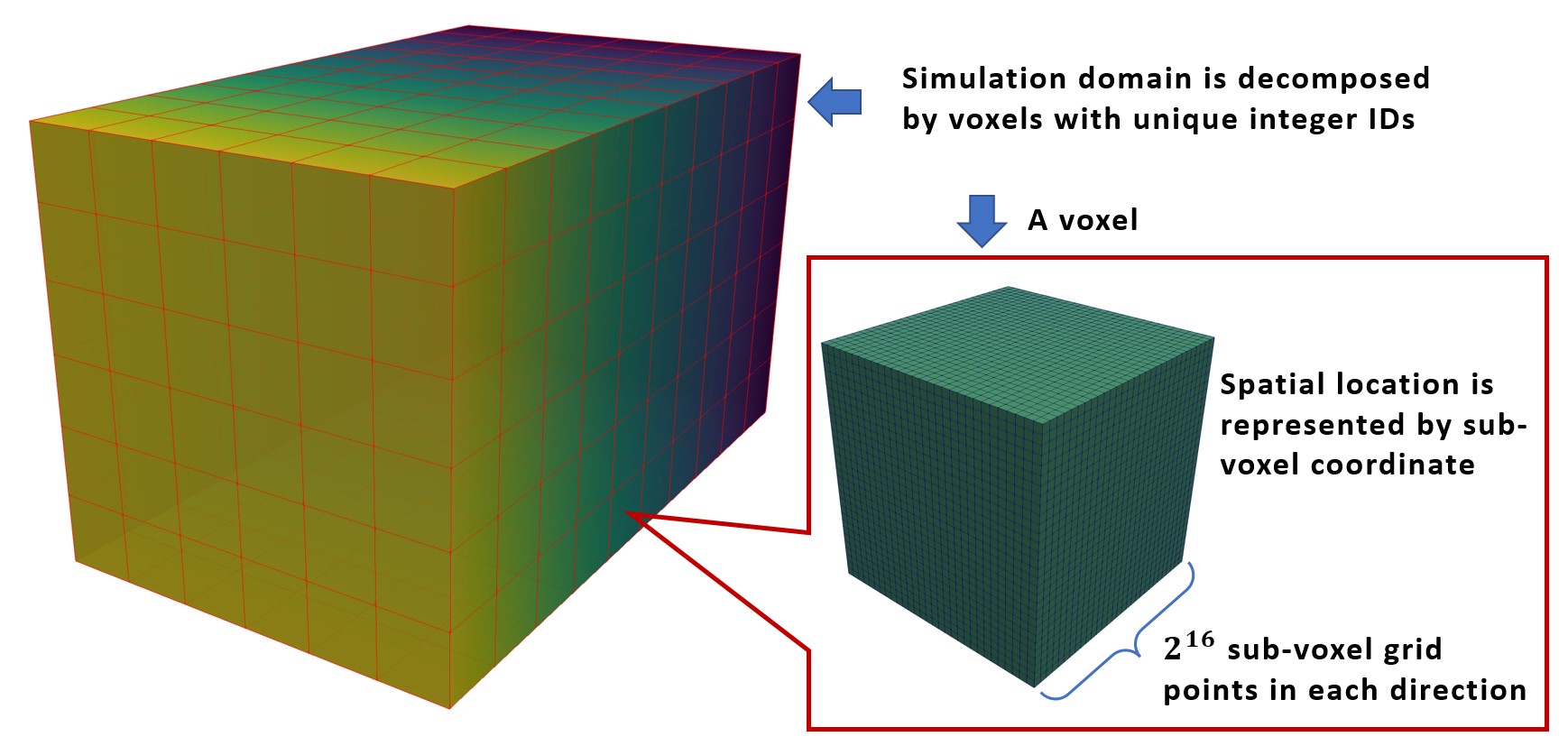}
	\caption{The domain decomposition that leads to a compressed spatial coordinate data type. The domain is decomposed into voxels with \sbelCode{uint64\_t} indices, then each voxel is split into $ 2^{16} \times 2^{16}  \times 2^{16}  $ sub-voxels. The typical precision is estimated to be around \SI{e-11}{m}.}
     \label{fig:voxels}
\end{figure}

For a cubic simulation domain with an edge length of \SI{1}{m}, the precision (i.e., the smallest discernible length unit within a voxel) is approximately \SI{e-11}{m}. This precision is adequate for capturing micro-deformations. Moreover, this compressed data type requires only 112 bits to represent a spatial location, which is more memory-efficient than using three \sbelCode{double}s that would require 192 bits in total. The voxel-based spatial coordinate data type is illustrated in Fig.~\ref{fig:voxels}.

The general rule used for the selection of mixed data types is that the data residing in the global memory take a 4-byte or compressed data type. The examples are the state variables such as the quaternions of the elements. The temporary variables used in kernel functions that are essential in governing the physics, on the other hand, use 8-byte data types, namely \sbelCode{double}. An example is the penetration depth between geometries in the Hertzian contact force calculation. The data type usage in DEM-Engine is summarized in Table~\ref{table:data_type}. Since data type conversion is essentially a free operation and the main bottleneck in GPU-based physics simulations is the memory bandwidth limit, the design choice enhances the performance without compromising the physics.

\begin{table}[h]
    \renewcommand{\arraystretch}{1}
    \caption{Various data types in DEM-Engine and their memory location.}
    \label{table:data_type}
	\begin{tabular}{ccc}
		\toprule
		\emph{Data Type}	                 & \emph{Variable}      & \emph{Memory Type} \\
		\midrule
		$\sbelCode{uint64\_t}$   &  Voxel index   & Global \\
		$\sbelCode{uint16\_t}$   &  Sub-voxel index  & Global \\
		$\sbelCode{int32\_t}$ or \sbelCode{float}   &   Kinematics quantities, friction history etc.  & Global \\
		$\sbelCode{double}$   &  Penetration  & Register \\
		$\sbelCode{float}$   &  Contact force calculation  & Register \\
		$\sbelCode{float}$   &  Clump types information  & Shared Memory \\
		\bottomrule
	\end{tabular}
\end{table}

Furthermore, DEM-Engine has a level of encapsulation of the data types in use. Most data types are specified in a file named \sbelCode{VariableTypes.h} using \sbelCode{typedef}, including the ones introduced in this section. If the user needs a different selection of data types, such as increasing the size of family tags from 1 byte to 2 bytes to allow for more varieties (see Sec.~\ref{sec:family} for context), or reducing the size of spatial coordinates to allow for faster computation in a low-accuracy setting, they can modify the data types in \sbelCode{VariableTypes.h} then recompile to conveniently get the updated executable.

\subsection{Geometry hierarchy and tracker}\label{sec:tracker}

DEM-Engine facilitates complex element geometries through a composition of multiple spheres, termed a ``clump''. This approach draws inspiration from~\cite{clumpSpheresPrice2007}. A clump denotes a collection of potentially overlapping spheres that together depict a specific element shape. Some examples of these clumps are visually presented in Fig.~\ref{fig:mixer_clumps} in Sec.~\ref{sec:mixer}. Throughout this paper, the terms ``element'' and ``clump'' are used interchangeably to discuss DEM elements with complex shapes. Beyond clumps, DEM-Engine supports integrating triangular meshes and analytical objects (such as rigid objects constructed from analytical planes or cylindrical surfaces) into the simulation framework. However, as a dedicated and performance-centric DEM package, DEM-Engine exclusively handles contacts between clumps and meshes, as well as between clumps and analytical geometries. Should there be a requirement for contacts between meshes or between analytical geometries, users can achieve this through co-simulation, as exemplified in Sec.~\ref{sec:rover}.

An important aspect of DEM-Engine's utilization is understanding its geometry hierarchy, delineating the roles of the ``owner'' versus the ``geometry''. An owner constitutes a simulation entity endowed with mass properties, hence governed by physics. In DEM-Engine's current implementation, an owner can manifest as a clump, a mesh, or an analytical entity. Conversely, the term geometry is associated with the constituent parts of an owner. A geometric entity can be a sphere (within a clump), a triangular facet (within a mesh), or an analytical component (like a plane in a multi-component analytical object). Each geometric entity carries associated material attributes, granting users flexibility in designing discrete element systems with simulation entities that have spatially varying material properties.

Further, DEM-Engine provides users the control over diverse simulation entities via ``tracker'' objects. Users can associate trackers with any owner, facilitating real-time status inquiries such as position and velocity or enforcing state modifications, from setting coordinates to applying external loads. Beyond basic operations, trackers offer advanced features: identifying clumps in contact with a tracked owner or, when monitoring a mesh, controlling its deformation. A practical demonstration of tracker usage is encapsulated in Sec.~\ref{sec:workflow}.

\subsection{Python wrapper}
\DEME\ has a Python wrapper, facilitated by the Pybind library. This allows users, irrespective of their CUDA expertise, to tap into {\DEME}'s features, all within Python's accessible library ecosystem and widely adopted science tools such as \sbelCode{numpy} and \sbelCode{scikit-learn}. The package has been made available on the Python Package Index (PyPI) and can be installed using the familiar {\sbelCode{pip}} command. Simply executing \sbelCode{pip install DEME} ensures that the computational capabilities and functionalities of the package become available within the Python environment, reducing the complexities often associated with software installations in high-performance computing scenarios. An example script is given in Sec.~\ref{sec:python_example}, where it is compared against its \Cpp\ counterpart.

\section{Sample script} \label{sec:workflow}
\begin{figure}[htp!]
	\centering
	\captionsetup{justification=centering}
	\includegraphics[width=.95\linewidth]{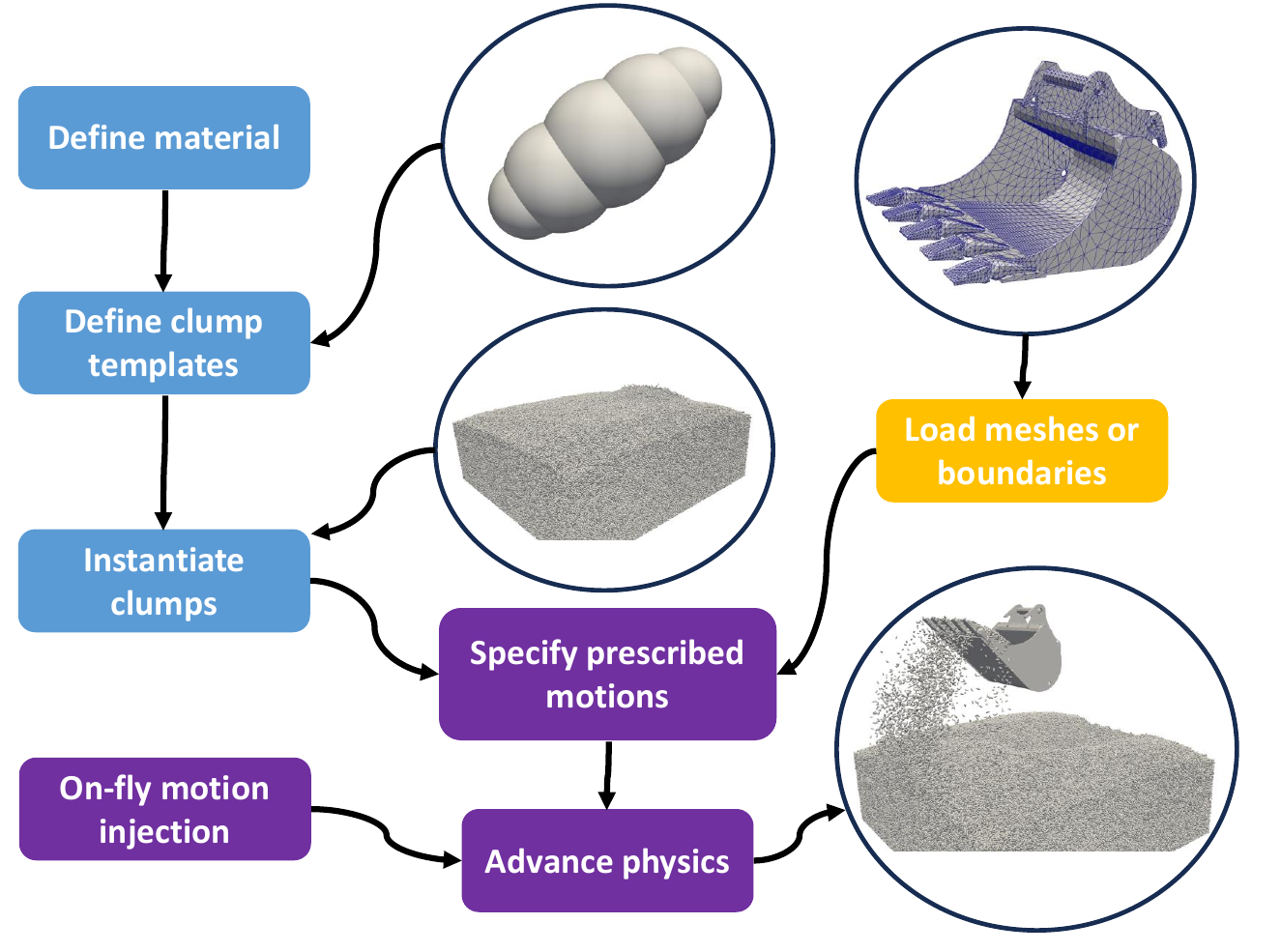}
	\caption{Typical workflow of running a DEM-Engine simulation.}
     \label{fig:user_workflow}
\end{figure}

This section discusses a script responsible for the mixer timing analysis discussed in Sec.~\ref{sec:mixer}. The focus is placed here on the code implementation. A visual representation of the simulation workflow is provided in Fig.~\ref{fig:user_workflow}. Examples are provided in both \Cpp\ and Python. The scripts corresponding to all simulations addressed in this paper can be located in the DEM-Engine's demo directory~\cite{RuochunDEMERepo}.

\subsection{\Cpp\ version}

The user should first create the \sbelCode{DEMSolver} object. While the solver comes with default meta-parameters, users have the flexibility to modify them, e.g., verbosity, output detail, and output format.
\begin{lstlisting}[style=customcpp]
DEMSolver DEMSim;
DEMSim.SetVerbosity("INFO");
DEMSim.SetOutputFormat("CSV");
DEMSim.SetOutputContent("ABSV");
DEMSim.SetMeshOutputFormat("VTK");
\end{lstlisting}

The following code snippet defines the material types for the mesh geometry and DEM elements. DEM-Engine will return a handle so this material can be used to define clump templates. If a material property, such as the frictional coefficient $\mu$, is defined between two materials, the method \sbelCode{SetMaterialPropertyPair} can be used to specify it.
\begin{lstlisting}[style=customcpp]
auto mat_type_mixer = DEMSim.LoadMaterial({{"E", 1e8}, {"nu", 0.3}, {"CoR", 0.6}, {"mu", 0.5}, {"Crr", 0.0}});
auto mat_type_granular = DEMSim.LoadMaterial({{"E", 1e8}, {"nu", 0.3}, {"CoR", 0.6}, {"mu", 0.2}, {"Crr", 0.0}});
DEMSim.SetMaterialPropertyPair("mu", mat_type_mixer, mat_type_granular, 0.5);
\end{lstlisting}

The following snippet defines the analytical boundaries of the simulation domain. 
\begin{lstlisting}[style=customcpp]
const double world_size = 1;
DEMSim.InstructBoxDomainDimension(world_size, world_size, world_size);
DEMSim.InstructBoxDomainBoundingBC("all", mat_type_granular);
auto walls = DEMSim.AddExternalObject();
walls->AddCylinder(make_float3(0), make_float3(0, 0, 1), world_size / 2., mat_type_mixer, 0);
\end{lstlisting}

The following snippet shows the mixer mesh being loaded into the simulation. The stock mixer mesh is then scaled to fit the size of the simulation domain. The mixer is assigned the family code 10, which is subsequently used to prescribe a constant angular velocity $\pi$ $\SI{}{rad/s}$ to the mixer. A ``tracker'' object is created for the mixer so that we can extract information in real time for this simulation entity, or apply fine-grain motion control, while the simulation is running. In this example, we use it to set the initial location of the mixer to obtain the torque exerted by the DEM elements.
\begin{lstlisting}[style=customcpp]
const float chamber_height = world_size / 3.;
auto mixer = DEMSim.AddWavefrontMeshObject((GET_DATA_PATH() / "mesh/internal_mixer.obj").string(), mat_type_mixer);
mixer->Scale(make_float3(world_size / 2, world_size / 2, chamber_height));
mixer->SetFamily(10);
DEMSim.SetFamilyPrescribedAngVel(10, "0", "0", "3.14159");
auto mixer_tracker = DEMSim.Track(mixer);
\end{lstlisting}

The next snippet creates a clump template. It contains mass, shape, and material information. There are stock clump shapes that the user can directly use to reproduce the examples we provide. The user can also easily scale or otherwise modify the template before using it to instantiate more DEM elements.
\begin{lstlisting}[style=customcpp]
float granular_rad = 0.005;
float mass = 2.6e3 * 5.5886717; 
float3 MOI = make_float3(2.928, 2.6029, 3.9908) * 2.6e3;
std::shared_ptr<DEMClumpTemplate> template_granular = DEMSim.LoadClumpType(mass, MOI, GetDEMEDataFile("clumps/3_clump.csv"), mat_type_granular);
template_granular->Scale(granular_rad);
\end{lstlisting}

When instantiating the DEM elements, the user has the option to leverage the sampler objects that come with the solver, as shown in the following snippet. A sampling region appropriate with respect to the simulation domain is defined, then the hexagonal close-packing sampler is used to create initial elements. These elements are duplicates of the clump template that has just been created.
\begin{lstlisting}[style=customcpp]
const float fill_height = chamber_height;
const float chamber_bottom = -world_size / 2.;
const float fill_bottom = chamber_bottom + chamber_height;
HCPSampler sampler(3.f * granular_rad);
float3 fill_center = make_float3(0, 0, fill_bottom + fill_height / 2);
const float fill_radius = world_size / 2. - 2. * granular_rad;
auto input_xyz = sampler.SampleCylinderZ(fill_center, fill_radius, fill_height / 2);
DEMSim.AddClumps(template_granular, input_xyz);
\end{lstlisting}

An initialization call is needed to instruct the solver to set up data structures on the GPUs. Before that, several simulation specs should be inputted, e.g., the time step size and metrics that the solver should watch in identifying a diverged simulation, as shown in the following snippet.
\begin{lstlisting}[style=customcpp]
float step_size = 5e-6;
DEMSim.SetInitTimeStep(step_size);
DEMSim.SetGravitationalAcceleration(make_float3(0, 0, -9.81));
DEMSim.SetErrorOutVelocity(20.);
DEMSim.SetForceCalcThreadsPerBlock(512);
DEMSim.Initialize();
\end{lstlisting}

Finally, the following code snippet shows the main simulation loop. The output directory is created, the simulation time length is indicated, and the mixer is translated to the correct initial position, before the main loop starts to iteratively make \sbelCode{DoDynamics} calls, advancing the simulation each time by a frame. The benefit of this design is that the user enjoys free interfacing with the simulation data while it is running. For example, the script writes the simulation status to a file, inspects the torque that the mixer is experiencing, and outputs the execution stats from the kinematics and dynamics threads at the frequency of 20 times per simulation second. Another opportunity this design brings is the ease of co-simulation. A related example is in Sec.~\ref{sec:rover}.
\begin{lstlisting}[style=customcpp]
std::filesystem::path out_dir = current_path();
out_dir += "/DemoOutput_Mixer";
create_directory(out_dir);

float sim_end = 10.0;
unsigned int fps = 20;
float frame_time = 1.0 / fps;
unsigned int currframe = 0;

mixer_tracker->SetPos(make_float3(0, 0, chamber_bottom + chamber_height / 2.0));
for (float t = 0; t < sim_end; t += frame_time) {
    std::cout << "Frame: " << currframe << std::endl;
    char filename[200], meshfilename[200];
    sprintf(filename, "%s/DEMdemo_output_%04d.csv", out_dir.c_str(), currframe);
    sprintf(meshfilename, "%s/DEMdemo_mesh_%04d.vtk", out_dir.c_str(), currframe++);
    DEMSim.WriteSphereFile(std::string(filename));
    DEMSim.WriteMeshFile(std::string(meshfilename));
    
    float3 mixer_moi = mixer_tracker->MOI();
    float3 mixer_acc = mixer_tracker->ContactAngAccLocal();
    float3 mixer_torque = mixer_acc * mixer_moi;
    std::cout << "Contact torque on the mixer is " << mixer_torque.x << ", " << mixer_torque.y << ", " << mixer_torque.z << std::endl;

    DEMSim.DoDynamics(frame_time);
    DEMSim.ShowThreadCollaborationStats();
}
\end{lstlisting}

\subsection{Python version} \label{sec:python_example}

A Python version of the same mixer simulation is given in this section. It follows the same workflow as the \Cpp\ version, including the material definition, template creation, clump instantiation, mesh loading and motion control, initialization, and a main simulation loop. The names of the methods are not changed in the Python version, and certain data structures are simply converted to their Python counterparts, streamlining the learning experience of the users switching between these programming languages. For example, the \Cpp\ version uses a \sbelCode{unordered\_map} to define the properties of a material, while the Python version uses a dictionary object; the \Cpp\ version takes a \sbelCode{float3} at some places to specify a coordinate, while the Python version uses a list or a \sbelCode{numpy} array of three \sbelCode{float}s.
\begin{lstlisting}[style=custompython]
import DEME
import numpy as np
import os
import time
if __name__ == "__main__":
    out_dir = "DemoOutput_Mixer/"
    out_dir = os.path.join(os.getcwd(), out_dir)
    os.makedirs(out_dir, exist_ok=True)

    DEMSim = DEME.DEMSolver()
    DEMSim.SetVerbosity("STEP_METRIC")
    DEMSim.SetOutputFormat("CSV")
    DEMSim.SetOutputContent(["ABSV", "XYZ"])
    DEMSim.SetMeshOutputFormat("VTK")

    # E, nu, CoR, mu, Crr... Material properties
    mat_type_mixer = DEMSim.LoadMaterial(
        {"E": 1e8, "nu":  0.3, "CoR":  0.6, "mu":  0.5, "Crr": 0.0})
    mat_type_granular = DEMSim.LoadMaterial(
        {"E":  1e8, "nu":  0.3, "CoR":  0.8, "mu":  0.2, "Crr":  0.0})
    DEMSim.SetMaterialPropertyPair(
        "CoR", mat_type_mixer, mat_type_granular, 0.5)

    # Now define simulation world size and add the analytical boundary
    step_size = 5e-6
    world_size = 1
    chamber_height = world_size / 3.
    fill_height = chamber_height
    chamber_bottom = -world_size / 2.
    fill_bottom = chamber_bottom + chamber_height
    DEMSim.InstructBoxDomainDimension(world_size, world_size, world_size)
    DEMSim.InstructBoxDomainBoundingBC("all", mat_type_granular)
    walls = DEMSim.AddExternalObject()
    walls.AddCylinder([0, 0, 0], [0, 0, 1], world_size / 2., mat_type_mixer, 0)

    # Define the meshed mixer and its prescribed motion
    mixer = DEMSim.AddWavefrontMeshObject(
        DEME.GetDEMEDataFile("mesh/internal_mixer.obj"), mat_type_mixer)
    print(f"Total num of triangles: {mixer.GetNumTriangles()}")
    mixer.Scale([world_size / 2, world_size / 2, chamber_height])
    mixer.SetFamily(10)
    DEMSim.SetFamilyPrescribedAngVel(10, "0", "0", "3.14159")
    # Track the mixer
    mixer_tracker = DEMSim.Track(mixer)

    # Define the clump template used in the simulation
    granular_rad = 0.005
    mass = 2.6e3 * 5.5886717
    MOI = np.array([2.928, 2.6029, 3.9908]) * 2.6e3
    template_granular = DEMSim.LoadClumpType(mass, MOI.tolist(), 
        DEME.GetDEMEDataFile("clumps/3_clump.csv"), mat_type_granular)
    template_granular.Scale(granular_rad)
    # Sampler uses hex close-packing
    sampler = DEME.HCPSampler(3.0 * granular_rad)
    fill_center = [0, 0, fill_bottom + fill_height / 2]
    fill_radius = world_size / 2. - 2. * granular_rad
    input_xyz = sampler.SampleCylinderZ(
        fill_center, fill_radius, fill_height / 2)
    DEMSim.AddClumps(template_granular, input_xyz)
    print(f"Total num of particles: {len(input_xyz)}")

    DEMSim.SetInitTimeStep(step_size)
    DEMSim.SetGravitationalAcceleration([0, 0, -9.81])
    DEMSim.SetErrorOutVelocity(20.)
    DEMSim.SetForceCalcThreadsPerBlock(512)
    DEMSim.Initialize()

    sim_end = 10.0
    fps = 20
    frame_time = 1.0 / fps

    # Keep a tab of the max velocity in the simulation
    max_v_finder = DEMSim.CreateInspector("clump_max_absv")

    print(f"Output at {fps} FPS")
    currframe = 0

    mixer_tracker.SetPos([0, 0, chamber_bottom + chamber_height / 2.0])

    t = 0.
    start = time.process_time()
    while (t < sim_end):
        print(f"Frame: {currframe}", flush=True)
        filename = os.path.join(out_dir, f"DEMdemo_output_{currframe:04d}.csv")
        meshname = os.path.join(out_dir, f"DEMdemo_mesh_{currframe:04d}.vtk")
        DEMSim.WriteSphereFile(filename)
        DEMSim.WriteMeshFile(meshname)
        currframe += 1

        max_v = max_v_finder.GetValue()
        print(
            f"Max velocity of any point in simulation is {max_v}", flush=True)
        print(
            f"Solver's current update frequency (auto-adapted): {DEMSim.GetUpdateFreq()}", flush=True)
        print(
            f"Average contacts each sphere has: {DEMSim.GetAvgSphContacts()}", flush=True)

        mixer_moi = np.array(mixer_tracker.MOI())
        mixer_acc = np.array(mixer_tracker.ContactAngAccLocal())
        mixer_torque = np.cross(mixer_acc, mixer_moi)
        print(
            f"Contact torque on the mixer is {mixer_torque[0]}, {mixer_torque[1]}, {mixer_torque[2]}", flush=True)

        DEMSim.DoDynamics(frame_time)
        DEMSim.ShowThreadCollaborationStats()

        t += frame_time

    elapsed_time = time.process_time() - start
    print(f"{elapsed_time} seconds (wall time) to finish this simulation")
\end{lstlisting}

\section{DEM model}\label{sec:force_model}

This section details the default force models in DEM-Engine and the implementation of the geometry representations.

\subsection{History-based Hertz--Mindlin model} \label{sec:hertz_model}

The default force model is anchored by the Hertzian contact model~\cite{hertz1882} and integrates the Mindlin friction model~\cite{mindlin53}. For a comprehensive analysis, readers may refer to \cite{luningFricModel2021}. For two bodies, namely \(i\) and \(j\), when they are in contact, the normal force, \(\vect{F}_n\), operates based on a spring--damper model. The tangential frictional force, \(\vect{F}_t\), is computed considering material attributes and microscopic deformations, ensuring it adheres to the Coulomb limit via the friction coefficient \(\mu\). The mathematical representation is as follows:
\begin{subequations}
	\begin{align}
		&\vect{F}_n = f(\bar{R},\delta_n) (k_n \vect{u}_n - \gamma_n \bar{m} \vect{v}_n),  \label{eq:normal_tangential_force_model1} \\
		&\vect{F}_t = f(\bar{R},\delta_n)(-k_t \vect{u}_t -  \gamma_t \bar{m} \vect{v}_t),  \quad	\|\vect{F}_t\| \leq \mu \|\vect{F}_n\| \; , \label{eq:normal_tangential_force_model2} \\
		&f(\bar{R},\delta_n) = \sqrt{\bar{R}\delta_n}, \\
		&\bar{R} = R_i R_j/(R_i + R_j),\\
		&\bar{m} = m_i m_j/(m_i + m_j),
	\end{align}
\end{subequations}
where the constants \(k_n\), \(k_t\), \(\gamma_n\), and \(\gamma_t\) are inferred from material characteristics, including Young's modulus \(E\), the Poisson's ratio \(\nu\), and the restitution coefficient, \(\text{CoR}\) \cite{jonJCND2015}. The terms \(\bar{m}\) and \(\bar{R}\) depict the effective mass and curvature radius for the specific contact. The foundational premise is that the geometries can undergo small penetration, \(\delta_n\), at the contact point. The normal penetration vector is \(\vect{u}_n = \delta_n \vect{n}\). The relative speed, \(\vect{v}_{rel} = \vect{v}_n + \vect{v}_t\), at the contact point is defined as:
\begin{subequations}
	\label{subeq:kinematicsContactInfo}
	\begin{align}
	\vect{v}_{rel}& = \vect{v}_j + \vect{\omega}_j \times \vect{r}_j - \vect{v}_i - \vect{\omega}_i \times \vect{r}_i, \\
	\vect{v}_n &= \left(\vect{v}_{rel} \cdot \vect{n} \right) \vect{n}, \\
	\vect{v}_t &= \vect{v}_{rel} - \vect{v}_n,
	\end{align}
where \(\vect{v}_{i}\), \(\vect{\omega}_{i}\) and \(\vect{v}_{j}\), \(\vect{\omega}_{j}\) denote the velocities at the mass centers and angular speeds of entities \(i\) and \(j\). The position vectors, \(\vect{r}_{i}\) and \(\vect{r}_j\), extend from the mass centers of bodies \(i\) and \(j\) to the shared contact point. The frictional force \(\vect{F}_t\) varies based on the historical tangential micro-displacement \(\vect{u}_t\), updated iteratively at each time interval throughout the interaction event based on \(\vect{v}_t\). Let $\vect{u}^\prime_t$ be the updated tangential micro-displacement, then
	\begin{align} \label{subsubeq:updateUt}
	\vect{u}^\prime &=\vect{u}_t  + h\vect{v}_t ,\\
	\vect{u}^\prime_t &= \vect{u}^\prime - (\vect{u}^\prime \cdot \vect{n})\vect{n},
	\end{align}
where $h$ is the time step size. The strategy adopted to update $\vect{u}^\prime_t$ is borrowed from \cite{jonJCND2015}.
After the update, we may need to clamp the updated tangential micro-displacement $\vect{u}^\prime_t$ to get the final $\vect{u}_t $ for the next time step in order to satisfy the capping condition $\|\vect{F}_t\| \leq \mu \|\vect{F}_n\|$:
	\begin{align} \label{subsubeq:cappingOfut}
	\vect{u}_t = \begin{cases} \vect{u}^\prime_t & \text{if } \|\vect{F}_t\| \leq \mu \|\vect{F}_n\|, \\
	 		 \frac{\mu\|\vect{F}_n\|}{k_t} \frac{\vect{u}^\prime_t}{\|\vect{u}^\prime_t\|} & \text{otherwise.} \end{cases}
	\end{align}
The rolling resistance arises from an asymmetric normal stress profile at the contact patch~\cite{johnson1987contact}. In DEM-Engine's default force model, it is implemented as the torque $\vect{\tau}_r$. This torque is induced by a force that has the magnitude of the rolling resistance coefficient $C_r$ times the normal force. The direction of this force is aligned with the rolling-contributed relative velocity at the contact point. This is summarized in the following equations:
	\begin{align} \label{subsubeq:Crr}
	\vect{F}_r &= \frac{\vect{\omega}_j \times \vect{r}_j - \vect{\omega}_i \times \vect{r}_i}{\| \vect{\omega}_j \times \vect{r}_j - \vect{\omega}_i \times \vect{r}_i \|}  C_r\vect{F}_t,\\
	\vect{\tau}_r &= \vect{r}_i \times \vect{F}_r.
	\end{align}
\end{subequations}

\begin{figure}[htp!]
	\centering
	\captionsetup{justification=centering}
	\includegraphics[width=.9\linewidth]{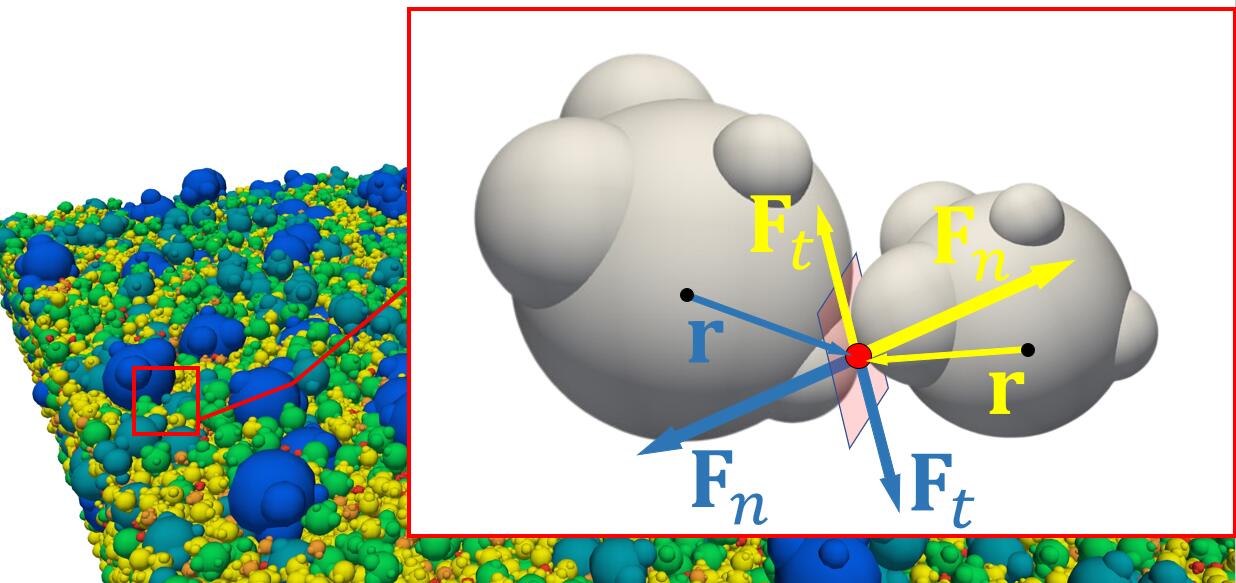}
	\caption{The normal and tangential contact forces between particles are calculated based on the penetration and displacement history of involved sphere components.} \label{fig:force}
\end{figure}

As discussed in Sec.~\ref{sec:tracker}, a clump has mass properties associated with it, whereas its component spheres have material properties associated with them -- in other words, each sphere of the clump that makes up an element can have different material properties.  Consequently, $\vect{F}_n$ and $\vect{F}_t$ in Eq.~(\ref{eq:newton_law_1})~and~(\ref{eq:newton_law_2}) need to be derived from the contacts between component spheres. Then a reduction process is invoked to use these contact forces to update the element $\vect{v}_i$ and $\vect{\omega}_i$, based on each clump's $m_i$ and $I_i$, as well as the location vector for the contact point, $\vect{r}_i$. This is visualized in Fig.~\ref{fig:force}, and the equations of motion for entity \(i\) assume the form
\begin{subequations}
	\begin{align}
		m_i \frac{d \vect{v}_i}{d t} &= m_i \vect{g} + \sum_{k=1}^{n_c} \vect{F}^k, \label{eq:newton_law_1} \\
		I_i \frac{d \vect{\omega}_i}{dt} &= \sum_{k=1}^{n_c} \left( \vect{r}^k \times \vect{F}^k + \vect{\tau}_r^k \right), \label{eq:newton_law_2} 
	\end{align}	
\end{subequations}
where $n_c$ is the number of contacts spheres that entity \(i\) has, and the the superscript $k$ iterates through each contact. In these equations, $\vect{F}^k = \vect{F}_n^k + \vect{F}_t^k$ means the total force, containing both the normal and tangential components.

\subsection{Providing a custom contact force model} \label{sec:custom_model}

To cater to diverse simulation needs, \DEME\ supports custom force models through user-provided scripts. This section delves further into this functionality, whose starting point is a custom force model provided as a \Cpp\ script. This script undergoes just-in-time compilation at the onset of the simulation (as detailed in Sec.~\ref{sec:jit}), replacing the default contact force model. The ``ingredients'' of a custom force model are called user-referable variables. A comprehensive list of these variables is provided in Table~\ref{tab:referables}. For each contact pair, the solver automatically determines the values for these variables. Users can then harness these referable variables to implement the customized contact force.

Central to scripting the force model is the modification of the user-referable variable \sbelCode{force}, analogous to $\vect{F}^k$ in Eqs.~\eqref{eq:newton_law_1}~and~\eqref{eq:newton_law_2}. This variable represents the force that geometry \textbf{A} experiences during contact in the global frame. The variable \sbelCode{force} takes the initial value of $(0,0,0)$. It is worth noting that the solver will auto-apply the corresponding reaction force to geometry \textbf{B}. In a similar vein, the user-referable variable \sbelCode{torque\_only\_force} can be adjusted to store an action--reaction force pair that solely produces torque (without affecting the linear velocity of contact geometries, but only their angular momentum). This is congruent to $\vect{F}_r$ in Eq.~\eqref{subsubeq:Crr}. In the default model, the implementation of rolling resistance hinges on this variable. As Eqs.~\eqref{eq:newton_law_1}~and~\eqref{eq:newton_law_2} indicate, a subroutine executed by the solver in each iteration, will integrate the motions of simulation entities post the force calculation. 

\begin{table}
	\centering
	\caption{The user-referable variables that can be used in composing the custom force model. All data types are the default data type. Some of the data types can be configured in \sbelCode{VariableTypes.h} upon compilation from the source to accommodate the user's specific needs, a concept introduced in Sec.~\ref{sec:data_type}.}
	\label{tab:referables}
	\renewcommand{\arraystretch}{1.2}
		\begin{tabular}{*3l}    \toprule
		\emph{Type} & \emph{Name} & \emph{Explanation}  \\ \toprule
			\sbelCode{double3} & \sbelCode{contactPnt} & Contact point coord in global  \\ \hline
			\sbelCode{float3} & \sbelCode{B2A} & \makecell{Unit vector pointing from \\ geometry \textbf{B} to geometry \textbf{A}} \\ \hline
			\sbelCode{double} & \sbelCode{overlapDepth} & The length of overlap \\ \hline
			\sbelCode{float} & \sbelCode{ts} & Time step size \\ \hline
			\sbelCode{float} & \sbelCode{time} & Current time in simulation \\ \hline
			\sbelCode{float3} & \sbelCode{locCPA}, \sbelCode{locCPB} & \makecell{Positions of the contact point \\ in the contact geometries' frames} \\\hline
			\sbelCode{double3} & \sbelCode{AOwnerPos}, \sbelCode{BOwnerPos} & Positions of both owners \\  \hline
			\sbelCode{double3} & \sbelCode{bodyAPos}, \sbelCode{bodyBPos} & Positions of both contact geometries \\ \hline
			\sbelCode{float4} & \sbelCode{AOriQ}, \sbelCode{BOriQ} & Quaternions of both owners  \\ \hline
			\sbelCode{float} & \sbelCode{AOwnerMass}, \sbelCode{BOwnerMass} & Masses of both owners \\ \hline
			\sbelCode{float3} & \sbelCode{AOwnerMOI}, \sbelCode{BOwnerMOI} & Moment of inertia for both owners \\\hline
			\sbelCode{float} & \sbelCode{ARadius}, \sbelCode{BRadius} & \makecell{Radius of curvature for both contact \\ geometries  at point of contact} \\ \hline
			\sbelCode{uint8\_t} & \sbelCode{bodyAMatType}, \sbelCode{bodyBMatType} & \makecell{Offset used to query the material properties \\ for both contact geometries} \\ \hline
			\sbelCode{uint8\_t} & \sbelCode{AOwnerFamily}, \sbelCode{BOwnerFamily} & Family number of both owners \\ \hline
			\sbelCode{float3} & \sbelCode{ALinVel}, \sbelCode{BLinVel} & Linear velocity of both owners \\ \hline
			\sbelCode{float3} & \sbelCode{ARotVel}, \sbelCode{BRotVel} & \makecell{Angular velocity of both owners, \\ in their local frames} \\ \hline
			\sbelCode{unsigned int} & \sbelCode{AOwner}, \sbelCode{BOwner} & \makecell{Offset for both owners \\ in system array} \\ \hline
			\sbelCode{unsigned int} & \sbelCode{AGeo}, \sbelCode{BGeo} & \makecell{Offset for both contact geometries \\ in system array} \\ \hline
			\sbelCode{float} & User-specified & \makecell{Contact wildcards: Extra properties \\ associated with contacts} \\\hline
			\sbelCode{float} & User-specified & \makecell{Owner wildcards: Extra properties \\ associated with owners} \\ \hline
			\sbelCode{float} & User-specified & \makecell{Geometry wildcards: Extra properties \\ associated with geometries} \\ \hline
			\sbelCode{float3} & \sbelCode{force} & Accumulator for contact force  \\\hline
			\sbelCode{float3} & \sbelCode{torque\_only\_force} & Accumulator for contact torque  \\
			\bottomrule
		\end{tabular}

\end{table}

Note that the three ``wildcard'' type variables in Table~\ref{tab:referables} are the custom properties that the user is allowed to associate with contacts, owners (clump, mesh, or analytical object), and geometries (sphere, triangle facet, or analytical component), respectively.
For the owner wildcards and geometry wildcards, the user can assign their values before or during the simulation, using trackers or family tags. These custom properties can then be used in the custom force model to derive force, or be modified so their values change during simulation according to a user-specified policy.
The contact wildcards, on the other hand, work differently. If the user chooses to associate a wildcard to contacts, then the memory space associated with a contact is allocated when this contact emerges, and deallocated when this contact vanishes. When it is allocated, it always takes the initial value of zero. This is useful for recording some quantities that evolve during the lifespan of a contact. For example, as shown in Sec.~\ref{sec:default_model}, the default force model uses contact wildcards to record the contact history needed for the history-based Hertz--Mindlin model.

\subsubsection{Default model implementation explained} \label{sec:default_model}

We elaborate on the implementation of the default Hertz--Mindlin model in the remainder of this section, which can be found in the file \sbelCode{FullHertzianForceModel.cu} from the repository~\cite{RuochunDEMERepo}. The code is an appropriate starting point for users to implement their own force model, potentially adding to the existing physics.

The preliminary step, as presented in the ensuing code snippet, involves extracting material properties of the contact geometries. Material property arrays adopt naming conventions consistent with the user-defined property names in the \sbelCode{LoadMaterial} function call. Consequently, if the default force model is employed, Young's modulus (\sbelCode{E}), Poisson's ratio (\sbelCode{nu}), coefficient of restitution (\sbelCode{CoR}), friction coefficient (\sbelCode{mu}), and rolling resistance coefficient (\sbelCode{Crr}) must be specified in the \sbelCode{LoadMaterial} invocation. For users implementing a custom force model, the material property names specified during the \sbelCode{LoadMaterial} function should align with the array names in the force model file. 
For properties associated singularly with a material type (e.g., Young's modulus), one should utilize the offset variables \sbelCode{bodyAMatType} or \sbelCode{bodyBMatType} to retrieve the property pertinent to the contact material. Conversely, for properties defined between two materials (like the friction coefficient), both offset variables are employed concurrently to obtain the appropriate value for the contact, as illustrated in the subsequent code snippet.
\begin{lstlisting}[style=customcpp]
// Material properties
float E_cnt, G_cnt, CoR_cnt, mu_cnt, Crr_cnt;
{
    // E and nu are associated with each material, so obtain them this way
    float E_A = E[bodyAMatType];
    float nu_A = nu[bodyAMatType];
    float E_B = E[bodyBMatType];
    float nu_B = nu[bodyBMatType];
    matProxy2ContactParam(E_cnt, G_cnt, E_A, nu_A, E_B, nu_B);
    // CoR, mu and Crr are pair-wise, so obtain them this way
    CoR_cnt = CoR[bodyAMatType][bodyBMatType];
    mu_cnt = mu[bodyAMatType][bodyBMatType];
    Crr_cnt = Crr[bodyAMatType][bodyBMatType];
}
\end{lstlisting}

In this implementation, because the \sbelCode{force} is set to be in the global frame, we do the calculation in the global frame. This requires us to compute the global angular velocity of the contact point on both contact geometries (albeit having the same location in space, the contact point on geometry \textbf{A} does not have the same velocity as that on geometry \textbf{B}, because of the intrinsic velocity that \textbf{A} and \textbf{B} have), since the user-referable variables \sbelCode{ARotVel} and \sbelCode{BRotVel} only give their angular velocity in local frames. This section of the code does this task. 
\begin{lstlisting}[style=customcpp]
float3 rotVelCPA, rotVelCPB;
{
    // This is local rotational velocity (the portion of linear vel contributed by rotation)
    rotVelCPA = cross(ARotVel, locCPA);
    rotVelCPB = cross(BRotVel, locCPB);
    // This is mapping from local rotational velocity to global
    applyOriQToVector3(rotVelCPA.x, rotVelCPA.y, rotVelCPA.z, AOriQ.w, AOriQ.x, AOriQ.y, AOriQ.z);
    applyOriQToVector3(rotVelCPB.x, rotVelCPB.y, rotVelCPB.z, BOriQ.w, BOriQ.x, BOriQ.y, BOriQ.z);
}
\end{lstlisting}

Then the model calculates the normal force. Readers are referred to Sec.~\ref{sec:hertz_model} to relate the implementation with the normal contact model. The material properties that are extracted previously, such as \sbelCode{E\_cnt}, are used here to derive the force. One extra task carried out in this part is the update of the ``wildcards'' \sbelCode{delta\_tan\_x}, \sbelCode{delta\_tan\_y}, \sbelCode{delta\_tan\_z} and \sbelCode{delta\_time}, which are used to record the friction history. The contact history is used in the friction and rolling resistance calculation. 
At the end of this snippet, the variable \sbelCode{force} is updated to record the normal force.
\begin{lstlisting}[style=customcpp]
// A few re-usable variables that might be needed for both the tangential and normal force
float mass_eff, sqrt_Rd, beta;
float3 vrel_tan;
float3 delta_tan = make_float3(delta_tan_x, delta_tan_y, delta_tan_z);

// Normal force calculation
{
    // The (total) relative linear velocity of A relative to B
    const float3 velB2A = (ALinVel + rotVelCPA) - (BLinVel + rotVelCPB);
    const float projection = dot(velB2A, B2A);
    vrel_tan = velB2A - projection * B2A;

    // Update contact history
    {
        delta_tan += ts * vrel_tan;
        const float disp_proj = dot(delta_tan, B2A);
        delta_tan -= disp_proj * B2A;
        delta_time += ts;
    }

    mass_eff = (AOwnerMass * BOwnerMass) / (AOwnerMass + BOwnerMass);
    sqrt_Rd = sqrt(overlapDepth * (ARadius * BRadius) / (ARadius + BRadius));
    const float Sn = 2. * E_cnt * sqrt_Rd;

    const float loge = (CoR_cnt < 1e-12) ? log(1e-12) : log(CoR_cnt);
    beta = loge / sqrt(loge * loge + deme::PI * deme::PI);

    const float k_n = 2. / 3. * Sn;
    const float gamma_n = 2. * sqrt(5. / 6.) * beta * sqrt(Sn * mass_eff);

    force += (k_n * overlapDepth + gamma_n * projection) * B2A;
}
\end{lstlisting}

The snippet below calculates the rolling resistance. At the end of this snippet, the variable \sbelCode{torque\_only\_force} is updated to record the rolling resistance. Recall that this imaginary ``force'' contributes only to the contact torque, in agreement with the rolling resistance model in Eq.~(\ref{subsubeq:Crr}).
\begin{lstlisting}[style=customcpp]
if (Crr_cnt > 0.0) {
    bool should_add_rolling_resistance = true;
    {
        float R_eff = sqrtf((ARadius * BRadius) / (ARadius + BRadius));
        float kn_simple = 4. / 3. * E_cnt * sqrtf(R_eff);
        float gn_simple = -2.f * sqrtf(5. / 3. * mass_eff * E_cnt) * beta * powf(R_eff, 0.25f);

        float d_coeff = gn_simple / (2.f * sqrtf(kn_simple * mass_eff));

        if (d_coeff < 1.0) {
            float t_collision = deme::PI * sqrtf(mass_eff / (kn_simple * (1.f - d_coeff * d_coeff)));
            if (delta_time <= t_collision) {
                should_add_rolling_resistance = false;
            }
        }
    }
    if (should_add_rolling_resistance) {
        // Tangential velocity (only rolling contribution) of B relative to A, at contact point, in global
        float3 v_rot = rotVelCPB - rotVelCPA;
        // This v_rot is only used for identifying resistance direction
        float v_rot_mag = length(v_rot);
        if (v_rot_mag > 1e-12) {
            torque_only_force = (v_rot / v_rot_mag) * (Crr_cnt * length(force));
        }
    }
}
\end{lstlisting}

The snippet below implements the friction force. The variable \sbelCode{force} is updated to record the friction force. Although the contact history variables (\sbelCode{delta\_tan\_x}, \sbelCode{delta\_tan\_y}, and \sbelCode{delta\_tan\_z}) are initially packed into a \sbelCode{float3} (\sbelCode{delta\_tan}) for cleaner code, they are unpacked in the end to allow the solver to detect their modifications and write them back to memory. The contact history variables need modifications due to the potential tangential micro-displacement clamping, as shown in Eq.~(\ref{subsubeq:cappingOfut}).
\begin{lstlisting}[style=customcpp]
if (mu_cnt > 0.0) {
    const float kt = 8. * G_cnt * sqrt_Rd;
    const float gt = -2. * sqrt(5. / 6.) * beta * sqrt(mass_eff * kt);
    float3 tangent_force = -kt * delta_tan - gt * vrel_tan;
    const float ft = length(tangent_force);
    if (ft > 1e-12) {
        // Reverse-engineer to get tangential displacement
        const float ft_max = length(force) * mu_cnt;
        if (ft > ft_max) {
            tangent_force = (ft_max / ft) * tangent_force;
            delta_tan = (tangent_force + gt * vrel_tan) / (-kt);
        }
    } else {
        tangent_force = make_float3(0, 0, 0);
    }
    force += tangent_force;
}

delta_tan_x = delta_tan.x;
delta_tan_y = delta_tan.y;
delta_tan_z = delta_tan.z;
\end{lstlisting}

The snippets provided combine to define the complete Hertz--Mindlin contact force model implemented in DEM-Engine. For a practical example of a custom force model in application, see Sec.~\ref{sec:breakage} for a material breakage simulation. Users can also refer to the \sbelCode{DEMdemo\_Electrostatic.cpp} demo within the repository~\cite{RuochunDEMERepo}. In that demo, elements are subjected to a contact force and an electrostatic force. 

\subsection{Contact model validation}
In this section, two small-scale tests are introduced to validate the implementation of the default force contact model. For notation brevity,  for the rest of the paper, variables have their scopes limited to the respective section.



\subsubsection{Sphere rolling on incline}

\begin{figure}[htp!]
\centering
\begin{minipage}{.4\textwidth}
	\centering
	\captionsetup{justification=centering}
	\includegraphics[width=.9\linewidth]{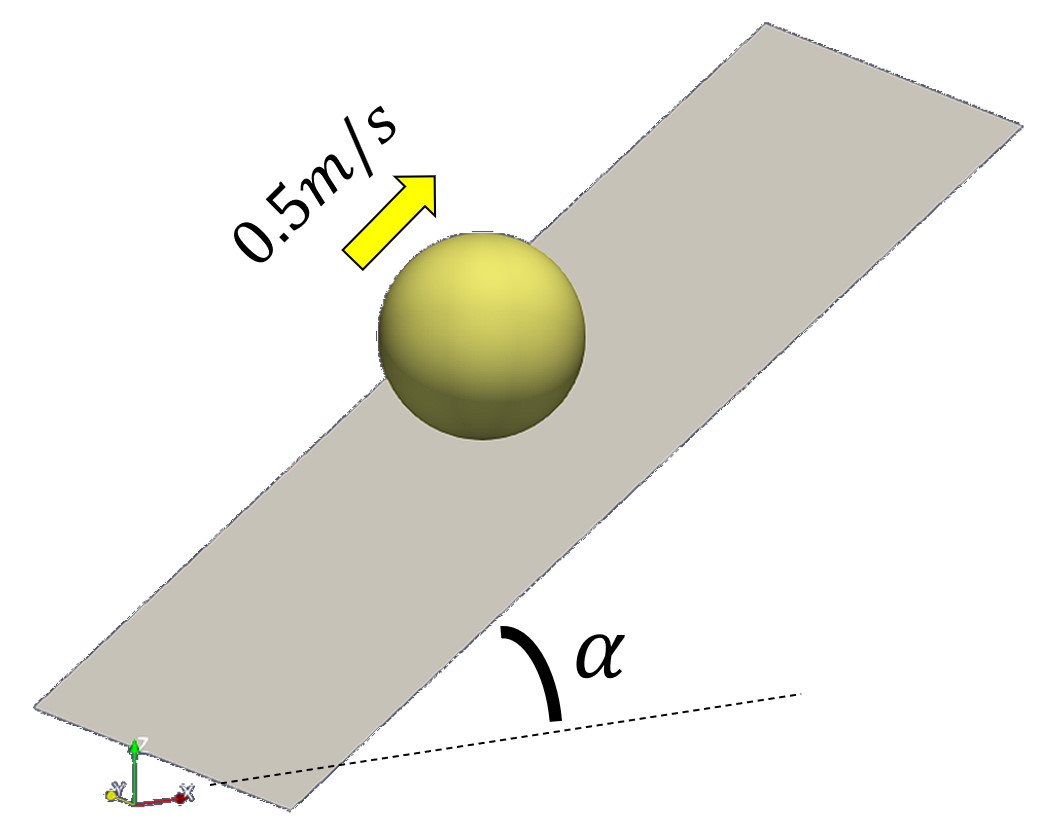}
	\caption{A rendering of the sphere moving up an incline.} \label{fig:up_incline}
\end{minipage}%
\hspace{.05cm}
\begin{minipage}{.56\textwidth}
	\centering
	\captionsetup{justification=centering}
	\includegraphics[width=.92\linewidth]{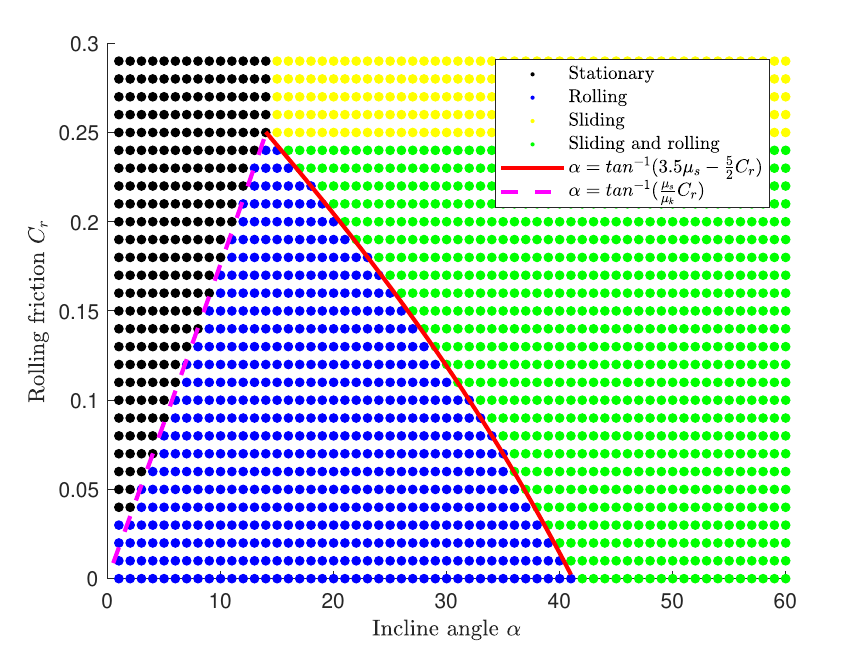}
	\caption{The end status of the sphere can be one of the following modes.} \label{fig:incline_res}
\end{minipage}
\end{figure}

This is a simple but insightful test borrowed from~\cite{luningFricModel2021}, in which a sphere rolls up an incline. The sphere of radius $r=\SI{0.2}{m}$ and mass $\SI{5}{kg}$ moves up on an incline with an initial velocity of $\SI{0.5}{m/s}$, parallel with the incline and pointing up. In~\cite{luningFricModel2021}, the static friction coefficient $\mu_s$ and kinetic friction coefficient $\mu_k$ are allowed to have different values; however, in the default force model that we are validating, they assume the same value, and in this test $\mu_s=\mu_k=0.25$.
A test scene is illustrated in Fig.~\ref{fig:up_incline}. The end status of the sphere can be one of the following modes depending on the incline angle $\alpha$ and rolling resistance $C_{r}$: stationary; sliding; rolling; sliding and rolling. These modes are defined by the final angular velocity $\omega$ and linear velocity $v$ of the sphere, and are summarized in Table~\ref{tab:modes}.

\begin{table}
	\centering
	\caption{The possible end status of the sphere in the rolling-on-incline test.}
	\label{tab:modes}
	\renewcommand{\arraystretch}{0.8}
		\begin{tabular}{*5l}    \toprule
			\emph{Mode} & Stationary & Sliding & Rolling & Sliding and rolling  \\\midrule
			\emph{Definition}& $\omega=0$, $v=0$  & $\omega=0$, $v>0$ & $v=\omega r$ & $\omega>0$, $v>\omega r$ \\ \bottomrule
			\hline
		\end{tabular}
\end{table}

The outcome of this set of simulations is plotted in Fig.~\ref{fig:incline_res}.
It is shown in~\cite{luningFricModel2021}~that for the sphere to be stationary on the incline, $\alpha\leq\text{tan}^{-1}(\frac{\mu_s}{\mu_k}C_r)$. For the sphere to roll down the incline without sliding, $\alpha\leq\text{tan}^{-1}(3.5\mu_s - \frac{5}{2}C_r)$. These two conditions are plotted in Fig.~\ref{fig:incline_res} as the dashed and solid lines respectively, which evidently separate the stationary region, pure rolling region, and sliding--rolling mixed region as the theory suggests. The DEM-Engine results confirm the results reported in \cite{luningFricModel2021}.

\subsubsection{Sphere stacking}
\begin{figure}[htp!]
\centering
\begin{minipage}{.4\textwidth}
	\centering
	\captionsetup{justification=centering}
	\includegraphics[width=.9\linewidth]{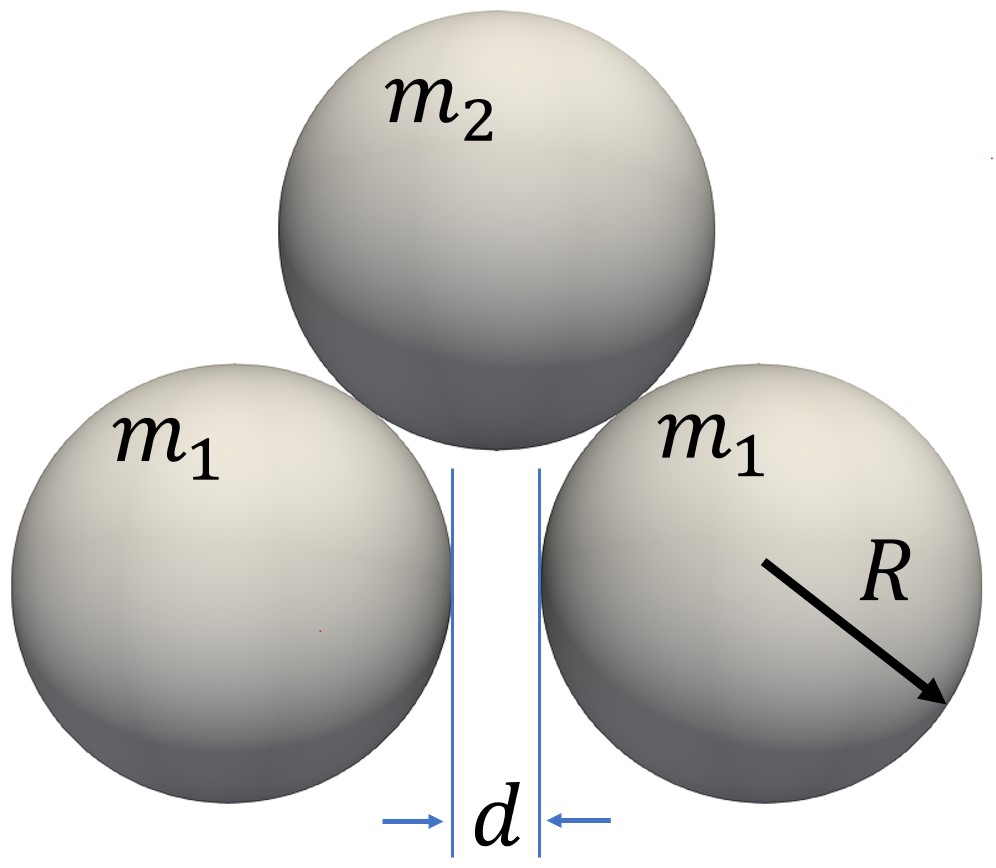}
	\caption{A rendering of the sphere-move-up-incline test.} \label{fig:stacking}
\end{minipage}%
\hfill
\begin{minipage}{.56\textwidth}
	\centering
	\captionsetup{justification=centering}
	\includegraphics[width=.92\linewidth]{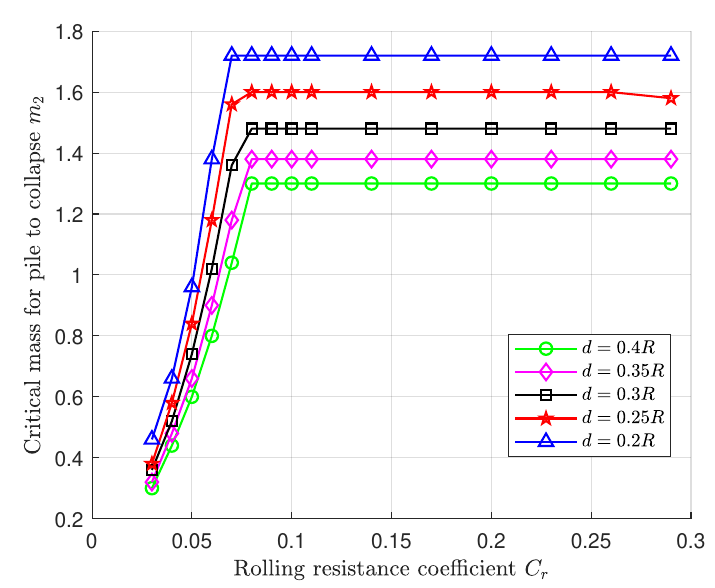}
	\caption{The end status of the sphere can be one of the following modes.} \label{fig:stacking_res}
\end{minipage}
\end{figure}

A set of three-sphere stacking tests were carried out to further validate the friction model implementation. This experiment is borrowed from~\cite{luningFricModel2021,chronoGranular2021}. For each test, two identical spheres of mass $m_1 = \SI{1}{kg}$ and radius $R = \SI{0.15}{m}$ with a small gap $d$ in between were settled on a flat surface. A third sphere of the same radius $R$ but a different mass was placed between and above the bottom spheres with zero initial velocity, as illustrated in Fig.~\ref{fig:stacking}. To minimize the influence of impact, the third sphere was initialized in contact with the bottom ones. Depending on $m_1$, the gap, and rolling resistance coefficient $C_r$, two scenarios can occur: the top sphere drops to the ground, or it moves down slightly but the structure eventually stabilizes with the bottom spheres supporting the top sphere. This is a type of physics that also comes into play on a larger scale in angle of repose experiments. For different selections of $C_r$, the mass of the top sphere was increased by \SI{0.02}{kg} to find the critical mass $m_2$ for the pile to collapse, and the result is demonstrated in Fig.~\ref{fig:stacking_res}. The critical masses found for all initial gap sizes show exact matches with the outcome reported in~\cite{luningFricModel2021}, validating DEM-Engine force model implementation.

\section{Simulator's performance}\label{sec:mixer}

\begin{figure}[htp!]
\centering
\begin{minipage}{.40\textwidth}
	\centering
	\captionsetup{justification=centering}
	\includegraphics[width=\linewidth]{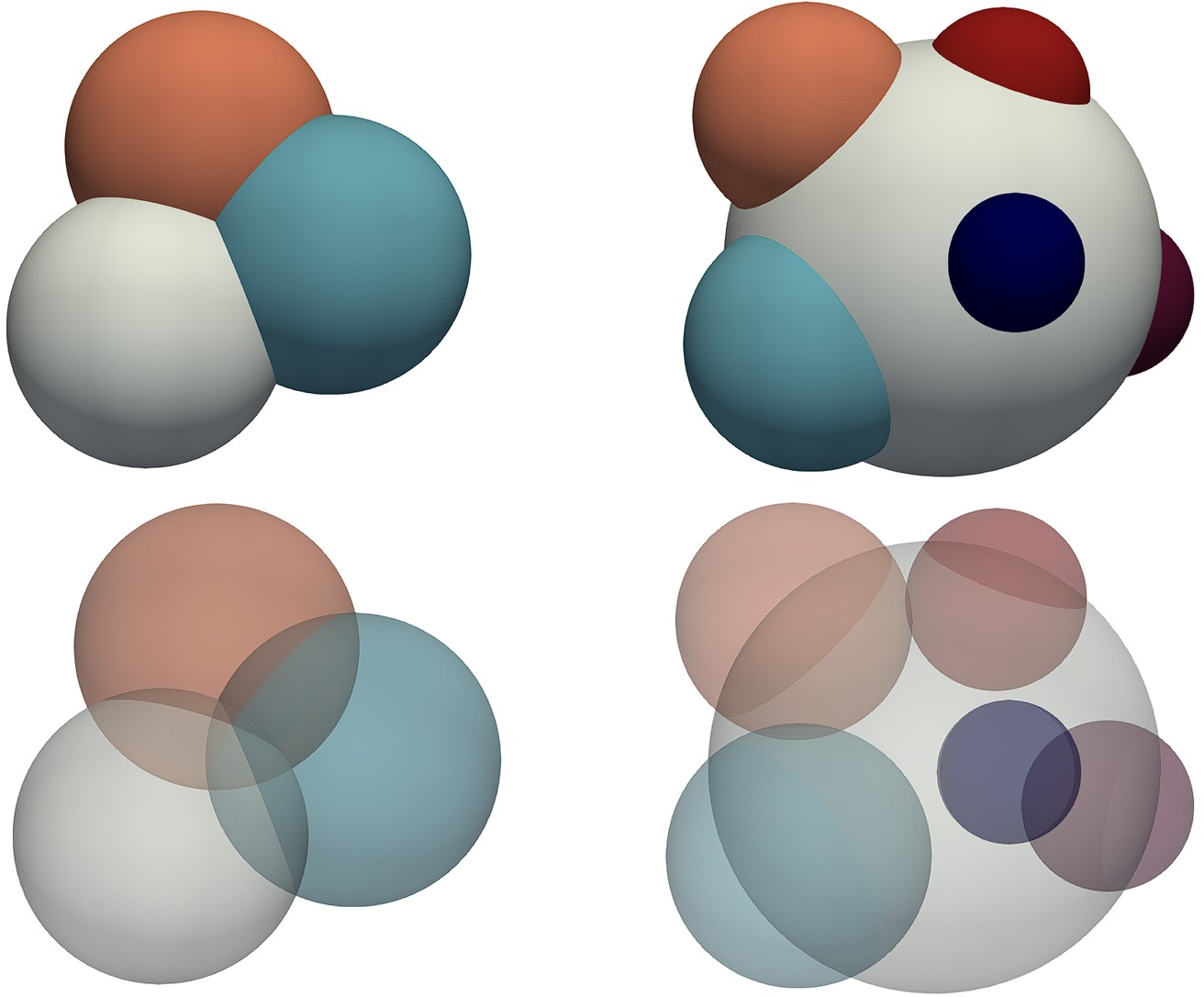}
	\caption{The element shapes for the three-sphere and six-sphere clumps.\label{fig:mixer_clumps}} 
\end{minipage}%
\hfill
\begin{minipage}{.52\textwidth}
	\centering
	\captionsetup{justification=centering}
	\includegraphics[width=\linewidth]{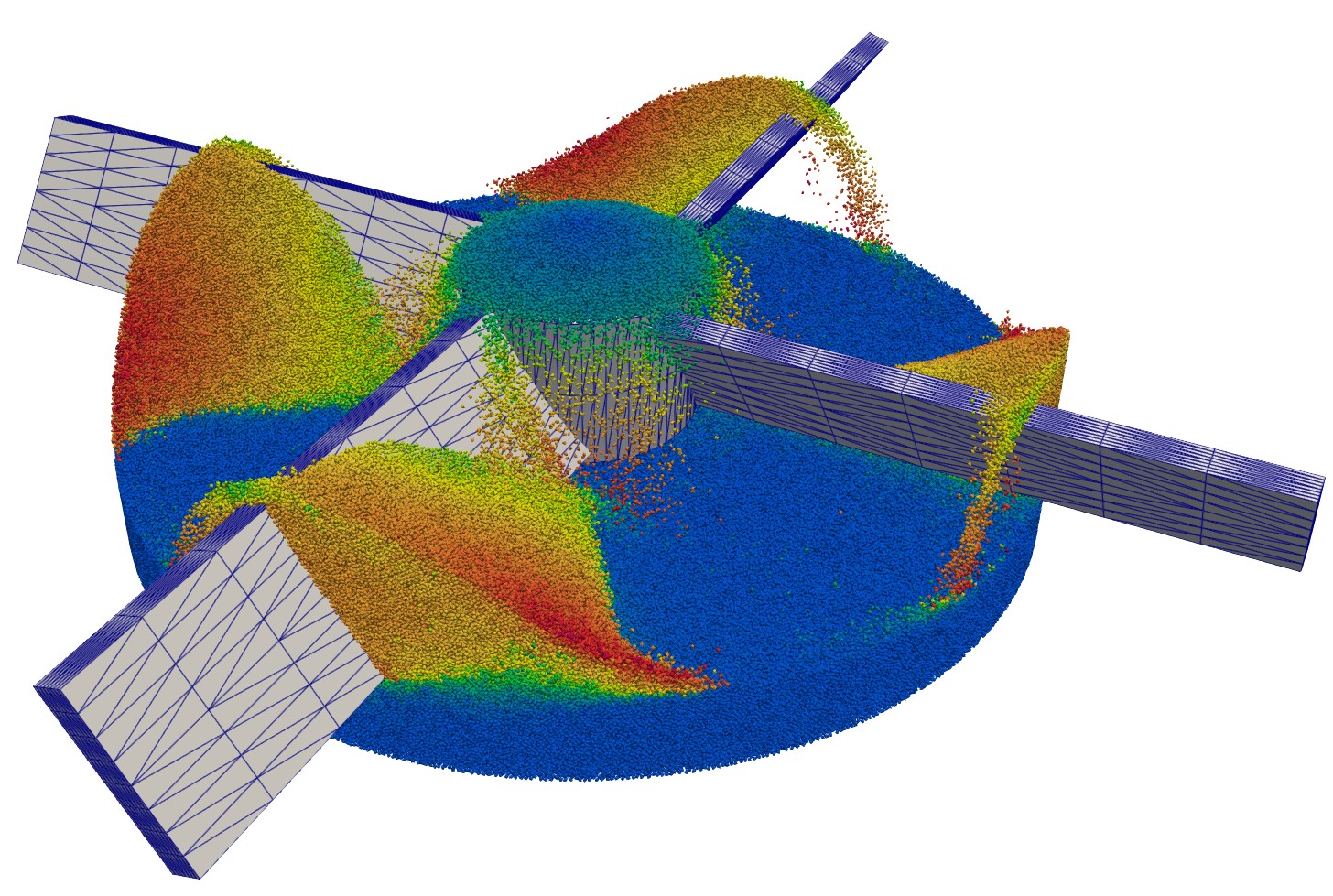}
	\caption{A rendering of the mixing process.\label{fig:mixer_scene}}
\end{minipage}
\end{figure}

The scaling analysis in this section seeks to offer insights into the expected simulation performance of DEM-Engine. The chosen test scenario involves a bladed mixer interacting with granular material, where the mixer is modeled using a triangular mesh. Throughout the simulation, the mixer blades maintain a constant angular velocity of \(2\pi \) \SI{}{\radian / s}. Initially, the elements are positioned within a cylindrical region with a radius of \SI{0.5}{\m} and a height of \(1/3\) \SI{}{\m} above the mixer, and are subsequently released at the simulation's onset. The test's selection is due to its intensive particle--particle and particle--mesh interactions, demonstrated in Fig.~\ref{fig:mixer_scene}. This puts the contact history preservation algorithm to the test, as contacts emerge and vanish in this highly dynamic problem. Material properties and simulation parameters can be found in Table~\ref{tab:mixer_mat}.

\begin{figure}[htp]
	\centering
	\captionsetup{justification=centering}
	\includegraphics[width=.65\linewidth]{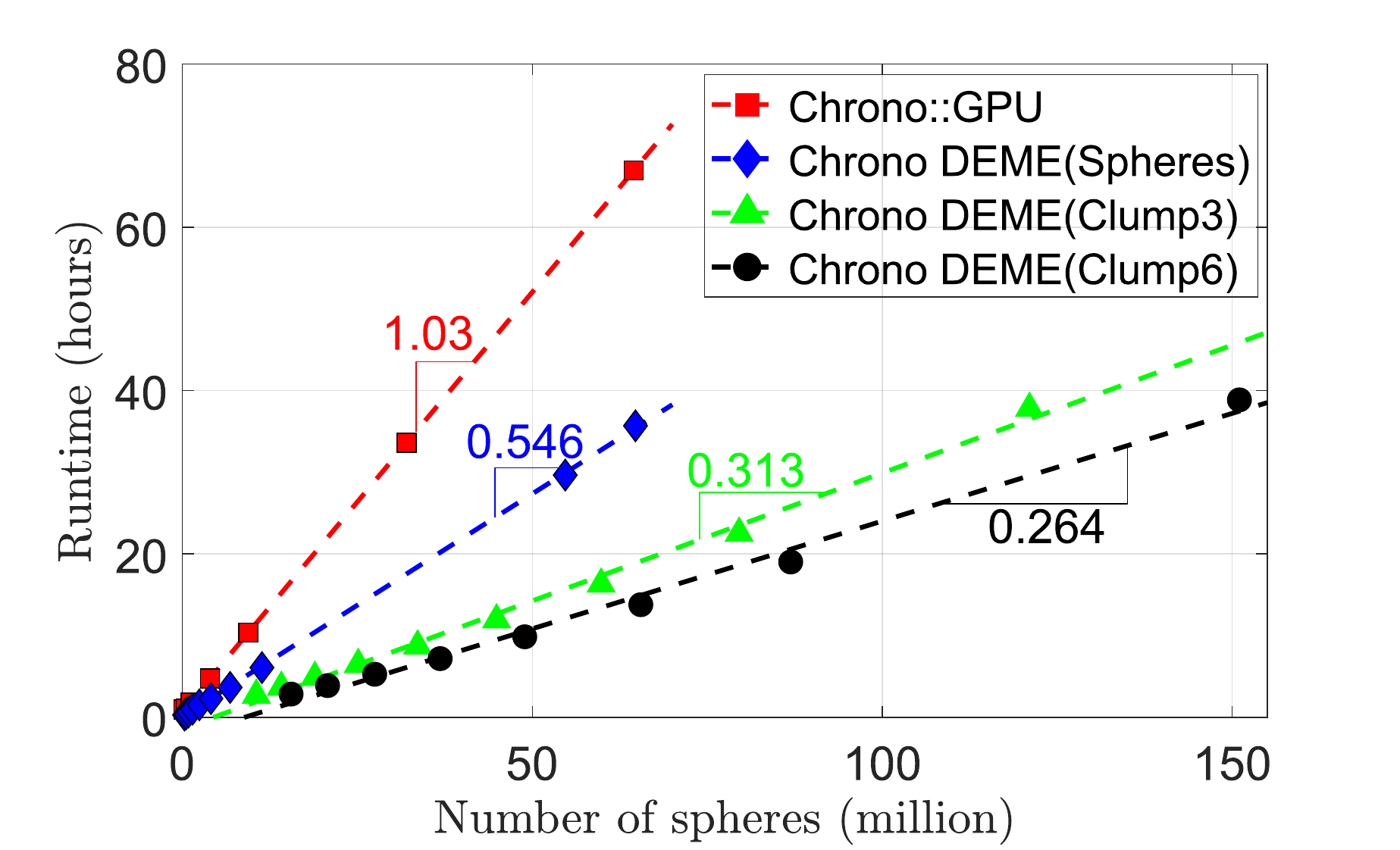}
	\caption{The scaling result of the mixer simulation using individual spheres, three-sphere clumps, and six-sphere clumps, on NVIDIA A100s. The wall time to finish simulating \SI{e6}{} steps is plotted against the number of component spheres in the simulation.}
	\label{fig:mixer_scaling}
\end{figure}

In this analysis, three clump types are employed: individual spheres, three-sphere clumps, and six-sphere clumps, depicted in Fig.~\ref{fig:mixer_clumps}. Element sizes are adjusted to regulate the total element count. The mesh representing the mixer blades remains consistent across simulations, comprising \SI{2892}{} triangular facets. Simulations are run until a pseudo-steady state is achieved at \SI{1}{s}, post which the wall time required to carry our \SI{e6}{} time steps is recorded. The time step size is \SI{5e-7}{s}. Figure~\ref{fig:mixer_scaling} displays the correlation between wall time and the total number of component spheres (distinct from the number of elements) via blue, green, and black markers. The simulations are performed on two NVIDIA Ampere A100 GPUs. On average, Chrono DEM-Engine takes 0.546, 0.313, and 0.264 hours to complete one million steps for every million component spheres in the simulations for the individual spheres, three-sphere clumps, and six-sphere clumps, respectively. The linear scaling persists to up to 150 million component spheres in the tests.

An identical simulation is also executed with Chrono::GPU (utilizing only one A100 as Chrono::GPU is limited to using a single GPU), and its scaling is represented with red markers. This juxtaposition is pertinent given a recent independent study's findings, which underscored that Chrono::GPU outperforms two other established DEM packages by two orders of magnitude~\cite{dem-pbrIdaho2023}. Therein, for a \SI{420000}{}-element pebble-packing simulation, Chrono::GPU running on a laptop GPU finished the simulation in an amount of time 261 times shorter than that required by LAMMPS, when the latter ran on 432 CPU cores of a cluster. For a \SI{660000}{}-element pebble-packing simulation, Chrono::GPU executed 501 times faster than \Starccm, which ran on 160 CPU cores. In both tests, Chrono::GPU ran on the RTX 2060 Mobile NVIDIA GPU card of a laptop. As indicated in Fig.~\ref{fig:mixer_scaling}, Chrono DEM-Engine demonstrates an additional twofold efficiency boost over Chrono::GPU in the test case of spherical elements. Owing to its ability to handle complex DEM particle shapes, Chrono DEM-Engine expands the modeling capacity of its predecessor without compromising per-GPU efficiency.

\begin{table}[h]
    \renewcommand{\arraystretch}{0.8}
    \centering
    \caption{The material and simulation properties used in the mixer scaling analysis.}
    \label{tab:mixer_mat}
    \begin{tabular}{*8c}
        \toprule
        Density [kg/m$^3$] & $E$ [Pa] & $\nu$ [-] & CoR [-] & Step size [s]\\
        \midrule
         \SI{2.6e3}{}   & \SI{1e9}{}         & 0.3        & 0.2  &   \SI{5e-7}{}      \\
        \bottomrule
    \end{tabular}
\end{table}

\begin{figure}[htp!]
    \centering
    \captionsetup{justification=centering}
    \includegraphics[width=.65\linewidth]{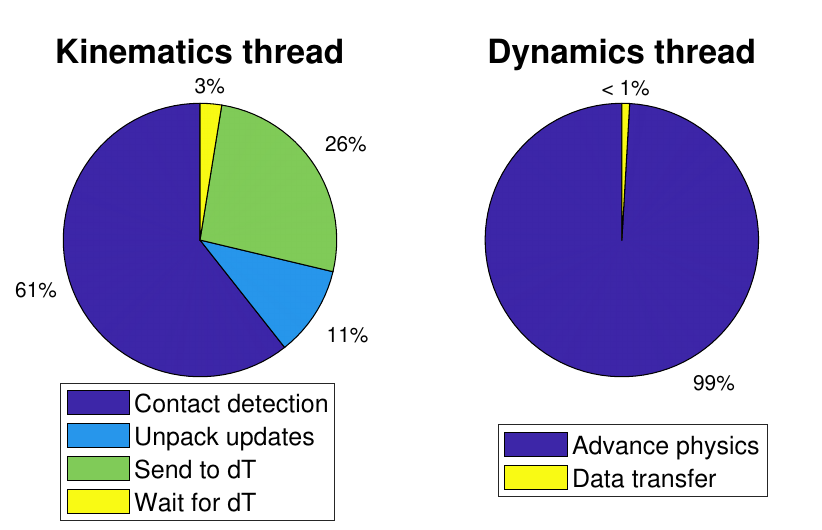}
    \caption{The runtime breakdown for the kinematics and dynamics threads, during the lifespan of the largest six-sphere-clump mixer simulation.}
    \label{fig:mixer_breakdown}
\end{figure}

\autoref{fig:mixer_breakdown} shows the time spent in the important steps of the kinematics and dynamics threads'  work cycles in the largest six-sphere-clump mixer simulation run in the scaling analysis. 
In that scenario, the amount of mutual contact data produced is relatively large, causing the kinematics thread to spend a large amount of time transferring it to the dynamics thread, reaching 26\% of the former thread's total runtime.  The dynamics thread spends minimal time on transferring data. This is done by design to enable the dynamics thread to almost exclusively focus on advancing the state of the system forward in time.

\section{Numerical experiments}\label{sec:num_exp}
This section introduces a series of numerical tests, from medium-sized hopper flow rate tests to large-scale co-simulation, designed to compare the DEM-Engine simulation results against experimental data.

\subsection{Ball impact test}

\begin{figure}[htp!]
    \centering
    \begin{minipage}{.45\textwidth}
        \centering
        \captionsetup{justification=centering}
        \includegraphics[width=\linewidth]{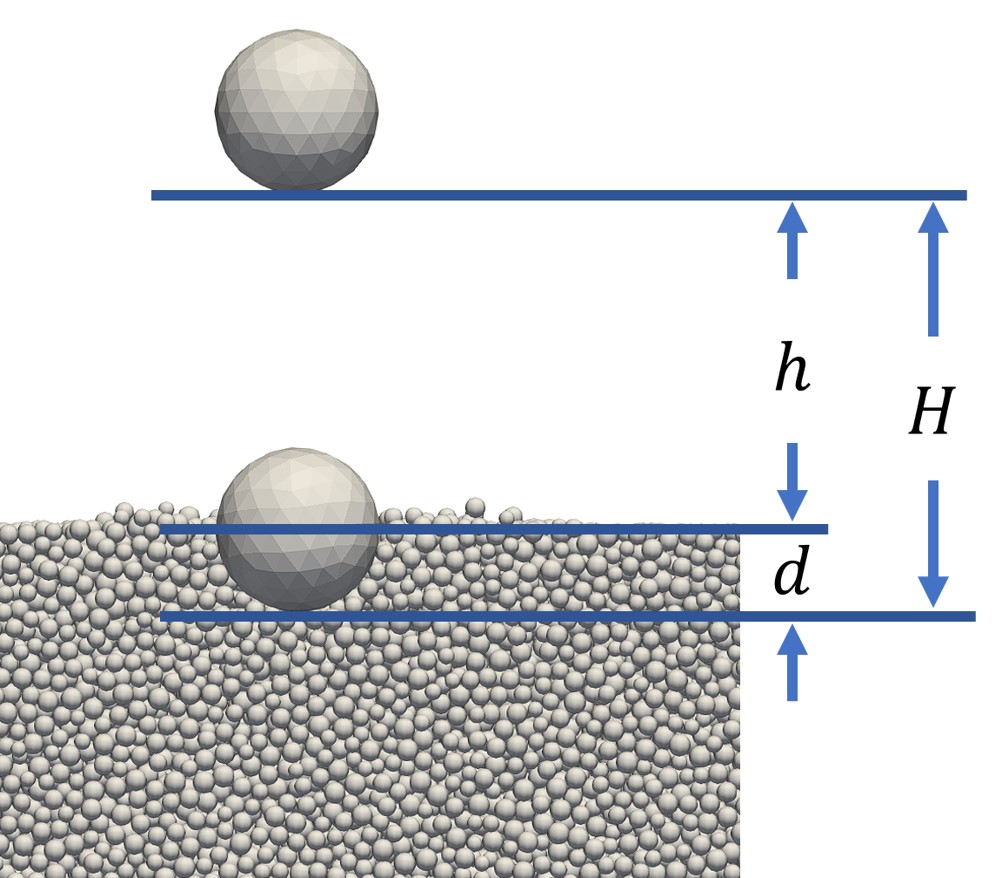}
        \caption{Diagram of the initial and final projectile positions.} \label{fig:balldrop_sketch}
    \end{minipage}%
    \hspace{.08cm}
    \begin{minipage}{.52\textwidth}
        \centering
        \captionsetup{justification=centering}
        \includegraphics[width=\linewidth]{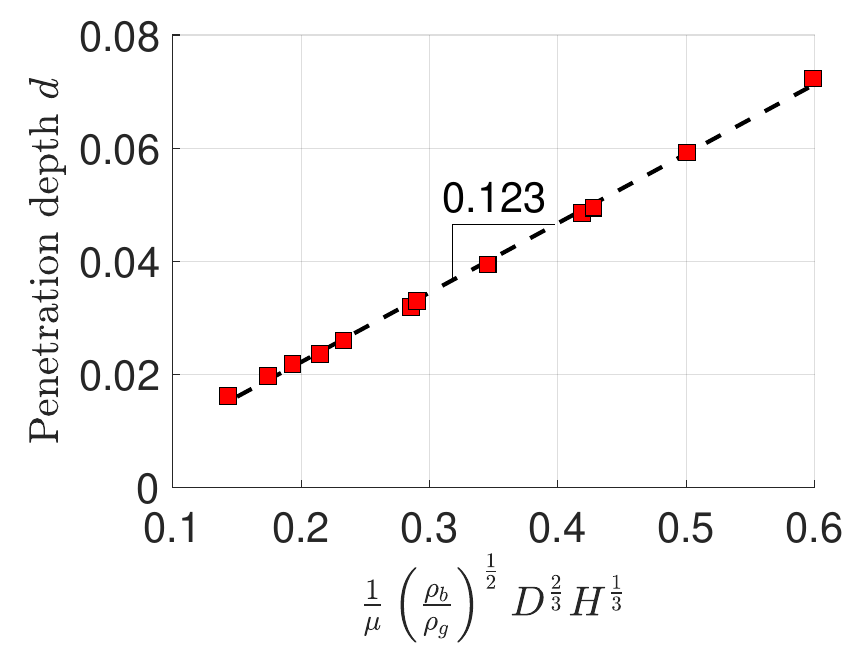}
        \caption{Penetration depth. Each red square represents a data point in the numerical test.}\label{fig:balldrop_res}
    \end{minipage}
\end{figure}

This experiment is described  in~\cite{durian2005}. A spherical projectile characterized by diameter $D$ and density $\rho_b$ was released from varying heights, $h$, onto a loosely packed pile of granular material, visualized in Fig.~\ref{fig:balldrop_sketch}. The resulting penetration depth $d$ of this sphere was gauged and set against the empirical model derived from the experimental data in~\cite{durian2005}
\begin{equation}
    \label{eq:crater_test_empirical_relation}
    d = \frac{C}{\mu}\left(\frac{\rho_b}{\rho_g}\right)^{\frac{1}{2}}D^{\frac{2}{3}}H^{\frac{1}{3}},
\end{equation}
where $\rho_g$ denotes the granular material's bulk density, and $H=h+d$ is the sum of penetration depth and drop height. In~\cite{durian2005}, the constant $C$ is estimated from experiments to be $C=0.14$.

Twelve numerical tests using DEM-Engine were run aiming to reproduce the experiment in~\cite{durian2005} as faithfully as possible. These tests incorporate combinations of projectile densities $\rho_b = 2.2, \, 3.8, \, 7.8, \, 15$ {\SI{}{\g/\cm\cubed}}, resembling Teflon, ceramic, steel, and tungsten, respectively. The diameter of the spherical projectile is $D=\SI{2.54}{\cm}$. The release heights take values $h = 5, \, 10, \, 20$ \SI{}{\cm}. Each simulation uses eleven types of spherical elements with diameters evenly distributed in the range between \SI{0.25}{\cm} and \SI{0.35}{\cm} (inclusive), and each DEM element has an even chance of spawning as one of them. The grain material in use has density $\rho_{\text{grain}}=\SI{2.5}{\g/\cm\cubed}$, resembling silica. This is to be differentiated from the bulk density of the granular bed, which is packed at $\rho_g=\SI{1.46}{\g/\cm\cubed}$, with a sliding friction coefficient of $\mu = 0.3$. 

The correlation between depth $d$ and the adjusted total release height $H$ can be observed in Fig.~\ref{fig:balldrop_res}. The line represents a linear regression of the numerical outcomes, showing a slope of $0.123$, which confirms the experimentally established empirical model in Eq.~(\ref{eq:crater_test_empirical_relation}). Comparable outcomes were also documented in~\cite{heynPhDThesis2013}~and~\cite{chronoGranular2021}, where both non-smooth and smooth contact dynamics approaches were leveraged for validating the same physics.

\subsection{Flow sensitivity test}
This section investigates the flow behavior exhibited by granular phases characterized by heterogeneous properties, encompassing variations in shape, density, and friction coefficient. Furthermore, relevant details regarding simulation runtimes are provided where applicable. The hardware configuration utilized for these numerical validations features an AMD Ryzen 9 5950X CPU in conjunction with a single NVIDIA A5000 GPU card.

\subsubsection{Drum tests}
\begin{figure}[htp!]
    \centering
    \captionsetup{justification=centering}
    \includegraphics[width=1\linewidth]{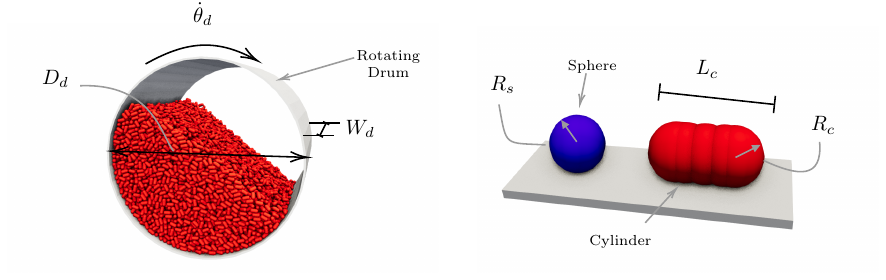}
    \caption{Schematic visualization of the drum rotating drum test.}
    \label{fig:flow:drum}
\end{figure}

The first test investigates the flowability of particle media comprising four typologies: plastic spheres, plastic cylinders, wooden spheres, and wooden cylinders. The reference data is presented in \citet{cui2023superDEM}, where experimental and numerical tests were performed on spherical and nonspherical particles. The experimental setup for the estimation of the angle of repose, a schematic of which is proposed in \autoref{fig:flow:drum}, comprised of a rotating drum made of transparent acrylic with an inner diameter ($D_d$) of \SI{0.19}{m} and a depth of \SI{0.20}{m} ($W_d$). For this investigation, the considered physical test outcomes refer to the test performing the drum rotating angular velocity, $\dot{\theta}_d$, of \SI{3.60}{} revolutions per minute (\SI{}{rpm}).

\begin{table}[h]
    \renewcommand{\arraystretch}{1}
    \caption{Properties of four different particle setups used in this numerical investigation.}
    \label{table:particle:definition}
    \begin{tabular}{*9c|cc}
        \toprule
        \emph{ID} & \emph{Material} & \emph{Shape} & Radius & Length & Density     & $E$     & $\nu$ & $CoR$  & Clumps & Spheres \\
                  &                 &              & [mm]   & [mm]   & [kg/m$^3$]  & [MPa] & [-]   & [-]  & [-]    & [-]     \\
        \midrule
        PS        & Plastic         & Sphere       & 3.0    & -      & \SI{1592}{} & 10.0  & 0.35  & 0.85 & 13024  & 13024   \\
        PC        & Plastic         & Cylinder     & 2.0    & 8.0    & \SI{1128}{} & 10.0  & 0.35  & 0.85 & 19036  & 95180   \\
        WS        & Wooden          & Sphere       & 2.95   & -      & 674         & 10.0  & 0.35  & 0.55 & 17112  & 17112   \\
        WC        & Wooden          & Cylinder     & 2.0    & 8.5    & 476         & 10.0  & 0.35  & 0.55 & 17016  & 85080   \\
        \bottomrule
    \end{tabular}
\end{table}

This test is also considered to assess the accuracy of DEM-Engine in simulating complex shapes, which are formed by a compound of spheres, and defined as clumps. In the following, as shown in Fig.~\ref{fig:flow:drum}, the two shapes that characterize the tested particles consist of pure spheres with uniform radii, and five sphere clumps to mimic the geometric outer shape of cylinders.

\begin{figure}[htp!]
    \centering
    \captionsetup{justification=centering}
    \includegraphics[width=1.0\linewidth]{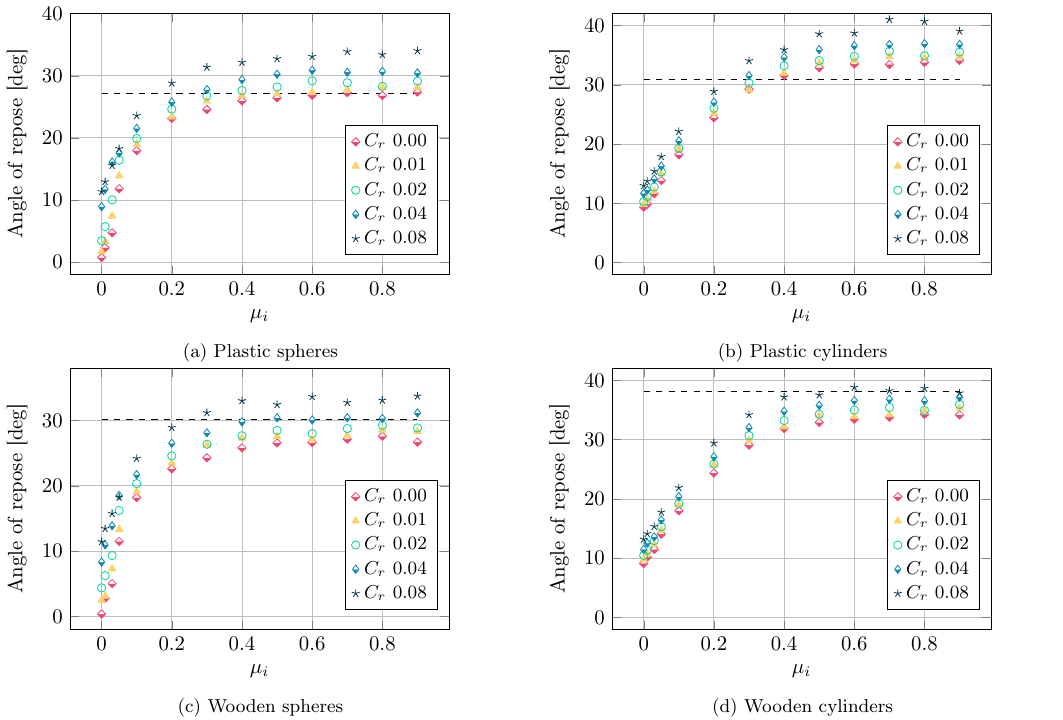}
    \caption{Sensitivity of the angle of repose to the inner ($\mu_i$) and rolling friction ($C_r$) for the four granular materials in Table \ref{table:particle:definition}. The dashed line in each chart reports the reference value for the corresponding experimental test \cite{cui2023superDEM}.}
    \label{fig:flow:sensitivity}
\end{figure}

Figure \ref{fig:flow:sensitivity} illustrates the sensitivity of the angle of repose for the rotating drum experiment. Each plot refers to a different material setup proposed in Table~\ref{table:particle:definition}, using a test matrix that uses 13 values $\in$ [0.00, 0.90] for the definition of the inner friction ($\mu_i$) and five values $\in$ [0.00, 0.08] for the definition of the rolling friction ($C_r$). 
The material is initialized to fill half of the volume of the drum; then, the drum initiates its rotation at a constant angular velocity of \SI{3.60}{rpm}, and let run for two seconds, after which it is assumed the system achieves a steady state. 
For the four different drum configurations, five seconds of simulations took approximately \SI{0.20}{h} for each case with spheres (i.e., PS and WS), whereas \SI{0.6}{h} hours for PC and WC.\@
The angle of repose, as reported in the charts, is computed as the mean value of thirty measurements taken at an interval during the three seconds of simulation. For all the cases, very little deviation was observed throughout the post-processing phase.

Figure \ref{fig:flow:sensitivity} illustrates some of the key characteristics exhibited by granular materials when simulated using a DEM-based numerical solver. Firstly, it is evident that as the internal friction assigned to the spheres approaches zero, the system response yields very small angles of repose, ultimately resulting in a near-horizontal surface in the absence of internal friction. Conversely, for cylindrical particles lacking internal friction, the shape itself contributes to the bearing capacity of the system, as expected. Moreover, rolling resistance influences the angle of repose. When $\mu_i$ is small, the disparity between cylindrical particles with and without $C_r$ remains consistently lower. Note that, for a given pair of ($\mu_i$, $C_r$), similar particle shapes yield comparable angles of repose, irrespective of particle size or density. These initial observations align with the findings reported in \cite{cui2023superDEM}, wherein the authors utilized the superquadratic DEM approach implemented in the open-source CFD suite MFiX \cite{gao2022superDEMimplementation} for simulating these same particles.

By contrasting the numerical solutions against the experimental data presented in \cite{cui2023superDEM} and illustrated in Fig.~\ref{fig:flow:sensitivity} through dashed black lines, one can assess the accuracy of the \DEME\ in simulating granular materials. First, when considering two simulated spherical particle materials (Fig.~\ref{fig:flow:sensitivity}a) and c)), in which the grain shapes align with their physical counterparts, the valid angles of repose exhibit a wide range of values in relation to internal friction (i.e., from $\mu_i$ $\in$ 0.25-0.90), while only minimal variability is linked to rolling friction. Secondly, employing 5-sphere clumps to emulate plastic and wooden cylinders, as reported in Fig.~\ref{fig:flow:sensitivity}b) and d), offers distinct operational domains for these un-physically consistent cylinders, where both shape and surface properties play pivotal roles. This analysis shows that the combined effects of internal and rolling frictions provide \DEME\ with greater versatility. 

\subsubsection{Hopper tests}
\begin{figure}[htp!]
    \centering
    \captionsetup{justification=centering}
    \includegraphics[width=0.5\linewidth]{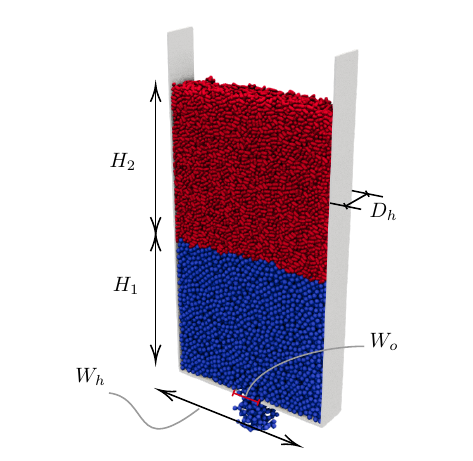}
    \caption{Schematic visualization of the flat-bottom hopper.}
    \label{fig:flow:hopper}
\end{figure}

This test assesses the dynamic properties exhibited by a flow of DEM particles when simulated using the DEM-Engine. As reference solutions for this task, data regarding the mass discharge rate for both single and binary component systems are targeted, as made available in \cite{jian2023shapeDEM}. The physical testing was conducted using a flat-bottom hopper, see Fig.~\ref{fig:flow:hopper}.

The hopper has a height of \SI{0.40}{\m}, width of \SI{0.20}{\m}, and depth of \SI{0.04}{\m}. An orifice of \SI{0.04}{\m} is symmetrically positioned on the lower surface. For the experimental campaign, various particle configurations were investigated. However, for this numerical validation, only four configurations, which precisely correspond to those outlined in Table~\ref{table:particle:definition}, are considered. Specifically, Table~\ref{table:hopper:definition} provides details on the hopper configuration for the tests presented in the subsequent sections. The parameters $\mu_i$ and \(C_r\) reported in the last two columns are set using the charts in Fig.~\ref{fig:flow:sensitivity}. Note that the first two tests consist of single-component discharge tests, whereas the remaining use binary particle compositions. Each simulation spans a physical time of \SI{7.50}{\s}, with approximate runtimes of: \SI{0.35}{h} for $ID$ 1; \SI{0.75} for $ID$ 2; and \SI{0.60} for $ID$s 3 and 4.

\begin{table}[h]
    \renewcommand{\arraystretch}{1.2}
    \caption{Properties of four different particle combinations used in the hopper numerical investigation.}
    \label{table:hopper:definition}
    \begin{tabular}{*7ccc|cc}
        \toprule
        \emph{Test ID} & \emph{Layer 1} & \emph{Layer 2} & $H_1$ & $H_2$ & $\mu_{i.1}$ & $C_{r.1}$ & $\mu_{i.2}$ & \(C_{r.2}\) & Clumps & Spheres \\
                       &                &                & [cm]  & [cm]  & [-]         & [-]       & [-]         & [-]         & [-]    & [-]     \\
        \midrule
        1              & PS             & -              & 36    & -     & 0.40        & 0.04      & -           & -           & 14058  & 14058   \\
        2              & WC             & -              & 36    & -     & 0.70        & 0.07      & -           & -           & 20014  & 100070  \\
        3              & PS             & PC             & 18    & 18    & 0.40        & 0.04      & 0.30           & 0.03           & 17545  & 59565   \\
        4              & PC             & PS             & 18    & 18    & 0.30        & 0.03      & 0.40        & 0.04        & 17904  & 61684   \\
        \bottomrule
    \end{tabular}
\end{table}

\begin{figure}[htp!]
    \centering
    \captionsetup{justification=centering}
    \includegraphics[width=0.6\linewidth]{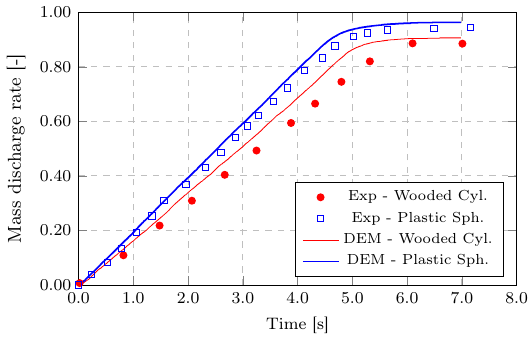}
    \caption{Experimental and numerical comparison of the mass discharge ratio for single component hoppers with plastic spheres (blue) and wooden cylinders (red).}
    \label{fig:flow:hopper:val}
\end{figure}

In Fig.~\ref{fig:flow:hopper:val}, the relative mass discharge is presented for Test IDs 1 and 2, which involve plastic sphere and wooden cylinder particles, respectively. The chart depicts a comparison between the experimental and numerical time evolution of the system, showcasing the mass discharge relative to the total mass. For both tests, \DEME\ demonstrates a fair level of accuracy in predicting the flow evolution. It exhibits an excellent match for purely spherical shapes (PS), while a slight overestimation is shown for the cylinders (PC). This discrepancy, leaning towards a more \textit{fluid} flow, can be attributed to the fact that the clumps of five spheres, used in place of actual cylindrical shapes, do not perfectly replicate the behavior of the physically consistent cylinders.

\begin{figure}[htp!]
    \centering
    \captionsetup{justification=centering}
    \includegraphics[width=0.95\linewidth]{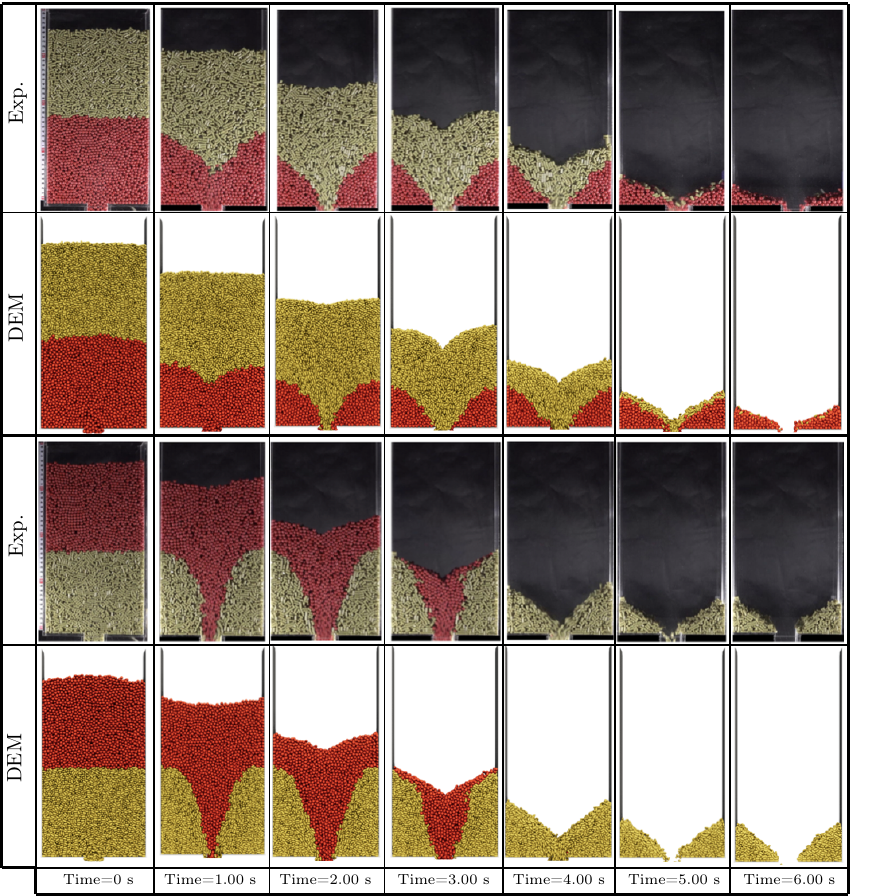}
    \caption{Experimental and numerical comparison of the discharging behavior of two different packing patterns of the Plastic Sphere and Cylinders. (reprinted from \cite{jian2023shapeDEM}; copyright (2023), LN 5657690415083, with permission from Elsevier).}
    \label{fig:flow:hopper:visual}
\end{figure}

In Fig.~\ref{fig:flow:hopper:visual}, a visual comparison is provided for the binary particle systems: Test IDs 3 and 4 as outlined in Table \ref{table:particle:definition}. This comparison contrasts snapshots from both experimental and numerical perspectives, offering lateral views of the hopper at one-second intervals, starting from the initial configuration at Time=\SI{0.00}{s}. The first and third rows respectively present data from \cite{jian2023shapeDEM}, while the second and fourth rows showcase the results from \DEME's simulation. The two timelines evolve in a remarkably similar fashion, highlighting that the numerical model accurately captures all the pertinent physical phenomena that unfold.

\subsection{Contact modeling for particle breakage}\label{sec:breakage}
DEM simulations have often been employed to characterize complex flows, factoring in not only the outer geometries of particles but also specialized features such as flexibility or particle breakage \cite{guo2015reviewComplexFlows}. The \DEME\ offers an open framework that allows users to implement user-defined constitutive laws. This example details a custom implementation to model the behavior of a cohesive yet highly brittle elastoplastic material. This test involves accounting for the failure of local bonds. To this end, the model outlined in \cite{scholtes2013DEMforFailure} is adopted for defining the constitutive laws and failure modes.

The following implementation leverages the variables presented in Sec.~\ref{sec:hertz_model} for the history-based Hertz--Mindlin model. Pivotal to this implementation is the capability of having stored information regarding the state of the system, as also detailed in Sec.~\ref{sec:default_model}. The material properties that are used to define, in this case, \textit{granite}, are used to define the contact forces. Concerning the parent contact method, two extra ``wildcards'' are defined: \sbelCode{unbroken} and \sbelCode{initialLength}, which are used to respectively record the contact state (i.e., broken or unbroken) and the initial length of the equivalent spring for the normal force. In the following, the general structure of the contact model is defined. Note that the value of the two wildcard variables are initialized to \SI{1.0}{} and \SI{0.0}{}, respectively.

\begin{lstlisting}[style=customcpp]
    // DEME force calculation for grain breakage.
    // The parameters required for the contact force computation are defined.
    
    if (unbroken > 1e-12) {
    // Computation of the contact force for the breakage model that accounts for normal and tangential forces, and bending moments.
    // Here goes the implementation
    } else {
        if (overlapDepth > 1e-12) {
    // The previously broken contact may still be engaged by compressive force, and this happens especially for compressive tests. The contact is treated with a Hertzian contact law.
    // Here goes the implementation
        }
    }
    \end{lstlisting}

The magnitude of the model calculates the normal interactive force $F_n$ using:
\begin{equation} \label{eq:break:kn}
    \vect{F}_n = k_n \vect{u}_n - \gamma_n \bar{m} \vect{v}_n,\\
\end{equation}
where $\gamma_n=0.01\sqrt{k_n/\bar{m}}$, $k_n$ is the normal stiffness and it is defined according to the following cases:
\begin{align} \label{eq:break:cappingOfkn}
    k_n =
    \begin{cases}
        E_{eq}\bar{R}               & \text{if } \sign(\vect{u}_n)\|\vect{u}_n\| > \delta_y,              \\
        \dfrac{-E_{eq}\bar{R}}{\xi} & \text{if } \delta_b \leq\sign(\vect{u}_n)\|\vect{u}_n\| < \delta_y, \\
        0                           & \text{otherwise,}
    \end{cases}
\end{align}
where $E_{eq}$ is the equivalent stiffness of the contact, $\xi$ is the degrading factor (softening) that accounts for the formation of initial cracks in the material, $\delta_y$ is the material yielding threshold, and $\delta_b$ is contact displacement failure, here assumes as three times $\delta_y$.
\begin{lstlisting}[style=customcpp]
    float tension = -9.3e6f;

    // Normal force calculation
    float deltaD = (overlapDepth - initialLength);
    float kn = Eeq * (ARadius * BRadius) /((ARadius + BRadius));    

    float intialArea = ((ARadius > BRadius) ? ARadius * ARadius : BRadius * BRadius) * deme::PI;

    float BreakingForce = tension * intialArea;
    float deltaY = BreakingForce / kn;
    float deltaU = 3.0f * deltaY;
    

    float force_to_A_mag = (deltaD > deltaY) ? kn * deltaD : ((deltaU - deltaD)-deltaY) * kn * 0.5f;
    
    float damping = 0.01 * sqrt(mass_eff * kn);

    force += B2A * force_to_A_mag - damping * velB2A;
    // breaking for excess of tensile force
    unbroken = (deltaD < deltaU) ? -1.0 : unbroken;
\end{lstlisting}

The tangential component of each contact follows from the following relationship:
\begin{equation}
    \vect{F}_t = -k_t \vect{u}_t -  \gamma_t \bar{m} \vect{v}_t,  \ \	\text{given:} \ \ \|\vect{F}_t\| \leq \;
    \begin{cases}
        \mu \|\vect{F}_n\| +c\, A_\text{int} & \text{if } \sign(\vect{u}_n)\|\vect{u}_n\| > \delta_y,    \\
        \mu \|\vect{F}_n\|                   & \text{if } \sign(\vect{u}_n)\|\vect{u}_n\| \leq \delta_y,
    \end{cases}
\end{equation}
where $k_t$ = $\nu_i k_n$, $c$ is the material cohesion, and $A_\text{int}$ is the interacting surface for the contact and defined as $\pi\cdot\min(Ri,Rj)^2$. The following snippet of code provides the specific details of the implementation:
\begin{lstlisting}[style=customcpp]
    float cohesion = 200e6;
    // Tangential force calculation

    float kt = nu_cnt * kn;    
    float Fsmax = (deltaD > deltaY) ? length(force) * mu_cnt + cohesion * intialArea : length(force) * mu_cnt;
    
    const float loge = (CoR_cnt < 1e-12) ? log(1e-12) : log(CoR_cnt);
    beta = loge / sqrt(loge * loge + deme::PI * deme::PI);
    float gt = 2. * sqrt(5. / 6.) * beta * sqrt(mass_eff * kt);
    
    float3 tangent_force = -kt * delta_tan - gt * vrel_tan;
    delta_tan = (tangent_force + gt * vrel_tan) / (kt);

    force += tangent_force;
    // breaking for excess of tangential stress
    unbroken = (length(tangent_force) > Fsmax) ? -1.0 : unbroken;
\end{lstlisting}

The bending resistance that arises at each contact, being representative of an element of finite size, is computed using Eq.~\eqref{subsubeq:Crr}, where the bending stiffness is defined as $k_r = R_i R_j k_t$ \cite{belheine2009DEMwithBending, liu2020DEMFailureBallast}. note that the maximum bending moment is capped by $\min(\eta_iRi,\eta_jR_j)\|\vect{F}_n\|$, where $\eta$ is a dimensionless coefficient that controls the rolling behavior of the contact. Lastly, here the code for the implementation of the fictitious bending resistance of the contact is listed. Note that no contact failure is associated with the bending moment value.
\begin{lstlisting}[style=customcpp]
    // Bending moment induced-force calculation
    float kr = ARadius * BRadius * kt;
    float eta = 0.1f;

    float var_1 = ts * kr / ARadius;
    float var_2 = eta * length(force);

    float3 torque_force;
    if (v_rot_mag > 1e-12) {
        float torque_force_mag = (var_1 < var_2) ? var_1 : var_2;
        torque_force = (v_rot / v_rot_mag) * torque_force_mag;
    }
    force += torque_force;
\end{lstlisting}

\begin{figure}[htp!]
    \centering
    \captionsetup{justification=centering}
    \includegraphics[width=0.50\linewidth]{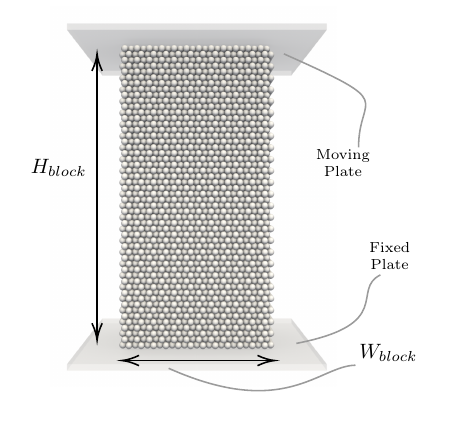}
    \caption{Numerical configuration for the axial compression test of a granite block.}
    \label{fig:breakage:scheme}
\end{figure}

The previous implementation has been validated against experimental data from a uniaxial compression test conducted on a granite block, as defined in \cite{Potyondy2004DEMforRock}. This particular test configuration is commonly utilized in the literature for code validation and calibration. In Fig.~\ref{fig:breakage:scheme} and Table \ref{table:properties:granite}, we present the numerical test rig along with the mechanical properties of the rock specimen, lower plate, and upper plate. These properties have also been reviewed and interpreted by other studies \cite{wang2009graniteDEM,  scholtes2013DEMforFailure, liu2020DEMFailureBallast}.
The test rig consists of two rigid plates, with the lower plate fixed to the reference system while the upper plate moves vertically at a constant velocity of \SI{5}{mm/s}. The tested specimen is constructed as a homogeneous assembly of spheres placed on a regular lattice arranged in a hexagonal close-packed (HCP) configuration, generated using an internal function provided by the \DEME\ package. The specimen has a base area of $W_{block} \times W_{block}$ with dimensions of \SI{5.0}{\cm} and a height $H_{block}$ of \SI{10.0}{\cm}. The chosen sphere radius of \SI{12}{\mm} ensures that there are 20 particles within the width of the specimen.

\begin{table}[h]
    \renewcommand{\arraystretch}{1}
    \caption{Mechanical properties of the granite block, as defined in \cite{Potyondy2004DEMforRock, scholtes2013DEMforFailure}}
    \label{table:properties:granite}
    \begin{tabular}{ccc}
        \toprule
        \emph{Material} & \emph{Parameter}   & \emph{Value (unit)} \\
        \midrule
        Rock            & Mass density       & \SI{2640}{kg/\m\cubed}    \\
                        & Young's modulus    & \SI{60e9}{Pa}        \\
                        & Poisson's ratio    & \SI{0.25}{}         \\
                        & Internal friction  & \SI{0.30}{}         \\
                        & Compressive Strength & \SI{200e6}{Pa}       \\
                        & Tensile strength & \SI{9.3e6}{Pa}       \\
        \midrule
        Plates & Young's modulus    & \SI{100e9}{Pa}      \\
        & Surface friction   & \SI{0.50}{}\\
        & Poisson's ratio    & \SI{0.30}{}\\
        \bottomrule
    \end{tabular}
\end{table}

A crucial parameter that significantly influences the accuracy of the proposed model for contact breaking is the particle interaction range, denoted as $\gamma_\text{int}R_i$, which defines the area of active links around each particle. Essentially, when a particle is initialized as part of the previously defined contact method, it is equipped with contacts that extend to the surrounding particles in accordance with the specified interaction range. In this study, three tests are conducted, considering different values of $\gamma_\text{int}$: [$0.70$, $0.90$, $1.10$]. Table~\ref{table:breakage:definition} provides a summary of the micro properties for these three tests, including the total number of potential contacts, denoted as $N$, and the statistical mode of the number of contacts for a single particle, denoted as $N_{i.\text{mod}}$.

\begin{table}[h]
    \renewcommand{\arraystretch}{1.2}
    \caption{Model parameters description for the simulation of the particle breakage in axial compressive tests.}
    \label{table:breakage:definition}
    \begin{tabular}{ccc|ccc}
        \toprule
        \emph{Test ID} & \emph{Radius} & \emph{$\gamma_{\text{int}}$} & \emph{Spheres}          & $\approx N$       & $N_{i.\text{mod}}$ \\
                       & [mm]          & [-]                          & [-] & [$\times$ 10$^3$] & [-]         \\
        \midrule
        1& \SI{12}{}     & 0.70  & 26754 & 154               & 6           \\
        2& \SI{12}{}     & 0.90  & 26754 & 200  & 8 \\
        3& \SI{12}{}     & 1.10   & 26754            & 230               & 9           \\
        \bottomrule
    \end{tabular}
\end{table}
 
\begin{figure}[htp!]
    \centering
    \captionsetup{justification=centering}
    \includegraphics[width=0.5\linewidth]{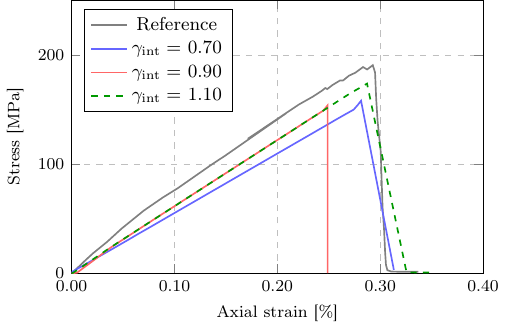}
    \caption{Strain--stress curves obtained from uniaxial compressive tests performed with three interaction range sizes. The reference solution corresponds to the numerical solution proposed in \cite{scholtes2013DEMforFailure} for $N_i=13.8$.}
\label{fig:breakage:validation}
\end{figure}

\begin{figure}[htp!]
    \centering
    \captionsetup{justification=centering}
    \includegraphics[width=0.7\linewidth]{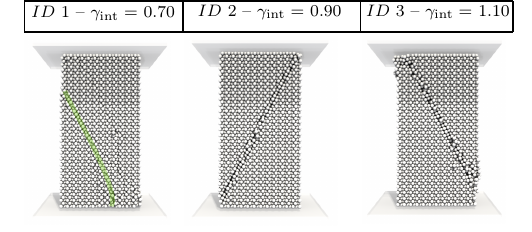}
    \caption{Visualization of the cracked configuration of the three specimens. For test $ID$ 1, the crack has been highlighted using a light green curve.}
    \label{fig:breakage:visual}
\end{figure}

Figure~\ref{fig:breakage:validation} displays the strain--stress curves for the three tests along with the numerical solution proposed in \cite{scholtes2013DEMforFailure}, where the average number of contacts per particle was $N_i=13.8$. The data presented in this chart suggests the excellent agreement achieved by the implemented model compared to the one from the literature. Particularly, an increase in the interaction range leads to a more accurate representation of the specimen's stiffness. Case emph{ID} 3 exhibits the highest level of agreement, with a relative error of less than 8\% on the material ultimate resistance and 5\% on the elastic modulus. One source of disagreement lies in the relatively small number of links (i.e., 9 compared to 13.8), which is a direct consequence of the uniform pattern used to initialize the particle arrangement and the uniform particle radius. Figure~\ref{fig:breakage:visual} proposes rendered visualizations for the final instants of the three tests.

\subsection{Rover mobility co-simulation} \label{sec:rover}

\begin{figure}[htp!]
    \centering
    \captionsetup{justification=centering}
    \includegraphics[width=\linewidth]{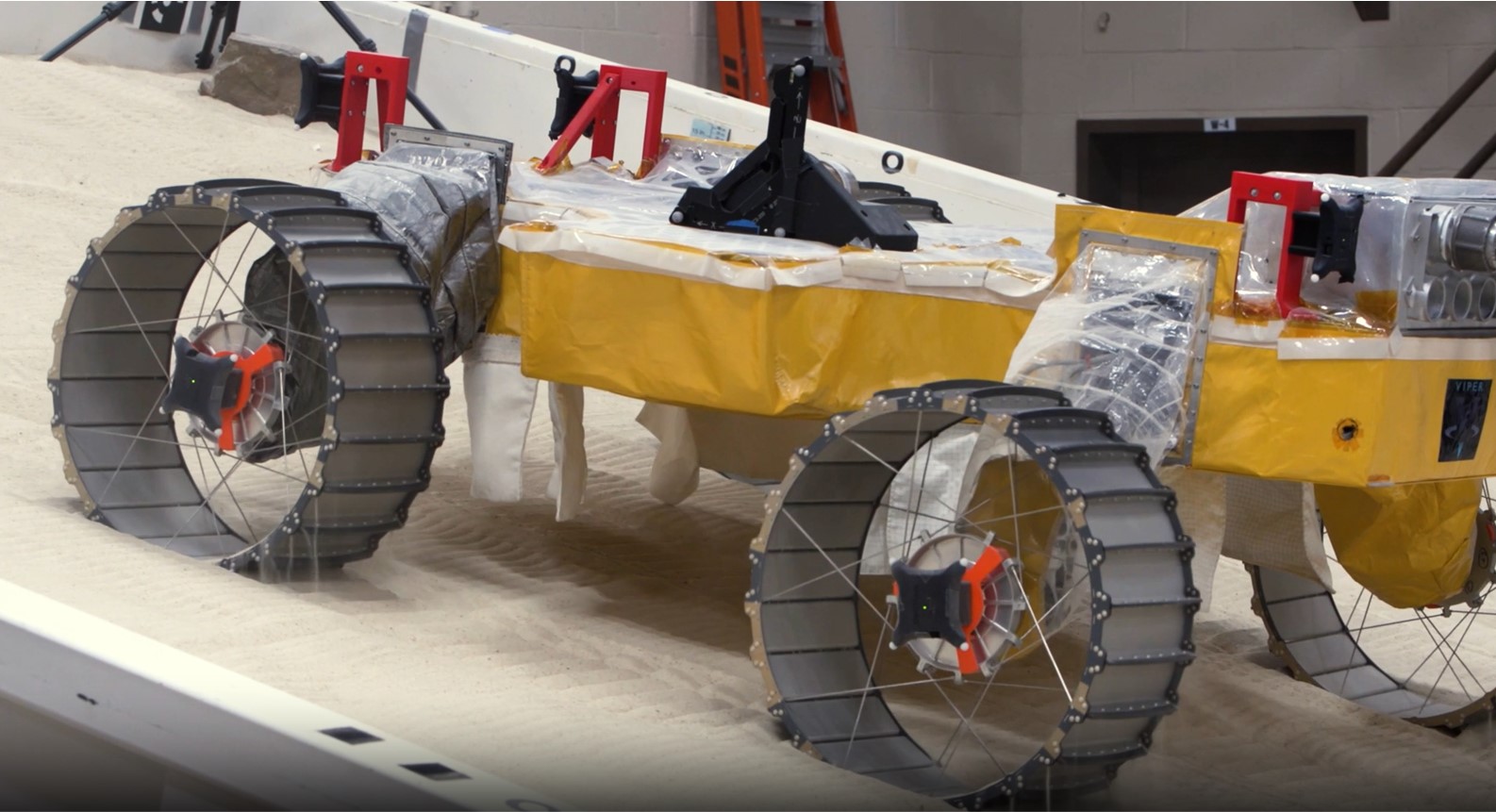}
    \vspace{5pt}
    \caption{MGRU3 climbing a ``tilt bed'' in NASA's Glenn Research Center testing facility~\cite{movieMGRU3Tilt}.}\label{fig:MGRU3movie}
    \vspace{-10pt}
\end{figure}

This section discusses a co-simulation between a multi-body system and a DEM system. The rover simulation originally presented in~\cite{zhang2023gpuaccelerated} is reproduced herein while adding the usage of the ``active box'' scheme introduced later in this section. The co-simulation aims to measure the slip ratios of a rover when operating on a ``tilt bed'' under Earth's gravitational pull. The experimental data used for comparison are obtained using NASA’s Moon Gravitation Representative Unit 3 (MGRU3), see Fig.~\ref{fig:MGRU3movie} (obtained from a publicly available video of the test~\cite{movieMGRU3Tilt}) for a photo of the test scene. However, in the co-simulation presented herein, since the MGRU3 CAD model is inaccessible, a similar VIPER rover model publicly available in the latest Chrono distribution~\cite{chronoOverview2016} is used. The rover moves around by prescribing all its four wheels a \SI{0.8}{rad/s} angular velocity on inclines of 0, 5, 10, 15, 20, and 25$^\circ$, where the inclines are modeled in simulation by adjusting the direction of the gravitational pull.

\begin{figure}[htp!]
    \centering
    \begin{minipage}{.45\textwidth}
        \centering
        \captionsetup{justification=centering}
        \includegraphics[width=.85\linewidth]{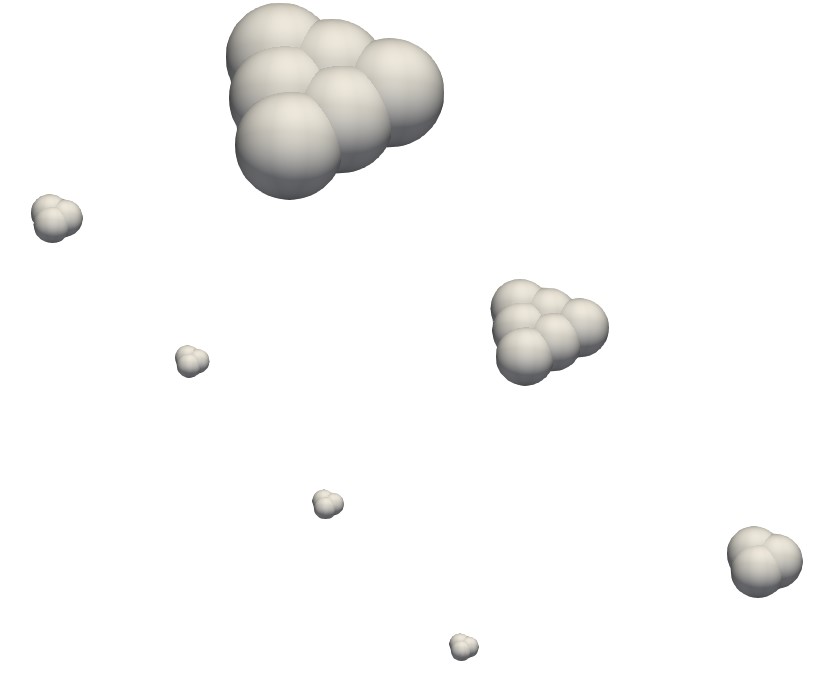}
        \caption{The seven clump shapes that are used in the rover co-simulation.} \label{fig:seven_types}
    \end{minipage}
    \hspace{.1cm}
    \begin{minipage}{.45\textwidth}
        \centering
        \captionsetup{justification=centering}
        \includegraphics[width=\linewidth]{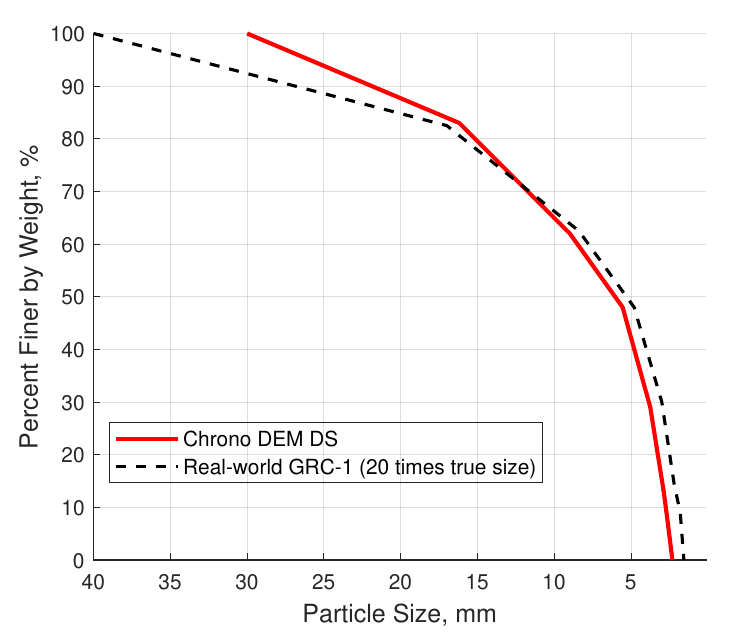}
        \caption{The size distribution of the DEM elements used in the rover co-simulation, plotted against a scaled real-world GRC-1 simulant size distribution.}\label{fig:GRCDS}
    \end{minipage}%
\end{figure}

The experiment shown in Fig.~\ref{fig:MGRU3movie} was done at Glenn Research Center, where the terrain simulant used is called GRC-1~\cite{ORAVEC2010361}. In the co-simulation presented herein, the numerical representation of the terrain is inherited from~\cite{zhang2023gpuaccelerated}, where seven different DEM element types are used (rendered in Fig.~\ref{fig:seven_types}), each with a specific size and percentage of the total weight, see Table~\ref{tab:GRCDS}. The size distribution is plotted in Fig.~\ref{fig:GRCDS}, showing the DEM representation is uniformly increased by a factor of 20 the actual particle sizes encountered in GRC-1. For more details and the validation of this terrain representation, see~\cite{zhang2023gpuaccelerated}. A rendering of the co-simulation is shown in Fig.~\ref{fig:rover_render}.

\begin{table}[htp!]
    \centering
    \caption{The weight distribution of the simulant used in the rover test, percent-wise, by clump size. For all element types, $E={\SI{e8}{N/m^2}}$, $\nu=0.3$, $\mu_s = 0.4$, and $\text{CoR}=0.5$ in this simulation.}
    \label{tab:GRCDS}
    \renewcommand{\arraystretch}{0.8}
    \begin{tabular}{*8l}    \toprule
        \emph{Type}                  & 1   & 2    & 3    & 4    & 5    & 6    & 7   \\\midrule
        Size [\SI{}{mm}]             & 21  & 11.4 & 6.6  & 4.5  & 3    & 2.75 & 2.5 \\
        Component radius [\SI{}{mm}] & 3.6 & 1.95 & 1.81 & 1.24 & 0.82 & 0.75 & 0.7 \\
        \%, by weight                & 17  & 21   & 14   & 19   & 16   & 5    & 8   \\ \bottomrule
        \hline
    \end{tabular}
\end{table}

\subsubsection{Co-simulation}

\begin{figure}[htp!]
    \centering
    \captionsetup{justification=centering}
    \includegraphics[width=.9\linewidth]{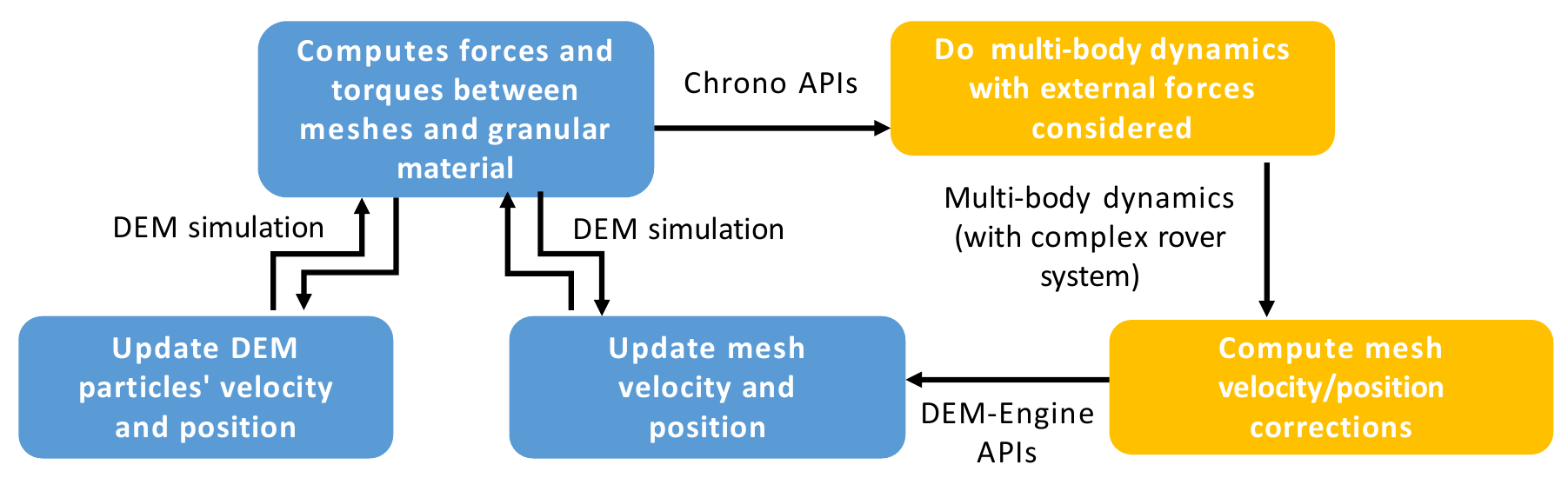}
    \caption{The co-simulation workflow between the multi-body system simulated by Chrono and DEM-Engine.} \label{fig:cosim_workflow}
\end{figure}

The co-simulation setup is depicted in Fig.~\ref{fig:cosim_workflow}. DEM-Engine handles the evolution of the granular terrain, while Chrono manages the rover dynamics. The two simulators are connected through the meshes representing the wheels. DEM-Engine calculates the force exerted by the terrain on the wheel mesh. This force information is employed when the Chrono numerical integrator propels the evolution of the meshes forward in time. Subsequently, the updated position of the wheels will serve as new boundary conditions for the granular material. The rover's mobility is also influenced by forces that originate in the chassis and suspension, independent of the motion of the granular terrain.
In this co-simulation, the rover system progresses with a time step size of \SI{2e-5}{s}, whereas the DEM system uses a smaller time step of \SI{2e-6}{s}. This means for every ten DEM time steps, the multi-body system in Chrono advances by just one step.

\subsubsection{Active box scheme}

\begin{figure}[htp!]
    \centering
    \captionsetup{justification=centering}
    \includegraphics[width=\linewidth]{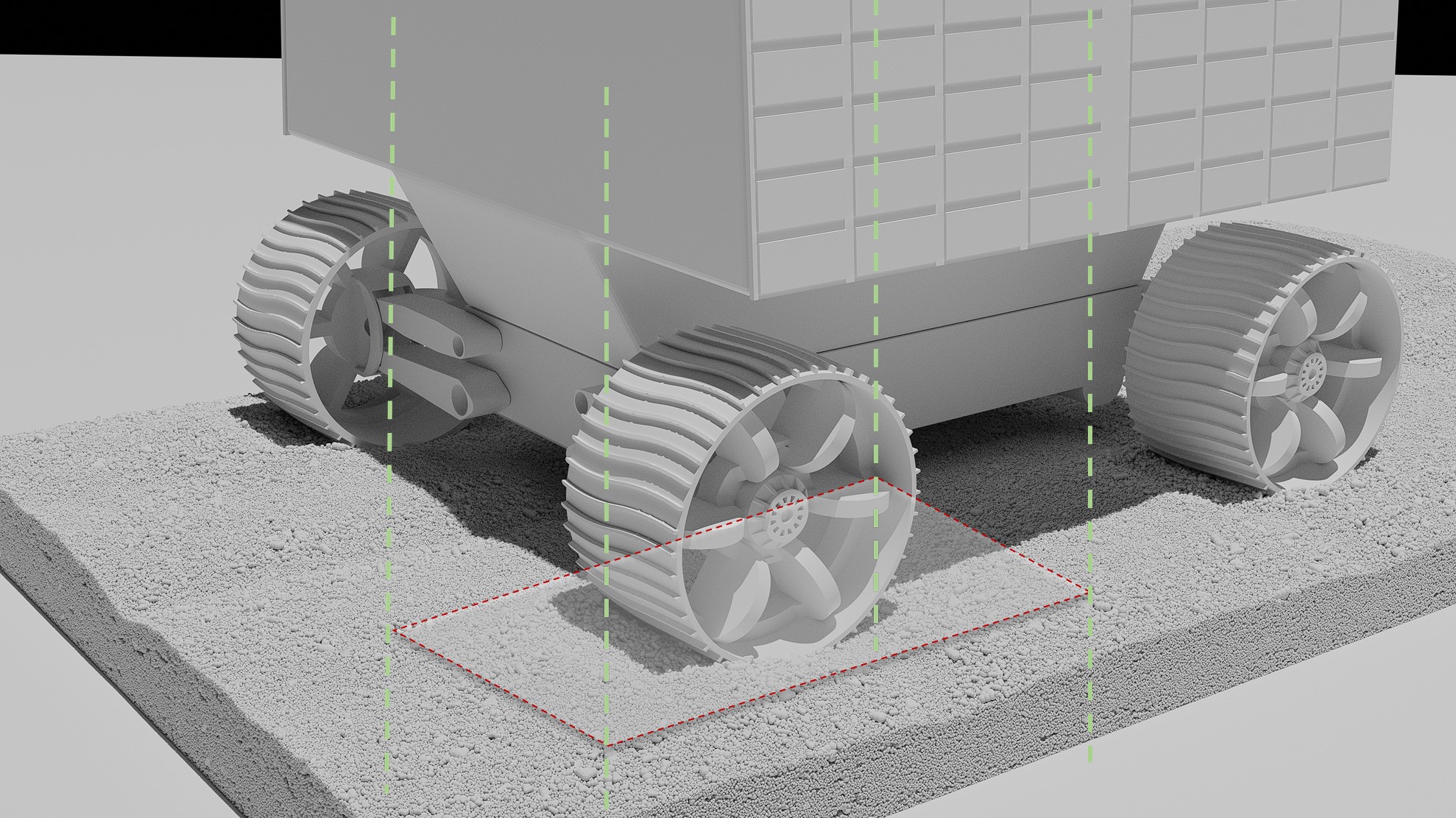}
    \caption{A rendering of the VIPER rover operating on a $20^\circ$ incline. The active box is marked and only the elements in that region are subject to the simulation physics; the rest are fixed.} \label{fig:rover_render}
\end{figure}

Using DEM-Engine's API, the user can implement a partially active simulation domain to reduce computational cost. The user can assign different family tags (introduced in Sec.~\ref{sec:family}) to the elements inside and outside certain regions in the simulation domain to distinguish them. In this use case, no assigned motions are prescribed to the elements inside the $\SI{1}{m}\times\SI{0.5}{m}$ boxes centered around each wheel, as shown in Fig.~\ref{fig:rover_render} -- their motion is to be determined by the simulator. These boxes are called active boxes. The DEM elements outside the active boxes are fixed in position and do not participate in the contact detection, i.e., remain dormant and contribute no computational cost. Note that the locations of the active boxes are updated (based on the locations of the wheel) 10 times per simulation second in this test.

\begin{figure}[htp!]
    \centering
    \captionsetup{justification=centering}
    \includegraphics[width=0.7\linewidth]{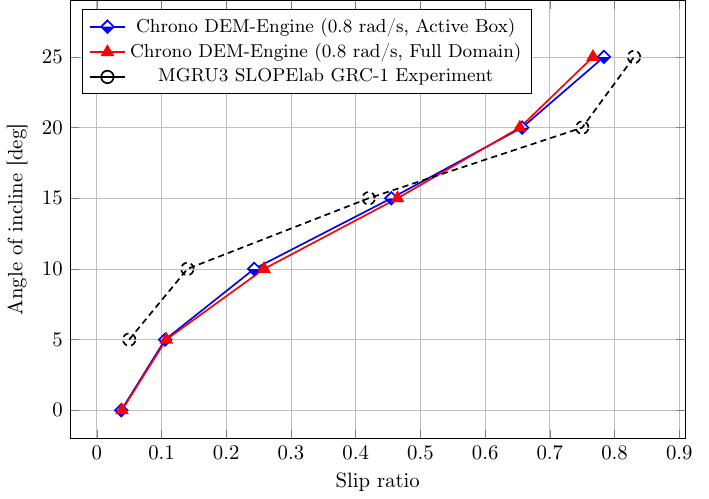}
    \caption{The comparison between the full-domain and active box-based VIPER rover slip test results. The experimental data used for comparison (black line) are from Glenn Research Center's MGRU3 experiments with the GRC-1 simulant.}\label{fig:rover_res}
\end{figure}

The full-simulation data shown in Fig.~\ref{fig:rover_res} displays no notable difference compared to the active box-based counterpart. In~\cite{zhang2023gpuaccelerated}, it is reported that the 15-second simulation requires approximately \SI{109}{} hours of run time on two NVIDIA A100 GPUs. At the same time, the active box-based simulation presented herein, which involves \SI{11336638}{} DEM elements (\SI{34691952}{} component spheres) takes around \SI{30}{} hours.
This suggests that the more expeditious active box-based tests are likely sufficient to gain insights into the rover's mobility attributes, while costing a fraction of the computational cost of a full simulation. The numerical simulations also show good agreement with the experimental data (black line) from NASA's Glenn Research Center MGRU3 experiments with the real-world GRC-1 simulant. The slip ratio increases relatively slowly with the slope angle in the interval between $0^\circ$ and $10^\circ$. Past $10^\circ$, this rate of increase escalates, and the rover almost fails to climb on a $25^\circ$ incline.

\section{Conclusions and future directions}\label{sec:conclusion}

This paper has introduced Chrono DEM-Engine, an open-source, physics-based, dual-GPU DEM package that supports complex element shapes, positioning it as an enhancement to the existing Chrono::GPU simulator. 
The most distinctive implementation feature of Chrono DEM-Engine is its partitioning of kinematic processes, such as contact detection, and dynamics computations, e.g., computation of the contact forces and carrying out numerical integrations. The resulting two computational threads operate asynchronously and share data only when necessary. Chrono DEM-Engine supports custom force models through just-in-time CUDA kernel compilation. 

This manuscript first presents the \Cpp\ and Python code implementations. They are detailed to highlight the primary code features and specialized software components. From the default force model, Hertz--Mindlin, which possesses the capability to trace the history of contact interactions, the paper focuses on the code structure. Emphasis is placed on the data handling, accompanied by an overview of the procedure for customizing the force model. Following a rigorous contact validation against analytical solutions, the computational performance of \DEME's core implementation is evaluated. This new DEM simulator can process tens of millions of elements on two A100 GPUs, achieving a throughput of one million time steps for one million DEM elements within an hour. In contrast to its predecessor, Chrono::GPU, which demonstrated in third-party studies to be two orders of magnitude faster than established DEM packages, the scaling analysis in this paper reveals that the new solver further increases this performance by a factor of 2$\times$. Furthermore, the new simulator demonstrates linear scalability for up to 150 million component spheres using two GPUs.

The paper validates the solver's implementation through a comprehensive set of tests, including fine-grain force model evaluations and macro-scale experiments, such as ball drop, hopper flow rate, and rover climbing. The software is designed to handle complex particle geometries using clump models. This feature is validated through comparisons with physical data for the flow discharge of spheres, cylinders, and combinations thereof from a rectangular hopper. Moreover, the software integrates with the multi-physics simulation engine Chrono, facilitating co-simulations with mechanical and multi-body systems, as evidenced by the proposed test case of simulating the rover operation.

Chrono DEM-Engine is an open-source, BSD3-distributed research code. As such, there is an inherent learning curve associated with its use. Users are required to sift through numerous APIs. Identifying and addressing the tool's limitations can also be daunting and may require time-consuming customization. This challenge becomes even more pronounced in modern cross-disciplinary research, where researchers are simultaneously handling a range of tools. However, the emergence of Large Language Models (LLMs)~\cite{chatgpt} offers a potential solution. As a future development thrust, it remains to investigate the use of LLMs to design assistant AIs that can translate users' natural language directives into executable DEM-Engine scripts. If this research trajectory proves successful, the resulting tool will be made available as open-source.

\section*{Code availability}

Chrono DEM-Engine is accessible as part of Project Chrono at \url{https://github.com/projectchrono/DEM-Engine}. All numerical examples discussed in this paper are provided as demo simulations.

\section*{Acknowledgments}

B. Tagliafierro gratefully acknowledges financial support for this publication by the Fulbright Schuman Program, which is administered by the Fulbright Commission in Brussels and jointly financed by the U.S. Department of State, and the Directorate-General for Education, Youth, Sport and Culture (DG.EAC) of the European Commission. The content of this manuscript does not represent the official views of the Fulbright Program, the Government of the United States, or the Fulbright Commission in Brussels. This work has been partially supported by NSF projects OAC2209791 and CISE1835674, and the US Army Research Office project W911NF1910431.

\bibliography{refsChronoSpecific,refsDEM,refsFSI,refsMBS,refsRobotics, refsSBELspecific, refsTerramech, refsCompSci,  refsNumericalIntegr, refsMachineLearning, refsStatsML, refsSensors}

\def\cprime{$'$}

\begin{thebibliography}{85}
\ifx \bisbn   \undefined \def \bisbn  #1{ISBN #1}\fi
\ifx \binits  \undefined \def \binits#1{#1}\fi
\ifx \bauthor  \undefined \def \bauthor#1{#1}\fi
\ifx \batitle  \undefined \def \batitle#1{#1}\fi
\ifx \bjtitle  \undefined \def \bjtitle#1{#1}\fi
\ifx \bvolume  \undefined \def \bvolume#1{\textbf{#1}}\fi
\ifx \byear  \undefined \def \byear#1{#1}\fi
\ifx \bissue  \undefined \def \bissue#1{#1}\fi
\ifx \bfpage  \undefined \def \bfpage#1{#1}\fi
\ifx \blpage  \undefined \def \blpage #1{#1}\fi
\ifx \burl  \undefined \def \burl#1{\textsf{#1}}\fi
\ifx \doiurl  \undefined \def \doiurl#1{\url{https://doi.org/#1}}\fi
\ifx \betal  \undefined \def \betal{\textit{et al.}}\fi
\ifx \binstitute  \undefined \def \binstitute#1{#1}\fi
\ifx \binstitutionaled  \undefined \def \binstitutionaled#1{#1}\fi
\ifx \bctitle  \undefined \def \bctitle#1{#1}\fi
\ifx \beditor  \undefined \def \beditor#1{#1}\fi
\ifx \bpublisher  \undefined \def \bpublisher#1{#1}\fi
\ifx \bbtitle  \undefined \def \bbtitle#1{#1}\fi
\ifx \bedition  \undefined \def \bedition#1{#1}\fi
\ifx \bseriesno  \undefined \def \bseriesno#1{#1}\fi
\ifx \blocation  \undefined \def \blocation#1{#1}\fi
\ifx \bsertitle  \undefined \def \bsertitle#1{#1}\fi
\ifx \bsnm \undefined \def \bsnm#1{#1}\fi
\ifx \bsuffix \undefined \def \bsuffix#1{#1}\fi
\ifx \bparticle \undefined \def \bparticle#1{#1}\fi
\ifx \barticle \undefined \def \barticle#1{#1}\fi
\bibcommenthead
\ifx \bconfdate \undefined \def \bconfdate #1{#1}\fi
\ifx \botherref \undefined \def \botherref #1{#1}\fi
\ifx \url \undefined \def \url#1{\textsf{#1}}\fi
\ifx \bchapter \undefined \def \bchapter#1{#1}\fi
\ifx \bbook \undefined \def \bbook#1{#1}\fi
\ifx \bcomment \undefined \def \bcomment#1{#1}\fi
\ifx \oauthor \undefined \def \oauthor#1{#1}\fi
\ifx \citeauthoryear \undefined \def \citeauthoryear#1{#1}\fi
\ifx \endbibitem  \undefined \def \endbibitem {}\fi
\ifx \bconflocation  \undefined \def \bconflocation#1{#1}\fi
\ifx \arxivurl  \undefined \def \arxivurl#1{\textsf{#1}}\fi
\csname PreBibitemsHook\endcsname

\bibitem[\protect\citeauthoryear{Cundall and
  Strack}{1979}]{cundall1979discrete}
\begin{barticle}
\bauthor{\bsnm{Cundall}, \binits{P.A.}},
\bauthor{\bsnm{Strack}, \binits{O.D.}}:
\batitle{A discrete numerical model for granular assemblies}.
\bjtitle{Geotechnique}
\bvolume{29}(\bissue{1}),
\bfpage{47}--\blpage{65}
(\byear{1979})
\end{barticle}
\endbibitem

\bibitem[\protect\citeauthoryear{P{\"o}schel and
  Schwager}{2005}]{poschelDEM-textbook2005}
\begin{bbook}
\bauthor{\bsnm{P{\"o}schel}, \binits{T.}},
\bauthor{\bsnm{Schwager}, \binits{T.}}:
\bbtitle{Computational Granular Dynamics: Models and Algorithms}.
\bpublisher{Springer},
\blocation{Berlin, Heidelberg}
(\byear{2005})
\end{bbook}
\endbibitem

\bibitem[\protect\citeauthoryear{Lemieux et~al.}{2008}]{lemieux2008large}
\begin{barticle}
\bauthor{\bsnm{Lemieux}, \binits{M.}},
\bauthor{\bsnm{L{\'e}onard}, \binits{G.}},
\bauthor{\bsnm{Doucet}, \binits{J.}},
\bauthor{\bsnm{Leclaire}, \binits{L.-A.}},
\bauthor{\bsnm{Viens}, \binits{F.}},
\bauthor{\bsnm{Chaouki}, \binits{J.}},
\bauthor{\bsnm{Bertrand}, \binits{F.}}:
\batitle{Large-scale numerical investigation of solids mixing in a v-blender
  using the discrete element method}.
\bjtitle{Powder Technology}
\bvolume{181}(\bissue{2}),
\bfpage{205}--\blpage{216}
(\byear{2008})
\end{barticle}
\endbibitem

\bibitem[\protect\citeauthoryear{Apostolou and
  Hrymak}{2008}]{apostolou2008discrete}
\begin{barticle}
\bauthor{\bsnm{Apostolou}, \binits{K.}},
\bauthor{\bsnm{Hrymak}, \binits{A.}}:
\batitle{Discrete element simulation of liquid-particle flows}.
\bjtitle{Computers \& Chemical Engineering}
\bvolume{32}(\bissue{4-5}),
\bfpage{841}--\blpage{856}
(\byear{2008})
\end{barticle}
\endbibitem

\bibitem[\protect\citeauthoryear{Tang et~al.}{2009}]{tang2009tsaoling}
\begin{barticle}
\bauthor{\bsnm{Tang}, \binits{C.-L.}},
\bauthor{\bsnm{Hu}, \binits{J.-C.}},
\bauthor{\bsnm{Lin}, \binits{M.-L.}},
\bauthor{\bsnm{Angelier}, \binits{J.}},
\bauthor{\bsnm{Lu}, \binits{C.-Y.}},
\bauthor{\bsnm{Chan}, \binits{Y.-C.}},
\bauthor{\bsnm{Chu}, \binits{H.-T.}}:
\batitle{{The Tsaoling landslide triggered by the Chi-Chi earthquake, Taiwan:
  insights from a discrete element simulation}}.
\bjtitle{Engineering Geology}
\bvolume{106}(\bissue{1-2}),
\bfpage{1}--\blpage{19}
(\byear{2009})
\end{barticle}
\endbibitem

\bibitem[\protect\citeauthoryear{Salciarini
  et~al.}{2010}]{salciarini2010discrete}
\begin{barticle}
\bauthor{\bsnm{Salciarini}, \binits{D.}},
\bauthor{\bsnm{Tamagnini}, \binits{C.}},
\bauthor{\bsnm{Conversini}, \binits{P.}}:
\batitle{Discrete element modeling of debris-avalanche impact on earthfill
  barriers}.
\bjtitle{Physics and Chemistry of the Earth, Parts A/B/C}
\bvolume{35}(\bissue{3-5}),
\bfpage{172}--\blpage{181}
(\byear{2010})
\end{barticle}
\endbibitem

\bibitem[\protect\citeauthoryear{{O'Sullivan}}{2011}]{OSullivan2011}
\begin{barticle}
\bauthor{\bsnm{{O'Sullivan}}, \binits{C.}}:
\batitle{Particle-based {D}iscrete {E}lement {M}odeling: Geomechanics
  perspective}.
\bjtitle{Int. J. Geomech.}
\bvolume{11}(\bissue{6}),
\bfpage{449}--\blpage{464}
(\byear{2011})
\end{barticle}
\endbibitem

\bibitem[\protect\citeauthoryear{S{\'a}nchez and
  Scheeres}{2011}]{sanchez2011simulating}
\begin{barticle}
\bauthor{\bsnm{S{\'a}nchez}, \binits{P.}},
\bauthor{\bsnm{Scheeres}, \binits{D.J.}}:
\batitle{Simulating asteroid rubble piles with a self-gravitating soft-sphere
  distinct element method model}.
\bjtitle{The Astrophysical Journal}
\bvolume{727}(\bissue{2}),
\bfpage{120}
(\byear{2011})
\end{barticle}
\endbibitem

\bibitem[\protect\citeauthoryear{Foldager et~al.}{2022}]{frederikOleDEM2022}
\begin{barticle}
\bauthor{\bsnm{Foldager}, \binits{F.F.}},
\bauthor{\bsnm{Munkholm}, \binits{L.J.}},
\bauthor{\bsnm{Balling}, \binits{O.}},
\bauthor{\bsnm{Serban}, \binits{R.}},
\bauthor{\bsnm{Negrut}, \binits{D.}},
\bauthor{\bsnm{Heck}, \binits{R.J.}},
\bauthor{\bsnm{Green}, \binits{O.}}:
\batitle{Modeling soil aggregate fracture using the discrete element method}.
\bjtitle{Soil and Tillage Research}
\bvolume{218},
\bfpage{105295}
(\byear{2022})
\end{barticle}
\endbibitem

\bibitem[\protect\citeauthoryear{Recuero
  et~al.}{2017}]{antonioVehicleTireGranMatSim2017}
\begin{barticle}
\bauthor{\bsnm{Recuero}, \binits{A.M.}},
\bauthor{\bsnm{Serban}, \binits{R.}},
\bauthor{\bsnm{Peterson}, \binits{B.}},
\bauthor{\bsnm{Sugiyama}, \binits{H.}},
\bauthor{\bsnm{Jayakumar}, \binits{P.}},
\bauthor{\bsnm{Negrut}, \binits{D.}}:
\batitle{A high-fidelity approach for vehicle mobility simulation: Nonlinear
  finite element tires operating on granular material}.
\bjtitle{Journal of Terramechanics}
\bvolume{72},
\bfpage{39}--\blpage{54}
(\byear{2017})
\doiurl{10.1016/j.jterra.2017.04.002}
\end{barticle}
\endbibitem

\bibitem[\protect\citeauthoryear{Johnson et~al.}{2015}]{iagnema2015}
\begin{barticle}
\bauthor{\bsnm{Johnson}, \binits{J.B.}},
\bauthor{\bsnm{Kulchitsky}, \binits{A.V.}},
\bauthor{\bsnm{Duvoy}, \binits{P.}},
\bauthor{\bsnm{Iagnemma}, \binits{K.}},
\bauthor{\bsnm{Senatore}, \binits{C.}},
\bauthor{\bsnm{Arvidson}, \binits{R.E.}},
\bauthor{\bsnm{Moore}, \binits{J.}}:
\batitle{Discrete element method simulations of {Mars} exploration rover wheel
  performance}.
\bjtitle{Journal of Terramechanics}
\bvolume{62},
\bfpage{31}--\blpage{40}
(\byear{2015})
\end{barticle}
\endbibitem

\bibitem[\protect\citeauthoryear{{OpenMP}}{2021}]{openMP}
\begin{botherref}
\oauthor{\bsnm{{OpenMP}}}:
{Specification Standard 5.2}.
Available online at \url{http://openmp.org/}
(2021)
\end{botherref}
\endbibitem

\bibitem[\protect\citeauthoryear{Amritkar et~al.}{2014}]{amritkar2014efficient}
\begin{barticle}
\bauthor{\bsnm{Amritkar}, \binits{A.}},
\bauthor{\bsnm{Deb}, \binits{S.}},
\bauthor{\bsnm{Tafti}, \binits{D.}}:
\batitle{Efficient parallel cfd-dem simulations using openmp}.
\bjtitle{Journal of Computational Physics}
\bvolume{256},
\bfpage{501}--\blpage{519}
(\byear{2014})
\end{barticle}
\endbibitem

\bibitem[\protect\citeauthoryear{Knuth et~al.}{2012}]{knuth12}
\begin{barticle}
\bauthor{\bsnm{Knuth}, \binits{M.A.}},
\bauthor{\bsnm{Johnson}, \binits{J.}},
\bauthor{\bsnm{Hopkins}, \binits{M.}},
\bauthor{\bsnm{Sullivan}, \binits{R.}},
\bauthor{\bsnm{Moore}, \binits{J.}}:
\batitle{Discrete element modeling of a mars exploration rover wheel in
  granular material}.
\bjtitle{Journal of Terramechanics}
\bvolume{49}(\bissue{1}),
\bfpage{27}--\blpage{36}
(\byear{2012})
\end{barticle}
\endbibitem

\bibitem[\protect\citeauthoryear{{Message Passing Interface
  Forum}}{2012}]{mpi-3.0}
\begin{botherref}
\oauthor{\bsnm{{Message Passing Interface Forum}}}:
{{MPI}: A Message-Passing Interface Standard Version 3.0}.
Chapter author for Collective Communication, Process Topologies, and One Sided
  Communications
(2012)
\end{botherref}
\endbibitem

\bibitem[\protect\citeauthoryear{Yan and Regueiro}{2018}]{yan2018comprehensive}
\begin{barticle}
\bauthor{\bsnm{Yan}, \binits{B.}},
\bauthor{\bsnm{Regueiro}, \binits{R.A.}}:
\batitle{A comprehensive study of mpi parallelism in three-dimensional discrete
  element method (dem) simulation of complex-shaped granular particles}.
\bjtitle{Computational Particle Mechanics}
\bvolume{5}(\bissue{4}),
\bfpage{553}--\blpage{577}
(\byear{2018})
\end{barticle}
\endbibitem

\bibitem[\protect\citeauthoryear{Checkaraou
  et~al.}{2018}]{checkaraou2018hybrid}
\begin{bchapter}
\bauthor{\bsnm{Checkaraou}, \binits{A.W.M.}},
\bauthor{\bsnm{Rousset}, \binits{A.}},
\bauthor{\bsnm{Besseron}, \binits{X.}},
\bauthor{\bsnm{Varrette}, \binits{S.}},
\bauthor{\bsnm{Peters}, \binits{B.}}:
\bctitle{Hybrid mpi+ openmp implementation of extended discrete element
  method}.
In: \bbtitle{2018 30th International Symposium on Computer Architecture and
  High Performance Computing (SBAC-PAD)},
pp. \bfpage{450}--\blpage{457}
(\byear{2018}).
\bcomment{IEEE}
\end{bchapter}
\endbibitem

\bibitem[\protect\citeauthoryear{{LIGGGHTS}}{2013}]{liggghts2013}
\begin{botherref}
\oauthor{\bsnm{{LIGGGHTS}}}:
Open Source Discrete Element Method Particle Simulation Code.
{http://cfdem.dcs-computing.com/?q=OpenSourceDEM}
(2013)
\end{botherref}
\endbibitem

\bibitem[\protect\citeauthoryear{{LAMMPS}}{2013}]{lammpsWebSite}
\begin{botherref}
\oauthor{\bsnm{{LAMMPS}}}:
A Molecular Dynamics Simulator.
{http://lammps.sandia.gov/}
(2013)
\end{botherref}
\endbibitem

\bibitem[\protect\citeauthoryear{}{2023}]{starccm}
\begin{botherref}
Simcenter {STAR-CCM+} software website.
\url{https://plm.sw.siemens.com/en-US/simcenter/fluids-thermal-simulation/star-ccm/}.
Accessed: 2023-09-25
(2023)
\end{botherref}
\endbibitem

\bibitem[\protect\citeauthoryear{Serban et~al.}{2017}]{raduNicDanGVSETS2018}
\begin{bchapter}
\bauthor{\bsnm{Serban}, \binits{R.}},
\bauthor{\bsnm{Olsen}, \binits{N.}},
\bauthor{\bsnm{Negrut}, \binits{D.}}:
\bctitle{High performance computing framework for co-simulation of
  vehicle-terrain interaction}.
In: \bbtitle{NDIA Ground Vehicle Systems Engineering and Technology Symposium}
(\byear{2017})
\end{bchapter}
\endbibitem

\bibitem[\protect\citeauthoryear{Xu et~al.}{2011}]{xu2011quasi}
\begin{barticle}
\bauthor{\bsnm{Xu}, \binits{J.}},
\bauthor{\bsnm{Qi}, \binits{H.}},
\bauthor{\bsnm{Fang}, \binits{X.}},
\bauthor{\bsnm{Lu}, \binits{L.}},
\bauthor{\bsnm{Ge}, \binits{W.}},
\bauthor{\bsnm{Wang}, \binits{X.}},
\bauthor{\bsnm{Xu}, \binits{M.}},
\bauthor{\bsnm{Chen}, \binits{F.}},
\bauthor{\bsnm{He}, \binits{X.}},
\bauthor{\bsnm{Li}, \binits{J.}}:
\batitle{Quasi-real-time simulation of rotating drum using discrete element
  method with parallel gpu computing}.
\bjtitle{Particuology}
\bvolume{9}(\bissue{4}),
\bfpage{446}--\blpage{450}
(\byear{2011})
\end{barticle}
\endbibitem

\bibitem[\protect\citeauthoryear{Govender
  et~al.}{2016}]{govender-BlazeDEMGPU2016}
\begin{barticle}
\bauthor{\bsnm{Govender}, \binits{N.}},
\bauthor{\bsnm{Wilke}, \binits{D.}},
\bauthor{\bsnm{Kok}, \binits{S.}}:
\batitle{{Blaze-DEMGPU}: Modular high performance {DEM} framework for the {GPU}
  architecture}.
\bjtitle{SoftwareX}
\bvolume{5},
\bfpage{62}--\blpage{66}
(\byear{2016})
\end{barticle}
\endbibitem

\bibitem[\protect\citeauthoryear{Gan et~al.}{2016}]{ganDEM-GPU2016}
\begin{barticle}
\bauthor{\bsnm{Gan}, \binits{J.}},
\bauthor{\bsnm{Zhou}, \binits{Z.}},
\bauthor{\bsnm{Yu}, \binits{A.}}:
\batitle{A {GPU}-based {DEM} approach for modeling of particulate systems}.
\bjtitle{Powder Technology}
\bvolume{301},
\bfpage{1172}--\blpage{1182}
(\byear{2016})
\end{barticle}
\endbibitem

\bibitem[\protect\citeauthoryear{He et~al.}{2018}]{he-powderGPU2018}
\begin{barticle}
\bauthor{\bsnm{He}, \binits{Y.}},
\bauthor{\bsnm{Evans}, \binits{T.}},
\bauthor{\bsnm{Yu}, \binits{A.}},
\bauthor{\bsnm{Yang}, \binits{R.}}:
\batitle{A {GPU}-based {DEM} for modeling large scale powder compaction with
  wide size distributions}.
\bjtitle{Powder Technology}
\bvolume{333},
\bfpage{219}--\blpage{228}
(\byear{2018})
\end{barticle}
\endbibitem

\bibitem[\protect\citeauthoryear{Kelly et~al.}{2019}]{conlainBillion2019}
\begin{bchapter}
\bauthor{\bsnm{Kelly}, \binits{C.}},
\bauthor{\bsnm{Olsen}, \binits{N.}},
\bauthor{\bsnm{Vanden~Heuvel}, \binits{C.}},
\bauthor{\bsnm{Serban}, \binits{R.}},
\bauthor{\bsnm{Negrut}, \binits{D.}}:
\bctitle{Towards the democratization of many-body dynamics: {B}illion degree of
  freedom simulation of granular material on commodity hardware}.
In: \bbtitle{Proceeding of the ECCOMAS Multibody Dynamics Conference},
\bconflocation{Duisburg, Germany}
(\byear{2019})
\end{bchapter}
\endbibitem

\bibitem[\protect\citeauthoryear{Iwashita and Oda}{1998}]{iwashitaRolling1998}
\begin{barticle}
\bauthor{\bsnm{Iwashita}, \binits{K.}},
\bauthor{\bsnm{Oda}, \binits{M.}}:
\batitle{Rolling resistance at contacts in simulation of shear band development
  by {DEM}}.
\bjtitle{Journal of Engineering Mechanics}
\bvolume{124}(\bissue{3}),
\bfpage{285}--\blpage{292}
(\byear{1998})
\end{barticle}
\endbibitem

\bibitem[\protect\citeauthoryear{Renzo and Maio}{2004}]{direnzo2004525}
\begin{barticle}
\bauthor{\bsnm{Renzo}, \binits{A.D.}},
\bauthor{\bsnm{Maio}, \binits{F.P.D.}}:
\batitle{Comparison of contact-force models for the simulation of collisions in
  {DEM}-based granular flow codes}.
\bjtitle{Chemical Engineering Science}
\bvolume{59}(\bissue{3}),
\bfpage{525}--\blpage{541}
(\byear{2004})
\end{barticle}
\endbibitem

\bibitem[\protect\citeauthoryear{da~Cruz et~al.}{2005}]{dacruz2005}
\begin{barticle}
\bauthor{\bsnm{Cruz}, \binits{F.}},
\bauthor{\bsnm{Emam}, \binits{S.}},
\bauthor{\bsnm{Prochnow}, \binits{M.}},
\bauthor{\bsnm{Roux}, \binits{J.N.}},
\bauthor{\bsnm{Chevoir}, \binits{F.}}:
\batitle{Rheophysics of dense granular materials: Discrete simulation of plane
  shear flows}.
\bjtitle{Physical Review E}
\bvolume{72},
\bfpage{021309}
(\byear{2005})
\doiurl{10.1103/PhysRevE.72.021309}
\end{barticle}
\endbibitem

\bibitem[\protect\citeauthoryear{Rycroft et~al.}{2006}]{bazant2006}
\begin{botherref}
\oauthor{\bsnm{Rycroft}, \binits{C.H.}},
\oauthor{\bsnm{Grest}, \binits{G.S.}},
\oauthor{\bsnm{Landry}, \binits{J.W.}},
\oauthor{\bsnm{Bazant}, \binits{M.Z.}}:
Analysis of granular flow in a pebble-bed nuclear reactor.
Physical Review E
\textbf{74 021306}
(2006)
\end{botherref}
\endbibitem

\bibitem[\protect\citeauthoryear{Kruggel-Emden et~al.}{2008}]{Emden2008}
\begin{barticle}
\bauthor{\bsnm{Kruggel-Emden}, \binits{H.}},
\bauthor{\bsnm{Sturm}, \binits{M.}},
\bauthor{\bsnm{Wirtz}, \binits{S.}},
\bauthor{\bsnm{Scherer}, \binits{V.}}:
\batitle{Selection of an appropriate time integration scheme for the discrete
  element method (dem)}.
\bjtitle{Computers \& Chemical Engineering}
\bvolume{32}(\bissue{10}),
\bfpage{2263}--\blpage{2279}
(\byear{2008})
\end{barticle}
\endbibitem

\bibitem[\protect\citeauthoryear{Wasfy et~al.}{2014}]{wasfy2014coupled}
\begin{bchapter}
\bauthor{\bsnm{Wasfy}, \binits{T.M.}},
\bauthor{\bsnm{Wasfy}, \binits{H.M.}},
\bauthor{\bsnm{Peters}, \binits{J.M.}}:
\bctitle{Coupled multibody dynamics and discrete element modeling of vehicle
  mobility on cohesive granular terrains}.
In: \bbtitle{ASME 2014 International Design Engineering Technical Conferences
  and Computers and Information in Engineering Conference},
pp. \bfpage{006}--\blpage{1005000610050}
(\byear{2014}).
\bcomment{American Society of Mechanical Engineers}.
\burl{http://proceedings.asmedigitalcollection.asme.org/proceeding.aspx?articleid=2091049}
\end{bchapter}
\endbibitem

\bibitem[\protect\citeauthoryear{Lommen}{2014}]{lommen2014}
\begin{botherref}
\oauthor{\bsnm{Lommen}, \binits{S.}}:
{DEM} speedup: Stiffness effects on behavior of bulk material.
Particuology,
107--112
(2014)
\end{botherref}
\endbibitem

\bibitem[\protect\citeauthoryear{Utili et~al.}{2015}]{Utili2015}
\begin{barticle}
\bauthor{\bsnm{Utili}, \binits{S.}},
\bauthor{\bsnm{Zhao}, \binits{T.}},
\bauthor{\bsnm{Houlsby}, \binits{G.T.}}:
\batitle{3{D} {DEM} investigation of granular column collapse: Evaluation of
  debris motion and its destructive power}.
\bjtitle{Engineering Geology}
\bvolume{186},
\bfpage{3}--\blpage{16}
(\byear{2015})
\end{barticle}
\endbibitem

\bibitem[\protect\citeauthoryear{Potticary et~al.}{2015}]{Potticary2015}
\begin{bchapter}
\bauthor{\bsnm{Potticary}, \binits{M.}},
\bauthor{\bsnm{Zervos}, \binits{A.}},
\bauthor{\bsnm{Harkness}, \binits{J.}}:
\bctitle{An investigation into the effect of particle platyness on the strength
  of granular material using the discrete element method}.
In: \bbtitle{IV International Conference on Particle-based Methods -
  Fundamentals and Applications}
(\byear{2015}).
\burl{https://eprints.soton.ac.uk/394117/1/particles2015.pdf}
\end{bchapter}
\endbibitem

\bibitem[\protect\citeauthoryear{Michael et~al.}{2015}]{michael2015}
\begin{botherref}
\oauthor{\bsnm{Michael}, \binits{M.}},
\oauthor{\bsnm{Vogel}, \binits{F.}},
\oauthor{\bsnm{Peters}, \binits{B.}}:
{DEM}-{FEM} coupling simulations of the interactions between a tire tread and
  granular terrain.
Computer Methods in Applied mechanics and engineering
(2015)
\end{botherref}
\endbibitem

\bibitem[\protect\citeauthoryear{Ciantia et~al.}{2016}]{Ciantia2016}
\begin{barticle}
\bauthor{\bsnm{Ciantia}, \binits{M.}},
\bauthor{\bsnm{Arroyo}, \binits{M.}},
\bauthor{\bsnm{Butlanska}, \binits{J.}},
\bauthor{\bsnm{Gens}, \binits{A.}}:
\batitle{{DEM} modelling of cone penetration tests in a double-porosity
  crushable granular material}.
\bjtitle{Computers and Geotechnics}
\bvolume{73},
\bfpage{109}--\blpage{127}
(\byear{2016})
\end{barticle}
\endbibitem

\bibitem[\protect\citeauthoryear{Zheng and Zang}{2017}]{zheng2017}
\begin{bchapter}
\bauthor{\bsnm{Zheng}, \binits{Z.}},
\bauthor{\bsnm{Zang}, \binits{M.}}:
\bctitle{Numerical simulations of the interactions between a pneumatic tire and
  granular sand by {3D} {DEM}-{FEM}}.
In: \bbtitle{7th International Conference on Discrete Element Methods},
pp. \bfpage{289}--\blpage{300}
(\byear{2017}).
\burl{https://link.springer.com/chapter/10.1007/978-981-10-1926-5_32}
\end{bchapter}
\endbibitem

\bibitem[\protect\citeauthoryear{Parteli and Poschel}{2016}]{parteli2016}
\begin{botherref}
\oauthor{\bsnm{Parteli}, \binits{E.}},
\oauthor{\bsnm{Poschel}, \binits{T.}}:
Particle-based simulation of powder application in additive manufacturing.
Powder Technology,
96--102
(2016)
\end{botherref}
\endbibitem

\bibitem[\protect\citeauthoryear{Kivugo}{2017}]{kivugo2017}
\begin{botherref}
\oauthor{\bsnm{Kivugo}, \binits{R.}}:
Tire-soil interaction for off-road vehicle applications.
Phd,
Politecnico di Milano
(2017).
\url{https://www.politesi.polimi.it/handle/10589/136229}
\end{botherref}
\endbibitem

\bibitem[\protect\citeauthoryear{Calvetti et~al.}{2016}]{calvetti2016}
\begin{botherref}
\oauthor{\bsnm{Calvetti}, \binits{F.}},
\oauthor{\bsnm{Prisco}, \binits{C.}},
\oauthor{\bsnm{Vairaktaris}, \binits{E.}}:
{DEM} assessment of impact forces of dry granular masses on rigid barriers.
Acta Geotechnica
(2016)
\end{botherref}
\endbibitem

\bibitem[\protect\citeauthoryear{Furuichi et~al.}{2018}]{japanDEMlarge2018}
\begin{botherref}
\oauthor{\bsnm{Furuichi}, \binits{M.}},
\oauthor{\bsnm{Nishiura}, \binits{D.}},
\oauthor{\bsnm{Kuwano}, \binits{O.}},
\oauthor{\bsnm{Bauville}, \binits{A.}},
\oauthor{\bsnm{Hori}, \binits{T.}},
\oauthor{\bsnm{Sakaguchi}, \binits{H.}}:
Arcuate stress state in accretionary prisms from real-scale numerical sandbox
  experiments.
Nature Scientific Reports - {\url{www.nature.com/scientificreports/}}
\textbf{8}
(2018)
\end{botherref}
\endbibitem

\bibitem[\protect\citeauthoryear{Henrich et~al.}{2018}]{LAMMPSDNA2018}
\begin{botherref}
\oauthor{\bsnm{Henrich}, \binits{O.}},
\oauthor{\bsnm{Gutierrez~Fosado}, \binits{Y.A.}},
\oauthor{\bsnm{Curk}, \binits{T.}},
\oauthor{\bsnm{Ouldridge}, \binits{T.}}:
Coarse-grained simulation of dna using lammps
(2018)
\end{botherref}
\endbibitem

\bibitem[\protect\citeauthoryear{Dias}{2021}]{dias2021molecular}
\begin{botherref}
\oauthor{\bsnm{Dias}, \binits{C.S.}}:
Molecular dynamics simulations of active matter using {LAMMPS}
(2021)
{\href{https://arxiv.org/abs/2102.10399}{{arXiv:2102.10399}}}
{[cond-mat.soft]}
\end{botherref}
\endbibitem

\bibitem[\protect\citeauthoryear{Li et~al.}{2022}]{MCDEM2022}
\begin{botherref}
\oauthor{\bsnm{Li}, \binits{R.}},
\oauthor{\bsnm{Liu}, \binits{Z.}},
\oauthor{\bsnm{Feng}, \binits{Z.}},
\oauthor{\bsnm{Liang}, \binits{J.}},
\oauthor{\bsnm{Zhang}, \binits{L.-G.}}:
High-fidelity {MC-DEM} modeling and uncertainty analysis of {HTR-PM} first
  criticality.
Frontiers in Energy Research
\textbf{9}
(2022)
\doiurl{10.3389/fenrg.2021.822780}
\end{botherref}
\endbibitem

\bibitem[\protect\citeauthoryear{Razavi et~al.}{2021}]{en14133797}
\begin{botherref}
\oauthor{\bsnm{Razavi}, \binits{F.}},
\oauthor{\bsnm{Komrakova}, \binits{A.}},
\oauthor{\bsnm{Lange}, \binits{C.F.}}:
{CFD–-DEM} simulation of sand-retention mechanisms in slurry flow.
Energies
\textbf{14}(13)
(2021)
\doiurl{10.3390/en14133797}
\end{botherref}
\endbibitem

\bibitem[\protect\citeauthoryear{Fang et~al.}{2021}]{chronoGranular2021}
\begin{botherref}
\oauthor{\bsnm{Fang}, \binits{L.}},
\oauthor{\bsnm{Zhang}, \binits{R.}},
\oauthor{\bsnm{Vanden~Heuvel}, \binits{C.}},
\oauthor{\bsnm{Serban}, \binits{R.}},
\oauthor{\bsnm{Negrut}, \binits{D.}}:
Chrono::{GPU}: An open-source simulation package for granular dynamics using
  the discrete element method.
Processes
\textbf{9}(10)
(2021)
\doiurl{10.3390/pr9101813}
\end{botherref}
\endbibitem

\bibitem[\protect\citeauthoryear{Tasora et~al.}{2016}]{chronoOverview2016}
\begin{bchapter}
\bauthor{\bsnm{Tasora}, \binits{A.}},
\bauthor{\bsnm{Serban}, \binits{R.}},
\bauthor{\bsnm{Mazhar}, \binits{H.}},
\bauthor{\bsnm{Pazouki}, \binits{A.}},
\bauthor{\bsnm{Melanz}, \binits{D.}},
\bauthor{\bsnm{Fleischmann}, \binits{J.}},
\bauthor{\bsnm{Taylor}, \binits{M.}},
\bauthor{\bsnm{Sugiyama}, \binits{H.}},
\bauthor{\bsnm{Negrut}, \binits{D.}}:
\bctitle{{Chrono}: An open source multi-physics dynamics engine}.
In: \beditor{\bsnm{Kozubek}, \binits{T.}} (ed.)
\bbtitle{High Performance Computing in Science and Engineering -- Lecture Notes
  in Computer Science},
pp. \bfpage{19}--\blpage{49}.
\bpublisher{Springer}, \blocation{???}
(\byear{2016})
\end{bchapter}
\endbibitem

\bibitem[\protect\citeauthoryear{Reger et~al.}{2023}]{dem-pbrIdaho2023}
\begin{barticle}
\bauthor{\bsnm{Reger}, \binits{D.}},
\bauthor{\bsnm{Merzari}, \binits{E.}},
\bauthor{\bsnm{Balestra}, \binits{P.}},
\bauthor{\bsnm{Stewart}, \binits{R.}},
\bauthor{\bsnm{Strydom}, \binits{G.}}:
\batitle{Discrete element simulation of pebble bed reactors on graphics
  processing units}.
\bjtitle{Annals of Nuclear Energy}
\bvolume{190},
\bfpage{109896}
(\byear{2023})
\doiurl{10.1016/j.anucene.2023.109896}
\end{barticle}
\endbibitem

\bibitem[\protect\citeauthoryear{Haustein et~al.}{2017}]{HAUSTEIN2017118}
\begin{barticle}
\bauthor{\bsnm{Haustein}, \binits{M.}},
\bauthor{\bsnm{Gladkyy}, \binits{A.}},
\bauthor{\bsnm{Schwarze}, \binits{R.}}:
\batitle{Discrete element modeling of deformable particles in {YADE}}.
\bjtitle{SoftwareX}
\bvolume{6},
\bfpage{118}--\blpage{123}
(\byear{2017})
\doiurl{10.1016/j.softx.2017.05.001}
\end{barticle}
\endbibitem

\bibitem[\protect\citeauthoryear{Romanova et~al.}{2020}]{snowYadeFoam}
\begin{bchapter}
\bauthor{\bsnm{Romanova}, \binits{D.}},
\bauthor{\bsnm{Strijhak}, \binits{S.}},
\bauthor{\bsnm{Kraposhin}, \binits{M.}}:
\bctitle{Development of {snowYadeFoam} solver for snow particles simulation}.
In: \bbtitle{2020 Ivannikov Ispras Open Conference (ISPRAS)},
pp. \bfpage{166}--\blpage{169}
(\byear{2020}).
\doiurl{10.1109/ISPRAS51486.2020.00032}
\end{bchapter}
\endbibitem

\bibitem[\protect\citeauthoryear{Ericson}{2005}]{Ericson05}
\begin{bbook}
\bauthor{\bsnm{Ericson}, \binits{C.}}:
\bbtitle{Real Time Collision Detection}.
\bpublisher{Morgan Kaufmann},
\blocation{San Francisco, CA}
(\byear{2005})
\end{bbook}
\endbibitem

\bibitem[\protect\citeauthoryear{Favier et~al.}{1999}]{favierClumpsSpheres1999}
\begin{botherref}
\oauthor{\bsnm{Favier}, \binits{J.}},
\oauthor{\bsnm{Abbaspour-Fard}, \binits{M.}},
\oauthor{\bsnm{Kremmer}, \binits{M.}},
\oauthor{\bsnm{Raji}, \binits{A.}}:
Shape representation of axi-symmetrical, non-spherical particles in discrete
  element simulation using multi-element model particles.
Engineering computations
(1999)
\end{botherref}
\endbibitem

\bibitem[\protect\citeauthoryear{Hilton and
  Cleary}{2011}]{particleShapeCleary2011}
\begin{barticle}
\bauthor{\bsnm{Hilton}, \binits{J.}},
\bauthor{\bsnm{Cleary}, \binits{P.}}:
\batitle{The influence of particle shape on flow modes in pneumatic conveying}.
\bjtitle{Chemical engineering science}
\bvolume{66}(\bissue{3}),
\bfpage{231}--\blpage{240}
(\byear{2011})
\end{barticle}
\endbibitem

\bibitem[\protect\citeauthoryear{Kiangi et~al.}{2013}]{particleShapeKiangi2013}
\begin{barticle}
\bauthor{\bsnm{Kiangi}, \binits{K.}},
\bauthor{\bsnm{Potapov}, \binits{A.}},
\bauthor{\bsnm{Moys}, \binits{M.}}:
\batitle{{DEM} validation of media shape effects on the load behaviour and
  power in a dry pilot mill}.
\bjtitle{Minerals Engineering}
\bvolume{46},
\bfpage{52}--\blpage{59}
(\byear{2013})
\end{barticle}
\endbibitem

\bibitem[\protect\citeauthoryear{Ren et~al.}{2013}]{unionOf2Spheres2013}
\begin{barticle}
\bauthor{\bsnm{Ren}, \binits{B.}},
\bauthor{\bsnm{Zhong}, \binits{W.}},
\bauthor{\bsnm{Jin}, \binits{B.}},
\bauthor{\bsnm{Shao}, \binits{Y.}},
\bauthor{\bsnm{Yuan}, \binits{Z.}}:
\batitle{Numerical simulation on the mixing behavior of corn-shaped particles
  in a spouted bed}.
\bjtitle{Powder technology}
\bvolume{234},
\bfpage{58}--\blpage{66}
(\byear{2013})
\end{barticle}
\endbibitem

\bibitem[\protect\citeauthoryear{Zhong
  et~al.}{2016}]{dem-cfdMonashAustralia2016}
\begin{barticle}
\bauthor{\bsnm{Zhong}, \binits{W.}},
\bauthor{\bsnm{Yu}, \binits{A.}},
\bauthor{\bsnm{Liu}, \binits{X.}},
\bauthor{\bsnm{Tong}, \binits{Z.}},
\bauthor{\bsnm{Zhang}, \binits{H.}}:
\batitle{{DEM/CFD-DEM} modelling of non-spherical particulate systems:
  theoretical developments and applications}.
\bjtitle{Powder technology}
\bvolume{302},
\bfpage{108}--\blpage{152}
(\byear{2016})
\end{barticle}
\endbibitem

\bibitem[\protect\citeauthoryear{Kawamoto et~al.}{2018}]{andradeDEM2018}
\begin{barticle}
\bauthor{\bsnm{Kawamoto}, \binits{R.}},
\bauthor{\bsnm{And{\`o}}, \binits{E.}},
\bauthor{\bsnm{Viggiani}, \binits{G.}},
\bauthor{\bsnm{Andrade}, \binits{J.E.}}:
\batitle{All you need is shape: Predicting shear banding in sand with ls-dem}.
\bjtitle{Journal of the Mechanics and Physics of Solids}
\bvolume{111},
\bfpage{375}--\blpage{392}
(\byear{2018})
\end{barticle}
\endbibitem

\bibitem[\protect\citeauthoryear{Marteau and
  Andrade}{2021}]{marteauExperimental2021}
\begin{botherref}
\oauthor{\bsnm{Marteau}, \binits{E.}},
\oauthor{\bsnm{Andrade}, \binits{J.E.}}:
An experimental study of the effect of particle shape on force transmission and
  mobilized strength of granular materials.
Journal of Applied Mechanics
\textbf{88}(11)
(2021)
\end{botherref}
\endbibitem

\bibitem[\protect\citeauthoryear{Zhang et~al.}{2022}]{RuochunDEMERepo}
\begin{botherref}
\oauthor{\bsnm{Zhang}, \binits{R.}},
\oauthor{\bsnm{Vanden~Heuvel}, \binits{C.}},
\oauthor{\bsnm{Negrut}, \binits{D.}}:
{DEM-Engine}, a multi-GPU {DEM} solver with complex geometry support.
\url{https://github.com/projectchrono/DEM-Engine}.
Simulation-Based Engineering Laboratory, University of Wisconsin-Madison
(2022)
\end{botherref}
\endbibitem

\bibitem[\protect\citeauthoryear{Mazhar et~al.}{2011}]{hammadTobyDan2012}
\begin{barticle}
\bauthor{\bsnm{Mazhar}, \binits{H.}},
\bauthor{\bsnm{Heyn}, \binits{T.}},
\bauthor{\bsnm{Negrut}, \binits{D.}}:
\batitle{A scalable parallel method for large collision detection problems}.
\bjtitle{Multibody System Dynamics}
\bvolume{26},
\bfpage{37}--\blpage{55}
(\byear{2011}).
\bcomment{10.1007/s11044-011-9246-y}
\end{barticle}
\endbibitem

\bibitem[\protect\citeauthoryear{Barsdell and Clark}{}]{NVRTC}
\begin{botherref}
\oauthor{\bsnm{Barsdell}, \binits{B.}},
\oauthor{\bsnm{Clark}, \binits{K.}}:
A single-header C++ library for simplifying the use of CUDA Runtime
  Compilation.
\url{https://github.com/NVIDIA/jitify}.
Accessed: 2023-08-24
\end{botherref}
\endbibitem

\bibitem[\protect\citeauthoryear{Berry et~al.}{2023}]{BERRY2023118209}
\begin{barticle}
\bauthor{\bsnm{Berry}, \binits{N.}},
\bauthor{\bsnm{Zhang}, \binits{Y.}},
\bauthor{\bsnm{Haeri}, \binits{S.}}:
\batitle{Contact models for the multi-sphere discrete element method}.
\bjtitle{Powder Technology}
\bvolume{416},
\bfpage{118209}
(\byear{2023})
\doiurl{10.1016/j.powtec.2022.118209}
\end{barticle}
\endbibitem

\bibitem[\protect\citeauthoryear{Coetzee and Scheffler}{2023}]{pr11010005}
\begin{botherref}
\oauthor{\bsnm{Coetzee}, \binits{C.J.}},
\oauthor{\bsnm{Scheffler}, \binits{O.C.}}:
Review: The calibration of dem parameters for the bulk modelling of cohesive
  materials.
Processes
\textbf{11}(1)
(2023)
\doiurl{10.3390/pr11010005}
\end{botherref}
\endbibitem

\bibitem[\protect\citeauthoryear{Price et~al.}{2007}]{clumpSpheresPrice2007}
\begin{botherref}
\oauthor{\bsnm{Price}, \binits{M.}},
\oauthor{\bsnm{Murariu}, \binits{V.}},
\oauthor{\bsnm{Morrison}, \binits{G.}}:
Sphere clump generation and trajectory comparison for real particles.
Proceedings of Discrete Element Modelling 2007
(2007)
\end{botherref}
\endbibitem

\bibitem[\protect\citeauthoryear{Hertz}{1882}]{hertz1882}
\begin{barticle}
\bauthor{\bsnm{Hertz}, \binits{H.}}:
\batitle{Ueber die verdunstung der fl{\"u}ssigkeiten, insbesondere des
  quecksilbers, im luftleeren raume}.
\bjtitle{Annalen der Physik}
\bvolume{253}(\bissue{10}),
\bfpage{177}--\blpage{193}
(\byear{1882})
\doiurl{10.1002/andp.18822531002}
{\href{https://arxiv.org/abs/https://onlinelibrary.wiley.com/doi/pdf/10.1002/andp.18822531002}{{https://onlinelibrary.wiley.com/doi/pdf/10.1002/andp.18822531002}}}
\end{barticle}
\endbibitem

\bibitem[\protect\citeauthoryear{Mindlin and Deresiewicz}{1953}]{mindlin53}
\begin{barticle}
\bauthor{\bsnm{Mindlin}, \binits{R.}},
\bauthor{\bsnm{Deresiewicz}, \binits{H.}}:
\batitle{Elastic spheres in contact under varying oblique forces}.
\bjtitle{Journal of Applied Mechanics}
\bvolume{20},
\bfpage{327}--\blpage{344}
(\byear{1953})
\end{barticle}
\endbibitem

\bibitem[\protect\citeauthoryear{Fang and Negrut}{2021}]{luningFricModel2021}
\begin{barticle}
\bauthor{\bsnm{Fang}, \binits{L.}},
\bauthor{\bsnm{Negrut}, \binits{D.}}:
\batitle{Producing {3D} friction loads by tracking the motion of the contact
  point on bodies in mutual contact}.
\bjtitle{Computational Particle Mechanics}
\bvolume{8},
\bfpage{905}--\blpage{929}
(\byear{2021})
\doiurl{10.1007/s40571-020-00376-9}
\end{barticle}
\endbibitem

\bibitem[\protect\citeauthoryear{Fleischmann et~al.}{2016}]{jonJCND2015}
\begin{barticle}
\bauthor{\bsnm{Fleischmann}, \binits{J.}},
\bauthor{\bsnm{Serban}, \binits{R.}},
\bauthor{\bsnm{Negrut}, \binits{D.}},
\bauthor{\bsnm{Jayakumar}, \binits{P.}}:
\batitle{On the importance of displacement history in soft-body contact
  models}.
\bjtitle{Journal of Computational and Nonlinear Dynamics}
\bvolume{11}(\bissue{4}),
\bfpage{044502}
(\byear{2016})
\end{barticle}
\endbibitem

\bibitem[\protect\citeauthoryear{Johnson}{1987}]{johnson1987contact}
\begin{bbook}
\bauthor{\bsnm{Johnson}, \binits{K.L.}}:
\bbtitle{Contact Mechanics}.
\bpublisher{{Cambridge University Press}}, \blocation{???}
(\byear{1987})
\end{bbook}
\endbibitem

\bibitem[\protect\citeauthoryear{Ambroso et~al.}{2005}]{durian2005}
\begin{barticle}
\bauthor{\bsnm{Ambroso}, \binits{M.A.}},
\bauthor{\bsnm{Santore}, \binits{C.R.}},
\bauthor{\bsnm{Abate}, \binits{A.R.}},
\bauthor{\bsnm{Durian}, \binits{D.J.}}:
\batitle{Penetration depth for shallow impact cratering}.
\bjtitle{Physical Review E}
\bvolume{71},
\bfpage{051305}
(\byear{2005})
\doiurl{10.1103/PhysRevE.71.051305}
\end{barticle}
\endbibitem

\bibitem[\protect\citeauthoryear{Heyn}{2013}]{heynPhDThesis2013}
\begin{botherref}
\oauthor{\bsnm{Heyn}, \binits{T.}}:
On the modeling, simulation, and visualization of many-body dynamics problems
  with friction and contact.
{PhD} thesis,
{Department of Mechanical Engineering, University of Wisconsin--Madison},
  \url{http://sbel.wisc.edu/documents/TobyHeynThesis_PhDfinal.pdf}
(2013)
\end{botherref}
\endbibitem

\bibitem[\protect\citeauthoryear{Cui et~al.}{2023}]{cui2023superDEM}
\begin{barticle}
\bauthor{\bsnm{Cui}, \binits{X.}},
\bauthor{\bsnm{Dai}, \binits{J.}},
\bauthor{\bsnm{Xu}, \binits{H.}},
\bauthor{\bsnm{Gao}, \binits{X.}}:
\batitle{Superdem simulation and experiment validation of nonspherical
  particles flows in a rotating drum}.
\bjtitle{Industrial and Engineering Chemistry Research}
\bvolume{62}(\bissue{16}),
\bfpage{6525}--\blpage{6535}
(\byear{2023})
\doiurl{10.1021/acs.iecr.3c00919}
\end{barticle}
\endbibitem

\bibitem[\protect\citeauthoryear{Gao
  et~al.}{2022}]{gao2022superDEMimplementation}
\begin{barticle}
\bauthor{\bsnm{Gao}, \binits{X.}},
\bauthor{\bsnm{Yu}, \binits{J.}},
\bauthor{\bsnm{Portal}, \binits{R.J.F.}},
\bauthor{\bsnm{Dietiker}, \binits{J.-F.}},
\bauthor{\bsnm{Shahnam}, \binits{M.}},
\bauthor{\bsnm{Rogers}, \binits{W.A.}}:
\batitle{Development and validation of superdem for non-spherical particulate
  systems using a superquadric particle method}.
\bjtitle{Particuology}
\bvolume{61},
\bfpage{74}--\blpage{90}
(\byear{2022})
\doiurl{10.1016/j.partic.2020.11.007}
\end{barticle}
\endbibitem

\bibitem[\protect\citeauthoryear{Jian and Gao}{2023}]{jian2023shapeDEM}
\begin{barticle}
\bauthor{\bsnm{Jian}, \binits{B.}},
\bauthor{\bsnm{Gao}, \binits{X.}}:
\batitle{Investigation of spherical and non-spherical binary particles flow
  characteristics in a discharge hopper}.
\bjtitle{Advanced Powder Technology}
\bvolume{34}(\bissue{5}),
\bfpage{104011}
(\byear{2023})
\doiurl{10.1016/j.apt.2023.104011}
\end{barticle}
\endbibitem

\bibitem[\protect\citeauthoryear{Guo and
  Curtis}{2015}]{guo2015reviewComplexFlows}
\begin{barticle}
\bauthor{\bsnm{Guo}, \binits{Y.}},
\bauthor{\bsnm{Curtis}, \binits{J.S.}}:
\batitle{Discrete element method simulations for complex granular flows}.
\bjtitle{Annual Review of Fluid Mechanics}
\bvolume{47}(\bissue{1}),
\bfpage{21}--\blpage{46}
(\byear{2015})
\doiurl{10.1146/annurev-fluid-010814-014644}
\end{barticle}
\endbibitem

\bibitem[\protect\citeauthoryear{Scholtès and
  Donzé}{2013}]{scholtes2013DEMforFailure}
\begin{barticle}
\bauthor{\bsnm{Scholtès}, \binits{L.}},
\bauthor{\bsnm{Donzé}, \binits{F.-V.}}:
\batitle{A dem model for soft and hard rocks: Role of grain interlocking on
  strength}.
\bjtitle{Journal of the Mechanics and Physics of Solids}
\bvolume{61}(\bissue{2}),
\bfpage{352}--\blpage{369}
(\byear{2013})
\doiurl{10.1016/j.jmps.2012.10.005}
\end{barticle}
\endbibitem

\bibitem[\protect\citeauthoryear{Belheine
  et~al.}{2009}]{belheine2009DEMwithBending}
\begin{barticle}
\bauthor{\bsnm{Belheine}, \binits{N.}},
\bauthor{\bsnm{Plassiard}, \binits{J.-P.}},
\bauthor{\bsnm{Donzé}, \binits{F.-V.}},
\bauthor{\bsnm{Darve}, \binits{F.}},
\bauthor{\bsnm{Seridi}, \binits{A.}}:
\batitle{Numerical simulation of drained triaxial test using 3d discrete
  element modeling}.
\bjtitle{Computers and Geotechnics}
\bvolume{36}(\bissue{1}),
\bfpage{320}--\blpage{331}
(\byear{2009})
\doiurl{10.1016/j.compgeo.2008.02.003}
\end{barticle}
\endbibitem

\bibitem[\protect\citeauthoryear{Liu et~al.}{2020}]{liu2020DEMFailureBallast}
\begin{barticle}
\bauthor{\bsnm{Liu}, \binits{G.-Y.}},
\bauthor{\bsnm{Xu}, \binits{W.-J.}},
\bauthor{\bsnm{Sun}, \binits{Q.-C.}},
\bauthor{\bsnm{Govender}, \binits{N.}}:
\batitle{Study on the particle breakage of ballast based on a gpu accelerated
  discrete element method}.
\bjtitle{Geoscience Frontiers}
\bvolume{11}(\bissue{2}),
\bfpage{461}--\blpage{471}
(\byear{2020})
\doiurl{10.1016/j.gsf.2019.06.006}
\end{barticle}
\endbibitem

\bibitem[\protect\citeauthoryear{Potyondy and
  Cundall}{2004}]{Potyondy2004DEMforRock}
\begin{barticle}
\bauthor{\bsnm{Potyondy}, \binits{D.O.}},
\bauthor{\bsnm{Cundall}, \binits{P.A.}}:
\batitle{A bonded-particle model for rock}.
\bjtitle{International Journal of Rock Mechanics and Mining Sciences}
\bvolume{41}(\bissue{8 SPEC.ISS.}),
\bfpage{1329}--\blpage{1364}
(\byear{2004})
\doiurl{10.1016/j.ijrmms.2004.09.011}
\end{barticle}
\endbibitem

\bibitem[\protect\citeauthoryear{Wang and Tonon}{2009}]{wang2009graniteDEM}
\begin{barticle}
\bauthor{\bsnm{Wang}, \binits{Y.}},
\bauthor{\bsnm{Tonon}, \binits{F.}}:
\batitle{Modeling lac du bonnet granite using a discrete element model}.
\bjtitle{International Journal of Rock Mechanics and Mining Sciences}
\bvolume{46}(\bissue{7}),
\bfpage{1124}--\blpage{1135}
(\byear{2009})
\doiurl{10.1016/j.ijrmms.2009.05.008}
\end{barticle}
\endbibitem

\bibitem[\protect\citeauthoryear{Zhang et~al.}{2023}]{zhang2023gpuaccelerated}
\begin{botherref}
\oauthor{\bsnm{Zhang}, \binits{R.}},
\oauthor{\bsnm{Heuvel}, \binits{C.V.}},
\oauthor{\bsnm{Schepelmann}, \binits{A.}},
\oauthor{\bsnm{Rogg}, \binits{A.}},
\oauthor{\bsnm{Apostolopoulos}, \binits{D.}},
\oauthor{\bsnm{Chandler}, \binits{S.}},
\oauthor{\bsnm{Serban}, \binits{R.}},
\oauthor{\bsnm{Negrut}, \binits{D.}}:
A GPU-accelerated Simulator for the DEM Analysis of Granular Systems Composed
  of Clump-shaped Elements
\end{botherref}
\endbibitem

\bibitem[\protect\citeauthoryear{\relax{Simulated Lunar Operations
  Laboratory}}{}]{movieMGRU3Tilt}
\begin{botherref}
\oauthor{\bsnm{\relax{Simulated Lunar Operations Laboratory}}}:
{NASA}'s {VIPER} Prototype Motors Through Moon-like Obstacle Course.
\url{https://www.nasa.gov/feature/ames/nasas-viper-prototype-motors-through-moon-like-obstacle-course}.
Accessed: 2023-04-02
\end{botherref}
\endbibitem

\bibitem[\protect\citeauthoryear{Oravec et~al.}{2010}]{ORAVEC2010361}
\begin{barticle}
\bauthor{\bsnm{Oravec}, \binits{H.A.}},
\bauthor{\bsnm{Zeng}, \binits{X.}},
\bauthor{\bsnm{Asnani}, \binits{V.M.}}:
\batitle{Design and characterization of {GRC}-1: A soil for lunar
  terramechanics testing in earth-ambient conditions}.
\bjtitle{Journal of Terramechanics}
\bvolume{47}(\bissue{6}),
\bfpage{361}--\blpage{377}
(\byear{2010})
\doiurl{10.1016/j.jterra.2010.04.006}
\end{barticle}
\endbibitem

\bibitem[\protect\citeauthoryear{}{}]{chatgpt}
\begin{botherref}
{OpenAI (2023), ChatGPT (Sep 25 version)}.
\url{https://chat.openai.com}
\end{botherref}
\endbibitem

\end{thebibliography}

\end{document}